\documentclass[useAMS,usenatbib]{mn2e}
\usepackage[utf8]{inputenc}

\usepackage{threeparttable,verbatim}
\usepackage[normalem]{ulem}
\usepackage[]{txfonts}
\usepackage{makecell}
\usepackage{cases}
\usepackage{color}
\usepackage[colorlinks,urlcolor=blue,citecolor=blue,linkcolor=blue]{hyperref}
\newcommand{\unit}[1]{\nobreak{\mathrm{\;#1}}} % For units of measure within math mode
\newcommand{\eqb}{\begin{eqnarray}}
\newcommand{\eqe}{\end{eqnarray}}

\newcommand{\eqn}[1]{eq.~(\ref{eq:#1})}
\newcommand{\fig}[1]{Fig.~\ref{fig:#1}}
\newcommand{\sect}[1]{Sect.~\ref{sec:#1}}
\newcommand{\app}[1]{Appendix \ref{sec:#1}}

\newcommand{\mel}{m_{\rm e}}
\newcommand{\mpr}{m_{\rm p}}

\newcommand{\af}{\alpha_{\rm f}}
\newcommand{\kep}{K_{\rm ep}}
\newcommand{\funp}{f_{\rm p}}
\newcommand{\fune}{f_{\rm e}}
\newcommand{\sth}{\sigma_T}
\newcommand{\spp}{\sigma_{\rm pp}}
\newcommand{\taut}{\tau_T}
\newcommand{\Bcr}{B_{\rm cr}}
\newcommand{\eB}{\epsilon_{\rm B}}
\newcommand{\eph}{\epsilon_{\rm ph}}
\newcommand{\ep}{\epsilon_{\rm p}}
\newcommand{\gpmin}{\gamma_{\rm p, m}}
\newcommand{\gemin}{\gamma_{\rm e, m}}

\newcommand{\gpmax}{\gamma_{\rm p, M}}
\newcommand{\gemax}{\gamma_{\rm e, M}}
\newcommand{\gmax}{\gamma_{\rm i,M}}
\newcommand{\gmin}{\gamma_{\rm i,m}}

\newcommand{\gcool}{\gamma_{\rm c}}
\newcommand{\vcool}{\nu_{\rm c}}

\newcommand{\rin}{r_{\rm in}}
\newcommand{\rbo}{r_{\rm bo}}
\newcommand{\rw}{r_{\rm w}}
\newcommand{\rd}{r_{\rm dec}}
\newcommand{\nw}{n_{\rm csm}}
\newcommand{\Aw}{A_{\rm w}}
\newcommand{\vw}{\upsilon_{\rm w}}
\newcommand{\vs}{\upsilon_{\rm s}}
\newcommand{\vo}{\upsilon_0}
\newcommand{\Msw}{M_{\rm sw}}
\newcommand{\Mej}{M_{\rm ej}}
\newcommand{\Qepp}{Q_{\rm e}^{(\rm pp)}}
\newcommand{\Nepp}{N_{\rm e}^{(\rm pp)}}
\newcommand{\Qe}{Q_{\rm e}}
\newcommand{\Qp}{Q_{\rm p}}
\newcommand{\qsyn}{q_{\rm syn}}
\newcommand{\qic}{q_{\rm ic}}

\newcommand{\pp}{{\sl pp\ }}
\newcommand{\ab}{a_{\rm B}}
\newcommand{\aph}{a_{\rm ph}}
\newcommand{\dpp}{q_{\rm pp}}
\usepackage[pdftex]{graphicx}
\usepackage{epsfig} 
\usepackage{epstopdf}
\title[Radio emission from pp collisions in SNe]{Radio synchrotron emission from secondary electrons in interaction-powered supernovae}
\author[Petropoulou, Kamble, Sironi]{M. Petropoulou$^{1}$\thanks{E-mail:
mpetropo@purdue.edu}\thanks{Einstein Postdoctoral Fellow}, A. Kamble$^{2}$ and L. Sironi$^{2}$\\
$^{1}$Department of Physics and Astronomy, Purdue University, 525 Northwestern
Avenue, West Lafayette, IN 47907, USA\\
$^{2}$ Harvard-Smithsonian Center for Astrophysics, 
60 Garden Street, Cambridge, MA 02138, USA}

\begin{document}

\maketitle
\begin{abstract}
Several supernovae (SNe) with an unusually dense circumstellar medium (CSM) have been recently observed at radio frequencies. Their radio emission is powered by relativistic electrons that can be either accelerated  at the SN shock (primaries) or injected as a by-product (secondaries) of inelastic proton-proton collisions. We investigate the radio signatures from secondary electrons, by detailing a semi-analytical model to calculate the temporal evolution of the distributions of protons, primary and secondary electrons.  With our formalism, we track the cooling history of all the particles that have been injected into the emission region up to a given time, and calculate the resulting radio spectra and light curves. For a SN shock propagating through the progenitor wind,  we find that secondary electrons control the early radio signatures, but their contribution decays faster than that of primary electrons.  This results in a { flattening} of the light curve at a given radio frequency that depends only 
upon the radial profiles of the CSM density and of the shock velocity, $\vo$. The relevant transition time { at the peak frequency} is $\sim {  190} \, {\rm d} \, K_{\rm ep,-3}^{-1} A_{\rm w,   16}{/  \beta_{0, -1.5}^2}$,  where $\Aw$ is the wind mass-loading parameter, $\beta_0=\vo/c$ and $K_{\rm ep}$ is the electron-to-proton ratio of accelerated particles.  We explicitly show that late peak times at 5~GHz  (i.e., $t_{\rm pk}\gtrsim300-1000$~d) suggest a
% fast ($\vo = 9\times 10^3 -3\times 10^4$~km s$^{-1}$) 
shock wave propagating in a dense wind ($\Aw \gtrsim 10^{16}-10^{17}$ gr cm$^{-1}$), where secondary electrons are likely to power the observed peak emission. 
\end{abstract}

\begin{keywords}
astroparticle physics  -- supernovae: general -- radiation mechanisms: non-thermal -- shock waves 
\end{keywords}

\section{Introduction}
\label{sec:intro}
The recent advance in wide-field surveys searching for optical transients (e.g. Palomar Transient Factory\footnote{http://www.ptf.caltech.edu/}, All Sky Automated Survey for Supernovae\footnote{http://www.astronomy.ohio-state.edu/~assassin/index.shtml}) has revealed a whole new class of supernovae (SNe) with atypical light curves and spectra that are often prominent in the first hours to days following the explosion \citep{Kasliwal2010,Drout2013}. Members of this class are super-luminous supernovae (SLSNe), which are at least ten times brighter (i.e., with peak luminosities $\gtrsim7\times10^{43}$~erg s$^{-1}$)  than typical ones  \citep{Gal-Yam2012}, but also normal-luminosity SNe with an unusually dense circumstellar medium (CSM), such as type IIn SNe \citep{Kiewe2012}. Several candidates for such SNe, that are powered by  the interaction with a dense CSM (hence classified as ``interaction-powered SNe''), were recently 
found \citep[][]{Gal-Yam2009, Quimby2011, Quimby2013, ofek14, Dong2015}\footnote{A couple of SLSNe were also suggested to be powered by interactions with a dense CSM \citep[e.g.][]{Quimby2011, chevalier_irwin11}}.

In typical SNe the first electromagnetic signal following the explosion emerges, as an X-ray flash, when the shock breaks out from the stellar surface, thus probing the properties of the progenitor star \citep[e.g.][]{klein_chevalier78, katzbudnik10, nakarsari10}. In interaction-powered SNe, the CSM is typically so dense as to be optically thick to radiation. Thus, a radiation-mediated shock propagates into the CSM and the shock breakout happens in the dense CSM, rather than in the progenitor atmosphere \citep{katz11, murase11}. In this case, the shock breakout signature carries information about the mass-loss history of the progenitor star prior to its explosion \citep[e.g.][]{chevalier_irwin11}.

Radio emission in typical SNe is believed to be powered by synchrotron radiation of relativistic electrons
accelerated at the SN shock wave (see e.g. \citealt{Chevalier1982,Chevalier1998}). In the case of interaction-powered SNe, where the shock is initially radiation-mediated \citep{weaver76}, particle acceleration at the shock front is suppressed at early times. After the shock breakout, a collisionless shock is formed \citep{katz11, murase11},  thus allowing for particle acceleration\footnote{The formation of a collisionless shock before the shock breakout and its implications on cosmic-ray acceleration have been investigated by \cite{giacinti_bell15}.}. Radio emission from accelerated electrons is routinely used as a probe of the immediate SN environment and of the particle acceleration efficiency \citep{Chevalier1998,Soderberg2005,Kamble2015}.
Indeed, several dozens of interaction-powered SNe have been detected at radio frequencies, displaying a wide distribution in luminosities due to the dispersion in their shock and CSM properties.

Together with electrons, protons (or ions, in general) will also be accelerated at the shock front \citep[e.g.][]{bell78, blandford_ostriker78}. Indeed, proton acceleration at SN remnant shocks has been invoked  to explain the production of Galactic cosmic-rays  (CRs) with energies up to  $\sim$few PeV
 \citep[see][for a review]{Bell13, blasi2013}. The possibility of CR acceleration beyond PeV energies in interaction-powered SNe has been recently addressed by \citet{katz11, murase14, cardillo2015, zirakashvili15}. The presence of CR protons in  dense environments  may lead to interesting multi-messenger signatures, such as GeV $\gamma$-ray emission and high-energy ($\sim100$ TeV) neutrino production, as a result of inelastic \pp collisions with the non-relativistic protons of the shocked CSM.  Although the smoking gun for CR acceleration in interaction-powered SNe would be the detection of high-energy neutrinos from this new class of SNe, a firm association of the IceCube neutrinos \citep{icecube13, aartsen14_3yr, aartsen14_4yr} with one (or more) astrophysical candidate classes of sources is still lacking \citep{padovaniresconi14,KopperICRC2015, aartsen15}. 
 Alternatively, the detection of photon signatures that are characteristic of \pp collisions would suggest that CR acceleration is at work in these sources.
%  Secondary electrons produced by the decay of charged pions may also leave their imprint on the multi-wavelength emission from interaction-powered SNe. 
 As we argue below, radio synchrotron emission from secondary electrons, which are produced in the  decay chain of charged pions, can be a valuable tool for identifying the signatures of proton acceleration. 

Recently, \cite{murase14} have argued that secondary electrons can emit detectable synchrotron radiation at high radio frequencies ($\gtrsim 100$~GHz)  or even at  far-infrared wavelengths, by deriving order-of-magnitude estimates of the peak luminosities and frequencies. Given the importance of radio observations as an indirect probe of CR acceleration in interaction-powered SNe, detailed model predictions on the radio emission are crucial. In this paper, we  present detailed semi-analytical calculations of radio light curves and spectra of synchrotron emission from primary and secondary electrons. 
For this, we adopt a semi-analytical formalism and calculate the temporal evolution of the non-thermal particle distributions that are contained in a thin shell of shocked CSM.  We follow the evolution of three species: protons and primary electrons, which are assumed to be accelerated at the SN forward shock, as well as secondary relativistic electrons, which result from inelastic \pp collisions of the shock-accelerated ions with the non-relativistic protons of the shocked CSM. 

Our analysis allows for the derivation of analytical expressions for various quantities that can serve as radio diagnostics, such as
the power-law decay slope of primary and secondary electron synchrotron light curves,  the peak synchrotron luminosities and the relevant break frequencies. We show that the peak time of the light curve, at a given radio frequency, is an important probe of the secondary electron contribution to the observed emission. In general, we find that early peak times ($\lesssim 100$~d) imply a dilute stellar wind and primary-dominated synchrotron emission. In contrast, late peak times  (i.e., $\gtrsim300-1000$~d) suggest a fast ($\vo = 9\times 10^3 -3\times 10^4$~km s$^{-1}$) shock wave propagating in a dense medium where secondary electrons are likely to power the peak flux. We also show that the transition from secondary-dominated to primary-dominated synchrotron emission is denoted by a change in the decay slope of the light curve. The {  flattening} in the decay slope depends only on the radial profiles of the CSM and of the shock velocity. For the specific case of a wind-like CSM, we  explicitly show that the 
transition 
time depends only on the mass-loading 
parameter, $\Aw$, and on the electron-to-proton ratio, $K_{\rm ep}$  as $t \propto K_{\rm ep}^{-1} A_{\rm w}$. 

This paper is structured as follows. In \sect{model} we describe the model under consideration. In \sect{evolution} we detail 
the semi-analytical formalism used to solve for the evolution of the non-thermal particle distributions with time, and continue in \sect{distributions} with the presentation of an indicative example. We focus on the radio synchrotron emission in \sect{synchrotron}, where we derive analytical expressions for various quantities that may serve as radio diagnostics. We present numerical results of synchrotron spectra and light curves, while discussing the effect
of various model parameters in \sect{numerical}. In \sect{radio} we demonstrate the relevance of our results to SN radio observations and discuss our results and the model predictions in \sect{discussion}. We conclude in \sect{summary}  with a  summary of our results. 
The  reader interested primarily in the astrophysical implications of our findings might want to skip the technical paragraphs (Sections~\ref{sec:evolution} and \ref{sec:distributions}) and move directly to \sect{synchrotron}.

Throughout this work we use the notation $G_x=G/10^{x}$ in cgs units, except for
the mass loss rate that is measured in $M_\odot\unit{/yr}$ and the masses of the SN ejecta  and CSM that are normalized to $M_\odot$. For the required transformations between the reference systems of the SNe and the observer, we have adopted a cosmology with $\Omega_{\rm m}=0.3$,
$\Omega_{\Lambda}=0.7$ and $H_0=70$ km s$^{-1}$ Mpc$^{-1}$.

\section{Model description}
\label{sec:model}
\begin{figure}
\centering
 \includegraphics[width=0.4\textwidth]{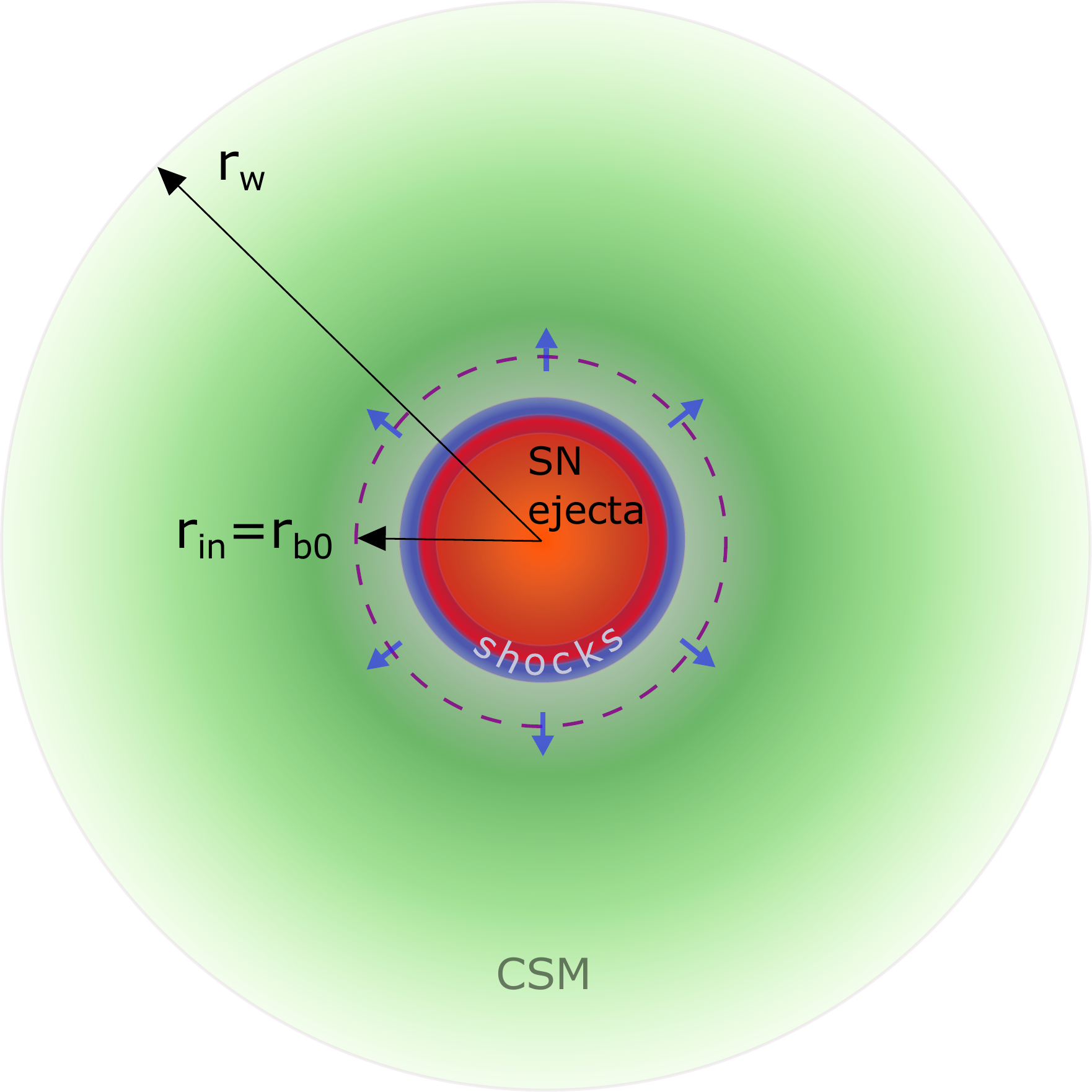}
 \caption{Sketch of the considered spherically symmetric geometry after the SN explosion. The outer edge of the CSM is marked as $\rw$. The dashed magenta line marks the position of the radiation-mediated shock breakout ($\rin=\rbo$). The interaction of the  SN ejecta (dark red colored region) with the CSM (light green  colored region) leads to the formation of shocks that propagate radially outwards. The color gradient represents the density gradient of the CSM. The light red and blue colored regions represent,  respectively, the shocked SN ejecta and the shocked CSM.}
 \label{fig:fig0}
\end{figure}

The interaction of the freely-expanding SN ejecta with the CSM gives rise to two shock waves, i.e. a fast shock wave in the circumstellar material (forward shock) and a  reverse shock in the outer parts of the SN ejecta. The CSM is modelled as an extended shell with inner and outer radii $r_{\rm csm}$ and $\rw$, respectively. This is illustrated in Fig.~\ref{fig:fig0}.  As long as the interaction between the SN ejecta and the CSM takes place within a region that is optically thick to Thomson scattering  ($\taut \gg 1$), the SN shock will be radiation-mediated and particle acceleration will be suppressed \citep[e.g.][]{murase11, katz11}. The shock wave is expected to break out at a radius $\rbo$ where $\taut\sim c/\vs$ and $\vs$ is the forward shock velocity \citep{weaver76}. {For dense CSM environments the shock is expected to break out in the wind ($\rbo > r_{\rm csm}$), whereas for dilute wind environments the shock breakout occurs in the stellar envelope. The second physical scenario is, however, not 
interesting for the present study, since for a dilute CSM the signature of secondary particles produced through \pp collisions will be quite weak.  At $r > \rbo$ the previously radiation-mediated shock becomes collisionless } and particle acceleration can, in principle, take place. Thus, $\rbo$ can be considered as an effective inner edge of the CSM, or $\rin\equiv\rbo$. Henceforth, this will be used as our normalization radius. 
The mass density profile of the CSM is written as \citep[e.g.][]{Chevalier1982}
\eqb
\rho_{\rm csm}(r) = \rho_0 \left (\frac{\rin}{r} \right)^{w},
\label{eq:rho-general}
\eqe
 where $\rho_0$ is the CSM mass density at $\rin$.  
% $\rho_0 = M_{\rm w}/ 4 \pi \vw \rin^w$. 
For $w=2$ the above expression simplifies into a wind density profile, and $\rho_0$ is related to the mass loading parameter $\Aw \equiv \dot{M}_{\rm w}/\vw$ as 
\eqb
\rho_0=\frac{\Aw}{4\pi \rin^2},
\label{eq:rho-wind}
\eqe
where $\dot{M}_{\rm w}$ and $\vw$ are, respectively, the  mass loss rate and velocity of the wind. 
Typical values lie in the range $\dot{M}_{\rm w}\approx10^{-5}-10^{-3}\, M_{\odot}\unit{/yr}$ with the high and low values corresponding respectively to Wolf-Rayet \citep[][]{crowther07} and red supergiant progenitors. The range in the velocities is accordingly wide, namely $\vw\approx10-10^3$ km s$^{-1}$ \citep[e.g.][]{chevalier00}, with slower winds being generally related to progenitors with higher mass loss rates.  Since the production rate of secondary particles through \pp collisions is proportional to the CSM number density, it is convenient to use $\rho_0$ --- or $\Aw$ in the case of a wind density profile --- as our main model parameter.

If $\mu$ is the mean number of nucleons per particle in the CSM, then the respective number density can be expressed as $n_{\rm csm}(r)=\rho_{\rm csm}(r)/\mu \mpr$. The Thomson optical depth due to photon scattering by electrons in the CSM is given by
\eqb
% \taut = \int_{r}^{\rw} dr' n_{\rm e}(r') \sth \approx \frac{\sth \rho_0 \rin}{\mu \mpr (w-1)}\left(\frac{\rin}{r} \right)^{w-1}.
\taut = \int_{r}^{\rw} {\rm d}r' n_{\rm e}(r') \sth \approx \frac{\sth \rho_0 \rin}{\mu_{\rm e} \mpr (w-1)}\left(\frac{\rin}{r} \right)^{w-1}.
\label{eq:tau-general}
\eqe
where $\mu_{\rm e}$ is the mean number of electrons per particle and  $\rw \gg r$. Henceforth, we adopt $\mu=\mu_{\rm e}=1$ for simplicity.
If $\vo$ is the the shock velocity at the breakout radius, the requirement that $\tau_{T}\sim c/\vo$ at breakout together with eq.~(\ref{eq:tau-general}) leads to a useful expression relating $\rho_0, \vo$ and $\rin$ \citep[see also][]{ofek14}, namely
\eqb
% \rho_0 \approx \frac{(w-1) \mu \mpr c}{\sth \rin \vo }.
\rho_0 \approx \frac{(w-1) \mpr c}{\sth \rin \vo }.
\label{eq:rho}
\eqe
For the case of a wind CSM density profile, the expected mass loading parameter is 
\eqb
\Aw \approx 10^{  16} \, r_{\rm in,13}\,\beta^{-1}_{0,{  -1.5}} \unit{gr \ cm}^{-1},
\label{eq:Aw-wind} 
\eqe
where $\beta_0 \equiv \vo/c$  was assumed.  As the breakout time of the SN shock and its respective velocity are the parameters to be first determined by SNe observations,  we will treat as primary model parameters $\rin$ and $\vo$ rather than  $\rho_0$ (or, $\Aw$  in the case of a wind density profile). For compact stars with tenuous fast winds the shock breakout may happen close to the progenitor radius, i.e. $\rin \sim$ a few times $10^{11}$ cm \citep[see][]{Waxman2007}. For  progenitors with high mass-loss rates, however, the shock breakout is expected to take place in the wind and well outside the progenitor, i.e. $\rin \sim 10^{13}-10^{14}$~cm \citep[see][]{Rabinak2011,Chevalier2012}.  

The ejecta-CSM interaction leads eventually to the formation of a shell {with high internal energy density} \citep[e.g.][]{Chevalier1982, chevalier94, truelove99}, whose equation of motion was studied by \cite{Chevalier1982} for the case of spherically expanding SN ejecta with a power-law density profile, i.e. $\rho_{\rm SN} \propto r^{-n_{\rm sn}}$. The density profiles of both the SN ejecta and the CSM are necessary for detailed calculations, since the time evolution of the shock front radius and velocity $\vs$ depends on these. For the purposes of our study it is sufficient to assume that the shock velocity has a radial power-law profile of the form 
\eqb
\vs(r)=\vo\left(\frac{r}{\rin}\right)^{\alpha},
\label{eq:vshock}
\eqe
The power-law index $\alpha\le0$ can be written in terms of the power-law indices of the CSM and SN ejecta density profiles \citep[e.g.][]{ofek14}. Typically, the forward shock is expected to be mildly decelerating, unless the SN ejecta have a flat density profile ($n_{\rm sn}<3$) \citep[e.g.][]{Chevalier1982, tan01}.  For example, $\alpha=-0.1$ for a wind-like CSM ($w=2$) and ejecta with $n_{\rm sn}=12$, whereas if both media are uniform the SN shock propagates with a constant velocity, at least in the free expansion phase \citep[e.g.][]{matzner99}. To keep our analysis as general as possible, we will treat the index $\alpha$ as a free parameter. Table~\ref{tab0} summarizes the free parameters of the model.
\begin{table}
\caption{Free model parameters and their default values used in the text.}
% \begin{threeparttable}
  \begin{tabular}{lcc}  
  \hline
Parameter & Symbol & Default value \\
\hline
           Input & & \\          
\hline
Power-law index of the CSM density profile    & $w$ & 2 \\
Power-law index of the shock velocity profile  & $\alpha$ & 0 \\
Power-law index of the SN luminosity profile  & $\aph$ & 0 \\
Power-law slope of shock accelerated particles & $s$ & 2\\
Magnetic energy density fraction & $\eB$ & $10^{-2}$\\
Accelerated proton energy fraction & $\ep$ & $10^{-1}$\\
Electron-to-proton ratio & $K_{\rm ep}$ & $10^{-3}$ \\
Peak supernova luminosity (erg s$^{-1}$) & $L_{\rm sn,pk}$ & $10^{41}$ \\
Breakout radius (cm) & $\rin$ & $10^{13}$\\
Shock velocity at breakout radius & $\vo$ & {0.03 c} \\
Wind velocity & $\vw$ & $0.01$ c \\
Electron temperature of unshocked CSM  (K) & $T_{\rm e}$  & $10^5$ \\
\hline
Output & & \\
\hline
$\ab$ & $w/2-\alpha$ & 1\\
$\qsyn$& $2\ab+\alpha$ & 2\\
$\qic$& $2+\aph+\alpha$ & 2\\
$\dpp$&  $w-1+\alpha$ & 1\\
\hline 
  \end{tabular}
\label{tab0}
\end{table} 

The free expansion phase of the SN ejecta lasts until the mass in the ejecta $\Mej$ is comparable to the swept-up mass $\Msw$. The transition from the free expansion phase to the so-called Sedov-Taylor phase will happen at a radius 
\eqb
\rd = \rin \left[\frac{(3-w) \Mej}{4\pi \rho_0 \rin^3}\right]^{1/(3-w)}. 
\label{eq:rdec-general}
\eqe
The above equation is valid as long as  $\rin \ll \rd$ and $w<3$. For the case of the wind, the deceleration radius may be written as 
\eqb
% \rd =\frac{\Mej}{\Aw} \simeq 9\times10^{20}\,{\rm cm} \, M_{\rm ej,0} \beta_{\rm w,-2} \dot{M}^{-1}_{\rm w,-5},
\rd =\frac{\Mej}{\Aw} \simeq {  1.8\times10^{18}}\, M_{\rm ej,1} \, \beta_{0,{  -1.5}} \, r_{\rm in,13}^{-1} \unit{cm} 
\label{eq:rdec-wind}
\eqe
where we used \eqn{rho-wind} and \eqn{Aw-wind}. In the above,  $M_{\rm ej,1}\equiv\Mej/{  10} M_\odot$.  For typical parameters, we thus find $\rd \simeq 0.3-3$~pc; these distances are much larger than the characteristic shock radii that we will consider in this work.

In the present work we will focus on the acceleration and non-thermal emission from the forward shock. As long as the density profile of the ejecta is steep, the   contribution of the reverse shock to the observed non-thermal radiation is expected to be small (\citealt{chevalier_fransson03}; see also Fig.~3 in \citealt{zirakashvili15}).  We further assume that all the material accreted by the forward shock is confined in a thin homogeneous layer downstream of the forward shock \citep[e.g.][]{Chevalier1982, sturner97}. At the early stages of the SN expansion, the forward shock is expected to have Mach numbers $\gg1$. We therefore assume that the compression ratio is $\approx 4$ \citep[e.g.][]{sturner97} and {the width of the shock is  then 25\% of the shock radius. The latter is in rough agreement with the findings of recent hydrodynamic calculations of the shock structure \citep[][]{zirakashvili15}. Similarly, \cite{Chevalier1982} found that for steep density profiles of the SN ejecta ($n_{\rm sn}=7-
12$) the width of the shocked CSM at radius $r$ is $h\sim 0.2 \,r$. 
 
 Under these considerations, the average density of the shocked CSM is a factor of four larger than the upstream  CSM density at that radius, i.e. $n(r)=4\, \nw(r)$ or
\eqb
\label{eq:n}
n(r)=n_0\left(\frac{\rin}{r}\right)^w,
\eqe
% with $n_0 \equiv 4\rho_0/\mu \mpr$. For a wind-type medium this reduces to 
with $n_0 \equiv 4\rho_0/\mpr$. For a wind-type medium this reduces to 
\eqb
\label{eq:nstar}
% n(r) \simeq 2\times 10^{12}\, \unit{cm}^{-3}\mu^{-1} {A}_{w,15} r_{\rm in, 13}^{-2}\left(\frac{\rin}{r}\right)^2. 
% n(r) \simeq 2\times 10^{12}\, \unit{cm}^{-3} {A}_{w,15} r_{\rm in, 13}^{-2}\left(\frac{\rin}{r}\right)^2. 
n(r) \simeq {  2\times 10^{13}}\,r_{\rm in, 13}^{-1}\, \beta_{0, {  -1.5}}^{-1}\left(\frac{\rin}{r}\right)^2 \unit{cm}^{-3} . 
\eqe
Strictly speaking, the Rankine-Hugoniot jump conditions across a strong shock determine the particle density right behind the shock as a function of the upstream properties. Here, we assume the same scaling for the volume-averaged post-shock density. 

For the calculation of the synchrotron emission, an estimate of the post-shock magnetic field strength is required. We assume that the magnetic energy density in the shell of shocked CSM is a fixed fraction $\eB$ of the post-shock thermal energy density\footnote{This assumption is also commonly made in studies of  gamma-ray burst (GRB) afterglows \citep[for a review, see][]{piran04,meszaros06}.}, which is written as  $U_{\rm th}(r)=(9/32) \mpr \vs^2(r) n(r)$.
Here, we considered that the upstream speed in the post-shock frame is $3\,\vs/4 $, so the mean post-shock
energy per particle is $(9/16)\mpr\vs^2/2$. The magnetic field strength is therefore given by
% $B(r) = (3/2) \vs(r) \sqrt{\pi\eB \mpr n(r)}$, or
\eqb
\label{eq:B}
B(r) = B_0 \left(\frac{\rin}{r}\right)^{\ab},
\eqe
where $\ab=w/2-\alpha$ and $B_0 \equiv (3/2)\sqrt{\pi \eB \mpr n_0 \vo^2}$. For a wind CSM density profile, the magnetic field is written as
\eqb
\label{eq:Bstar}
% B(r) \simeq 1.4\times10^3 \unit{G}\, \epsilon_{\rm B, -2}^{1/2} r_{\rm in, 13}^{-1}\beta_{0, -1} A_{\rm w, 15}^{1/2} \left(\frac{\rin}{r}\right),
B(r) \simeq {  1.4\times10^3}\, \epsilon_{\rm B, -2}^{1/2} r_{\rm in, 13}^{-1/2}\beta_{0, {  -1.5}}^{1/2} \left(\frac{\rin}{r}\right)  \unit{G},
\eqe
where we used \eqn{nstar} and $\beta_0\equiv \vo/c$, while $\eB$ is considered a free parameter of the model.  Typical values inferred from radio observations of interaction-powered SNe lie in the range $\eB \sim 10^{-2}-10^{-1}$ \citep[][]{Chevalier1998, bjornsson04,Chevalier2006,kamble15}, in contrast to GRB afterglows where $\eB \ll 10^{-2}$ are usually inferred from the observations \citep[][]{santana14,barniolduran14,zhangetal15}. 
{  We note that at times relevant for $\sim$GHz radio observations \footnote{This implies that synchrotron emission at the observing frequency is not free-free absorbed, i.e.  $\nu_{\rm ff}\lesssim \nu_{\rm obs}$. For more details, see Sec.~5.1} the magnetic field strength has already reached $\sim$G values:
\eqb
B \simeq 3.6 \, \epsilon_{\rm B, -2}^{1/2}r_{\rm in, 13}^{1/2} A_{\rm w, 16}^{3/2} \beta_{\rm 0, -1.5}^{1/2} T_{\rm e, 5}^{1/2}\nu_{\rm obs, 9}^{2/3}\unit{G},
\eqe
where $T_{\rm e}$ is the electron temperature of the unshocked CSM and $\nu_{\rm obs}$  is the observing frequency.}

\subsection{The injection of primary particles}
\label{sec:primary}
In this section we determine the injection distributions of primary electrons and protons,  focusing only on relativistic particles.
We assume that a fraction of the incoming particles at the SN shock will be accelerated into a power-law distribution in energy\footnote{If particle acceleration is governed by the first order Fermi process \citep[e.g.][]{axford77, bell78, webb84}, then strictly speaking, the accelerated particle distribution will be a power-law in in {\sl momentum}, not {\sl energy}. 
However, by focusing on relativistic particles  with $\gamma \gtrsim 2$, we can 
use interchangeably the terms particle energy and momentum (see also \app{app1}).}, with the nature of the acceleration process itself not being crucial for the present study. {  CR that have accelerated at the shock front will subsequently escape in the donwstream region of the shock where they will lose energy via various physical processes (see below). It is in this region where the production of secondary particles due to \pp collisions takes place \citep[see e.g.][]{mastichiadis96, sturner97}.}

If $s$ is the power-law index of the distribution, the injection function 
for both primary electrons and protons, i.e.  the number of relativistic 
particles injected in the shell per unit Lorentz factor and per unit radius, $Q_{\rm i}(\gamma, r) \equiv {\rm d}^2 N_{\rm i}/{\rm d} \gamma  {\rm d}r$, 
can be written as 
\eqb
Q_{\rm i}(\gamma, r)= Q_{\rm 0i}\,g_{\rm i}(r)\,\gamma^{-s}H[\gamma-\gmin] H[\gmax-\gamma]
\eqe
where ${\rm i=e,p}$, $H[x]=1$ for $x>0$ and 0 otherwise, $g_{\rm i}$ is an arbitrary function of radius, and $\gmin, \gmax$ are the minimum and maximum Lorentz factors of particles at injection (to be specified below). The exact value of the power-law index depends on the details of shock acceleration, such as non-linear effects at the shock front and shock obliquity  \citep[e.g.][and references therein]{jonesellison91}, which are not treated in this study. Thus, we will consider $s$ to be a free parameter of the model and assume that is the same for electrons and protons. We remark that the formalism we present below does not hold only for a power-law injection function, but it can be applied to injection distributions with an arbitrary dependence on energy.

\subsubsection{The injection rate and minimum energy at injection}
The normalization factor $Q_{\rm 0i}$ can be determined based on recent kinetic simulations, that
can capture the acceleration of protons at non-relativistic shocks from first principles \citep[e.g.][]{caprioli_spitkovsky14b, park2015}.
It has been found that, if the shock is quasi-parallel, i.e., 
the magnetic field in the upstream is nearly aligned (within $\lesssim 45^\circ$) with 
the direction of shock propagation, a fraction $\ep \sim 5-10\%$ of the shock energy  is channeled into
relativistic protons ($\gamma \gtrsim 2$). In the following, we assume $\ep$ to be constant in time and treat it as a free parameter (see also \app{app1}).
The incoming kinetic energy per unit radius is $\mathcal{E}_{\rm k}=(9\pi/32) \mpr \vs^2(r)\, r^2 n(r)$,
where we used the same considerations as for the derivation of $U_{\rm th}$. From the requirement that relativistic protons acquire a fraction $\ep$ of the incoming energy per unit radius, we can determine the normalization 
$Q_{\rm 0p}$ as
\eqb
\label{eq:Qop}
% Q_{\rm 0p}= \frac{9\pi}{32}\ep \rin^2 \,\mu n_0 \beta_0^2 \left\{
 Q_{\rm 0p}= \frac{9\pi}{32}\ep \rin^2 \, n_0 \beta_0^2 \left\{
\begin{array}{cc}
 (s-2), & s>2 \\ \\
 \ln^{-1}\left(\frac{\gpmax}{\gpmin}\right), & s=2
 \end{array}
\right.
\eqe
and the radial dependence of the injection rate as 
\eqb
\label{eq:gp}
g_{\rm p}(r) = \left(\frac{r}{\rin}\right)^{2\alpha+2-w}
\eqe
The above relations are derived  using eqs.~(\ref{eq:vshock}) and (\ref{eq:n}) under the assumptions of $\gpmax\gg\gpmin$ and $\gpmin=2$ (see also \app{app1}). In the case of a wind-like CSM ($w=2$) and constant shock velocity ($\alpha=0$) the injection rate is constant, i.e. $g_{\rm p}=1$.

% We now evaluate the injection term for primary electrons. 
Unlike protons, shock-accelerated electrons are all likely to be ultra-relativistic, which is indeed consistent with modeling the radio emission signature of interaction-powered supernovae: SN\,2009ip \citep{margutti14},  SN\,2010jl \citep{fransson14, ofek14}, SN\,1988Z \citep{chugai94}, and 2006jd \citep{chandra12}. Assuming energy equipartition between thermal post-shock electrons and protons, we find that the Lorentz factor of thermal electrons is 
\eqb
% \gamma_{\rm th,e} \sim \frac{9}{64} \frac{\mpr  \mu }{\mel}\frac{\vs^2}{c^2}\sim 2.5  \mu \beta_{0,-1}^2,
\gamma_{\rm th,e}(r) \simeq \frac{9}{64} \frac{\mpr}{\mel}\left(\frac{\vs(r)}{c}\right)^2\simeq 2.5 \beta_{0,-1}^2 \left(\frac{r}{\rin}\right)^{2\alpha},
\label{eq:gth}
\eqe
{  where here the shock velocity was normalized to $0.1c$. Because of the quadratic dependence on the shock velocity, the thermal electrons turn out to be non-relativistic at slower shocks. Recent particle-in-cell (PIC) simulations \citep[e.g.][]{park2015,guo_14a,guo_14b} show that the minimum momentum of the electron distribution is three times larger than the thermal one. For our calculations, we will therefore adopt $\gemin=3\gamma_{\rm e, th}$, if the latter is mildly relativistic (see eq.~(\ref{eq:gth})). Otherwise, we will assume that $\gemin=1$  (see also \app{app1}).  }
% where for simplicity we assume a constant shock velocity. 
% Thus, thermal electrons are mildly relativistic, and the low-energy cutoff of the distribution of 
% accelerated electrons will also fall in the relativistic regime. 
% For our calculations, we will adopt a minimum Lorentz factor that is three times larger than the thermal one, which is also supported recent particle-in-cell (PIC) simulations \citep[e.g.][]{park2015,guo_14a,guo_14b}. 

For the electron normalization, we adopt the common assumption that the ratio of the electron to proton spectrum at a given energy $E\gg m_p c^2$ is $\kep$, namely
\eqb
\kep = \frac{Q_{\rm e}\left(\frac{E}{\mel c^2}, r\right) }{Q_{\rm p}\left(\frac{E}{\mpr c^2},r\right) } \frac{\mpr}{\mel},
\eqe
with $\kep$ ranging typically between $10^{-3}-5\times10^{-3}$ according to PIC simulations \citep[e.g.][]{park2015}. These values are in rough agreement with  those inferred from observations. A value of $\kep \simeq 0.01$ is determined by direct measurements of cosmic-rays at Earth at 10~GeV \citep[e.g.][]{cohen_ramaty73,picozza2013}. Observations of young SN remnants at X-ray and GeV $\gamma$-rays imply $\kep \lesssim 10^{-3}$ \citep[e.g.][]{volk2005, morlino_caprioli2012, yuan2012}, although somewhat higher values $\sim 0.01$ have also been reported \citep[e.g.][]{gaisserprotheroe98}. In the following, we adopt $\kep=10^{-3}$ as the default value (see Table~\ref{tab0}).  The above expression leads finally to
\eqb
Q_{\rm 0e}=\kep \left(\frac{\mpr}{\mel}\right)^{s-1} Q_{\rm 0p},
%\frac{Q_{0e}}{Q_{0p}} = \frac{\tilde{Q}_{0e}}{\tilde{Q}_{0p}}\left(\frac{m_e}{m_p} \right)^{-p+1}
\label{eq:Qoe}
\eqe
while the radial dependence of the electron injection function is the same as for the protons, namely $g_{\rm e}(r) = g_{\rm p}(r)$.

\subsubsection{The maximum energy at injection}
The maximum energy of particles being accelerated at the shock is determined by the competition of acceleration and various cooling as well as loss processes \citep[e.g.][]{gaisser90, mastichiadis96,sturner97}. In this paragraph, we estimate the maximum Lorentz factor of accelerated electrons and protons by comparing the relevant acceleration  timescales with the dynamical and cooling timescales.
\begin{itemize}
\item {Acceleration timescale:}
the scattering at the basis of the Fermi acceleration process is usually assumed to proceed in the Bohm limit, leading to $t_{\rm acc, i} \sim 6 \gamma m_{\rm i} c^{3}/e B \vs^{2}$ \citep[e.g.][]{protheroeclay04, rieger07, tammiduffy09}. 
\item {Dynamical timescale:} by requiring that the acceleration timescale is shorter than the dynamical timescale $t_{\rm dyn}(r)\sim r/\vs$ (which is the typical timescale of adiabatic losses) we derive 
\eqb
\label{eq:gmax1}
\gmax^{(1)}(r)= \frac{e B_0 \beta_0 \rin}{6 m_{\rm i} c^2}\left(\frac{r}{\rin}\right)^{1-\ab+\alpha},
\eqe
	which is independent of the shock radius for $\ab=1$ and $\alpha=0$. As $t_{\rm dyn}$ is also a measure of the age
	of the system, we will refer to this criterion as the ``age criterion''. For the wind CSM density profile and typical parameter values, we find that the age constraint limits the maximum particle Lorentz factor to
\eqb
\label{eq:gmax1-wind}
% E^{(1)}_{\rm M} \simeq 70\,\unit{PeV} \, \epsilon_{\rm B, -2}^{1/2}\,\beta_{0,-1}^2 A_{\rm w, 15}^{1/2},
\gamma^{(1)}_{\rm i, M} {m_{\rm i}c^2} \simeq {  1.2\times 10^{16} \unit{eV}} \, \epsilon_{\rm B, -2}^{1/2}\,\beta_{0,{  -1.5}}^2 A_{\rm w, {  16}}^{1/2}
\eqe
where we used \eqn{Bstar}, while $A_{\rm w}$ is related to $\beta_0$, $\rin$ through \eqn{Aw-wind}. Protons in the shocks of type IIn SNe can, in principle, achieve multi-PeV energies due to the high-density CSM, as it has been already noted by \citet{katz11, murase11, murase14, zirakashvili15}.

The age criterion is closely related to the so-called confinement criterion, which requires the particle gyro-radius to be smaller than the maximum wavelength $\lambda_{\max}$ of the scattering  turbulence, to avoid the particle escape from the acceleration region. If $\lambda_{\max} \sim r$, the respective maximum Lorentz factor is larger by a factor of $c/\vs$ than that derived from the age criterion. In this regard, the age criterion is more constraining.
However, if $\lambda_{\max} \ll r$, then the maximum proton energy may be limited to energies much below the PeV energy range, as it has been recently demonstrated by \cite{metzger_caprioli15} for the case of gamma-ray novae. The cosmic-ray confinement is non-trivial, since it depends on the details of the acceleration process, the magnetic field amplification and the ionization properties of the upstream medium\footnote{The details of the microphysical processes could be incorporated in a dimensionless parameter $\xi_\lambda\le 1$ by parameterizing $\lambda_{\max}=\xi_\lambda r$. Then, the maximum particle Lorentz factor would given by \eqn{gmax1-wind} multiplied by the term $\min[1, \xi_\lambda c/\vs]$. }. The exact value of the maximum particle energy is not crucial for the purposes of the present study. Thus, \eqn{gmax1-wind} will be used in the following.
\item {Cooling timescales:} the most relevant cooling processes for relativistic electrons are synchrotron and inverse Compton (IC) scattering energy losses. The corresponding losses for relativistic protons are negligible. Inelastic \pp collisions and photohadronic interactions (photopion and photopair production) are the most relevant processes for determining the maximum energy of relativistic protons \citep[e.g.][]{katz11, murase11}. Coulomb and bremsstrahlung energy losses become important only at low energies and do not play role in setting the maximum particle energy.
\begin{enumerate}  
\item {Synchrotron cooling:}  by setting $t_{\rm acc,e}=t_{\rm syn}$, where $t_{\rm syn}(\gamma,r)=6 \pi \mel c/\sth \gamma B^2(r)$, 
we derive the respective maximum electron Lorentz factor. For the general case, this is given by
\eqb
\gemax^{(2)}(r)=\left(\frac{\pi e \beta_0^2}{\sth B_0}\right)^{1/2}\left(\frac{r}{\rin}\right)^{\ab/2+\alpha}
\label{eq:gmax2}
\eqe
and reduces to the expression below for a wind CSM density and constant shock velocity
\eqb
\gemax^{(2)}(r)\simeq {  3.8}\times10^4 \, \beta_{0,{  -1.5}}^{1/2}\, r_{\rm in,13}^{1/2}\epsilon_{\rm B, -2}^{-1/4} A_{\rm w, {  16}}^{-1/4}\left(\frac{r}{\rin}\right)^{1/2}.
\label{eq:gmax2-wind}
\eqe
\item {Inverse Compton cooling:}
relativistic electrons also lose energy due to IC scattering of soft photons provided by the SN\footnote{For simplicity, we consider the SN itself as the main source of soft photons. In principle, the  hot shell of shocked CSM can also be an efficient radiation, via free-free emission. See Appendix \ref{sec:app4} for details.}. To take into account a possible decay of the SN luminosity with time (or radius) we model the SN luminosity as $L_{\rm SN}=L_{\rm SN,pk}(\rin/r)^{\aph}$ with $\aph \ge 0$. 
The energy density of the SN photon field in the shell is then written as 
\eqb
U_{\rm ph} = \frac{L_{\rm SN}}{4\pi c (r+h)^2}\approx U_0 \left(\frac{\rin}{r}\right)^{2+\aph}
\label{eq:uph1}
\eqe
where $U_0=L_{\rm SN,pk}/4\pi c \rin^2$. We note that ambient photon fields, such as the cosmic microwave background and the galactic optical/infrared photon fields, are not included in the present analysis, since their energy density is typically lower than that given in \eqn{uph1}. For example, if $L_{\rm SN}=$const\,$\sim10^{41}$~erg s$^{-1}$, the energy density of SN photons is higher than that of ambient photon fields as long as $r\lesssim 0.3$~kpc (see also \citealt{gaisserprotheroe98}).  Assuming that the IC scattering takes place in  the Thomson regime, the  maximum  electron Lorentz factor is given by
% \eqb
% t_{\rm IC}(\gamma,r)=\tau_{\rm IC}\gamma^{-1}\left(\frac{r}{\rin}\right)^{2+s},
% \label{eq:tic}
% \eqe
% 	where $\tau_{\rm IC}=3\mel c / 4\sth U_0$. The respective
\eqb
\gemax^{(3)}(r)=\left(\frac{e \beta_0^2 B_0}{8 \sth U_0} \right)^{1/2}\left(\frac{r}{\rin}\right)^{(2+\aph-\ab+2\alpha)/2}.
\label{eq:gmax3}
\eqe
which, for the case of wind-type medium, constant shock velocity and SN luminosity, simplifies to
\eqb
\gemax^{(3)}(r)\simeq {  3\times 10^5}\,\beta_{0,{  -1.5}}^{3/2} \epsilon_{\rm B, -2}^{1/4}A_{\rm w, {  16}}^{1/4} r_{\rm in, 13}^{1/2} L_{\rm sn,pk,41}^{-1/2} \left(\frac{r}{\rin}\right)^{1/2}.
\label{eq:gmax3-wind}
\eqe
The above hold as long as the IC scatterings take place in the Thomson regime. However, the Klein-Nishina  suppression of the scattering rate becomes important for $\gamma \gtrsim \mel c^2/ \eph$, or $\gamma \gtrsim 5\times 10^5$ for $\eph=1$~eV, which is the typical photon energy for the SN emission. Although the Klein-Nishina suppression may be not relevant for primary electrons, the IC scatterings of optical photons by secondary electrons will take place deep in the Klein-Nishina regime (see \sect{evolution} for more details). 

At each shock radius, the highest energy electrons will be injected with a Lorentz factor  $\gemax=\min[\gemax^{(1)}, \gemax^{(2)}, \gemax^{(3)}]$. 
 In general, synchrotron (and IC) cooling will dominate at early times, whereas the age criterion becomes relevant at larger shock radii, where both
the magnetic and photon field energy densities have considerably decreased.
\item {Inelastic \pp collision losses:} let us consider next the effect of \pp collisions on the accelerated protons. The corresponding timescale is 
\eqb
\label{eq:tpp}
t_{\rm pp}\simeq \left(\kappa_{\rm pp} \sigma_{\rm pp} n(r) c\right)^{-1}
\eqe
where we assumed a constant inelasticity $\kappa_{\rm pp} \simeq 0.5$ and neglected the weak energy-dependence of the cross section $\sigma_{\rm pp}\simeq 3\times 10^{-26}$~cm$^2$; however, the energy-dependent cross section is taken into account in our numerical calculations (for more details, see \sect{evolution}). {  In eq.~(\ref{eq:tpp}) it is the density of the shocked CSM that appears, since \pp collisions are assumed to take
place in the post-shock region.} By requiring that $t_{\rm acc,p} = t_{\rm pp}$, we find that the maximum proton Lorentz factor is limited to
\eqb
\label{eq:gmax4}
\gpmax^{(2)} = \frac{eB_0\beta_0^2}{6 \kappa_{\rm pp} \sigma_{\rm pp} n_0 \mpr c^2}\left(\frac{r}{\rin}\right)^{w-\ab+2\alpha}
\eqe
which increases linearly in radius for a shock propagating with a constant velocity in a wind-like CSM, namely
\eqb
\label{eq:gmax4-wind}
\gpmax^{(2)} \simeq {  2.7\times10^5} \, \beta_{0,{  -1.5}}^3 r_{\rm in, 13} A_{\rm w, {  16}}^{-1/2} \left(\frac{r}{\rin}\right)
\eqe
Inspection of eqs.~(\ref{eq:gmax1-wind}) and (\ref{eq:gmax4-wind}) shows that \pp losses dominate over adiabatic proton cooling at small radii, but since the former increases with radius, the maximum energy of protons at  later times is set by the age criterion.  
\item {Photohadronic energy losses:}
in the presence of photon fields, energetic protons may lose energy through photohadronic interactions. Although the energy threshold for photopion production on $\eph=1$~eV photons is high ($\gamma_{\rm p, p\pi} \gtrsim m_{\pi} c^2/\eph \simeq 1.5\times10^8/(\eph/1\,{\rm eV})$) and thus not relevant for our study,  protons with Lorentz factors $\gamma_{\rm p, BH} \gtrsim  \mel  c^2/\eph = 5\times 10^5/(\eph/1\,{\rm eV})$ may lose energy through Bethe-Heitler pair production. It can be shown that the respective timescale, which can be recast in a similar form as $t_{\rm pp}$ (see \app{BHloss}), is the longest one for a wide range of parameter values and can be safely neglected.
Same as for the electrons, the maximum proton Lorentz factor at injection will be $\gpmax=\min[\gpmax^{(1)},\gpmax^{(2)}]$. 
\end{enumerate}
        
\end{itemize}
\subsection{The injection of secondary electrons}
\label{sec:secondary}
Secondary electrons are the by-product of charged pion decays, i.e. $\pi^\pm \rightarrow \mu^\pm + \nu_\mu(\bar{\nu}_\mu)$ and $\mu^\pm \rightarrow e^\pm + \bar{\nu}_\mu(\nu_\mu) + \nu_{\rm e}(\bar{\nu}_{\rm e})$. Their injection rate  ($\Qepp$) depends on the number density of ``thermal'' (i.e., non-relativistic) protons that are the targets for \pp collisions, or equivalently on $n(r)$, and on the  distribution of accelerated protons, that evolves as the shock propagates in the CSM. Thus, the evolution of the proton distribution is necessary for the calculation of the  injection rate $\Qepp$. 
Aim of this paragraph is to derive a rough estimate of the ratio $\Qepp/Q_{\rm e}$ and investigate its dependence on the model parameters, such as $A_{\rm w}$. An accurate treatment of the secondary injection distributions will be presented in \sect{radio}, where we employ the expressions by \cite{kelner06}.
\subsubsection{The injection rate}
As an indicative example we assume a power-law proton distribution of the form $N_{\rm p}(\gamma, r) = Q_{\rm 0p}\,r\,\gamma^{-s}\,H[\gpmax-\gamma]H[\gamma-\gpmin]$, where the radial dependence is relevant to the case of a wind-type CSM and a constant shock velocity (for the general case of $w\ne 2$ and $\alpha \ne 0$, see \app{app3}). We furthermore use approximate expressions for the pion production rate, as presented  in  \cite{mannheim94}. The following calculations are based on several assumptions that we list for clarity:
\begin{itemize}
 \item we ignore the (weak) energy dependence of the \pp cross section, i.e. $\sigma_{\rm pp} \simeq 3\times 10^{-26}$~cm$^2$. 
 \item the multiplicity of pion production is taken to be $\xi \sim 1/3$ independent of the proton energy in the laboratory frame.
 \item for a single proton, the rate of increase of pion energy is approximated by a $\delta$-function centered at the average pion energy $\langle E_{\pi}\rangle \simeq (1/6) E_{\rm k,p}$, where  $E_{\rm k,p}=E_{\rm p}-\mpr c^2$ is the proton kinetic energy
\eqb
\label{eq:power_pi}
\frac{{\rm d}E_\pi}{{\rm d}t} = 1.3\spp c n \xi E_\pi\delta\left(E_\pi-\langle E_{\pi}\rangle\right),
\eqe
for $E_{\rm p}>E_{\rm p, thr} \simeq 1.2$~GeV and the factor 1.3 accounts for the chemical composition of the matter\footnote{The contribution of nuclei heavier than protons can be accounted for by a multiplication factor of 1.45 in  the injection rates of gamma-rays from neutral pion decay  and secondary electrons \citep{sturner97}.}.
\item the secondary electron injection rate is given by 
 \eqb
 \label{eq:qe_pp}
 \Qepp(\gamma)\equiv \frac{{\rm d} N_{\rm e}^{(\rm pp)}}{{\rm d} r {\rm d}\gamma} =\frac{1}{A} Q_\pi\left( \frac{\gamma}{A}\right) 
 \eqe
 where $A=70 \simeq m_\pi/(4 \mel)$.
 \end{itemize}
The pion injection rate for a proton distribution is then given by
\eqb
\label{eq:qpion}
Q_\pi(\gamma_\pi) \equiv \frac{{\rm d} N_\pi}{{\rm d}r {\rm d}\gamma_\pi} \simeq \frac{m_\pi c^2} {E_\pi \vs}\int_1^{\infty}\!\!{\rm d}\gamma N_p(\gamma,r) \frac{{\rm d}E_\pi}{{\rm d}t},
\eqe
where we made use of $\vs={\rm d}r/{\rm d}t$ and neglected the factor 1.26 that accounts for contributions from $\alpha-p$ and $\alpha-\alpha$ collisions \citep{mannheim94}. Using \eqn{power_pi} we find
\eqb
\label{eq:qpion-inst}
Q_\pi(\gamma_\pi) \simeq \frac{2.5\spp n c}{\vs}\frac{m_\pi}{\mpr}N_{\rm p}\left(6\gamma_\pi\frac{m_\pi}{\mpr}+1, r\right),
\eqe
for pions with Lorentz factors in the range
\eqb
\frac{\mpr}{6m_\pi}\gamma_{\rm p, thr} \le \gamma_\pi \lesssim \frac{\mpr}{6m_\pi} \gpmax,
\eqe
where $\gamma_{\rm p, thr}=E_{\rm p, thr}/\mpr c^2 \simeq 1.22$.
Finally, using \eqn{qe_pp} and substitution of $N_{\rm p}(\gamma,r)$ we find
\eqb
\label{eq:qe-inst}
\Qepp (\gamma,r) \simeq \left(\frac{\mel}{\mpr}\right)^{-s+1}\frac{10 Q_{\rm 0p}\spp n(r)rc}{\vs(r)}\left(24\gamma\right)^{-s}
\eqe
% \ls{see my comment above, I think for uniform medium we should have $r^4$.} 
for electron Lorentz factors in the range
\eqb
\frac{\mpr}{24 \mel} \gamma_{\rm p, thr} \le  \gamma \lesssim \frac{\mpr}{24\mel}\gpmax,
\label{eq:minmax}
\eqe
where $\gpmax$ is, in principle, a function of radius. Thus, the typical Lorentz factor of the injected secondary electrons is 
\eqb
\label{gesec}
\gamma_{\rm e}^{(\rm pp)} \simeq \frac{\gamma_{\rm p}\mpr}{24\mel}.
\eqe 
Using \eqn{qe-inst} and \eqn{Qoe} the ratio of secondary to primary injection rates
at a given Lorentz factor is written as (here, we generalize to any values of $w$ and $\alpha$)
\eqb
\label{eq:ratio}
\frac{\Qepp}{Q_{\rm e}} =24^{-s}\frac{10\spp n_0 \rin}{ \beta_0 \kep}\left( \frac{r}{\rin}\right)^{1-w-\alpha},
\eqe
where we made use of eqs.~(\ref{eq:n}) and (\ref{eq:vshock}). The injection ratio is a decreasing function of radius for $w=2$ and $\alpha\lesssim0$, suggesting that the contribution of the secondaries to the synchrotron emission from the SN shock is expected to be most important at small radii. Interestingly, if the shock propagates in a thick CSM shell with uniform (high) density ($w=0$), then the ratio of secondary to primary electrons will increase as the shock propagates, provided that it is not significantly decelerated. 
Plugging into \eqn{ratio} typical values for the parameters, as well as $w=2$ and $\alpha=0$, we find 
\eqb
\frac{\Qepp}{Q_{\rm e}} \simeq {  3\times 10^3}\, A_{\rm w, {  16}}  r_{\rm in, 13}^{-1} K^{-1}_{\rm ep,-3} \beta_{0,{  -1.5}}^{-1}\left(\frac{\rin}{r} \right),
\label{eq:ratio-wind}
\eqe
or 
\eqb
\frac{\Qepp}{Q_{\rm e}} \simeq {  3\times 10^3}\, K^{-1}_{\rm ep,-3} \beta_{0,{  -1.5}}^{-2} \frac{\rin}{r},
\label{eq:ratio-wind-2}
\eqe
where we substituted $\Aw$ in the above expression using \eqn{Aw-wind}.
Interestingly, at $r=\rin$ the ratio depends only on $\kep$ and the shock velocity. According to the above, slower shocks favour the production of secondaries, since the ratio ${\Qepp}/{Q_{\rm e}}$ increases. This result may seem counterintuitive at first sight but, in retrospect, it is not surprising. The ratio of injection between secondary and primary electrons will be proportional to $(Q_{\rm p}/Q_{\rm e}) \times  (t_{\rm age}/t_{\rm pp})\propto K^{-1}_{\rm ep} A_{\rm w}\beta_{0}^{-1} r^{-1}$, where $t_{\rm age}\sim r/\beta_0 c$ is the age of the system. Slower shocks allow for \pp interactions to act for a longer time, thus leading to a larger ${\Qepp}/{Q_{\rm e}}$ ratio.

\subsection{The parameter space $\Aw-r$}
\label{sec:parspace}
In the case of wind CSM density profile, a parameter space of the mass loading  parameter $\Aw$ versus the shock radius $r$ can be constructed using the following considerations:
\begin{enumerate}
\item Particle acceleration at radiation mediated shocks is suppressed \citep[e.g.][]{waxmanloeb01,katz11}. Roughly speaking, the 
shock is not radiation mediated if $\taut \lesssim c/\vs$ or
\eqb
 \label{eq:cond1}
 \Aw \le \frac{4\pi \mpr c \rin^{\alpha}}{\sth \vo}r^{1-\alpha}
\eqe
 where we used eqs.~(\ref{eq:tau-general}) and (\ref{eq:vshock}). 
\item As the magnetic field in the shocked shell depends on the CSM density, it is expected that for high enough $\Aw$ values the magnetic energy density $U_{\rm B}$ in the shell will exceed that of the SN radiation field $U_{\rm ph}$. A lower limit on $\Aw$ can be derived by requiring  $U_{\rm B} \ge U_{\rm ph}$, namely
 \eqb
 \label{eq:cond2}
 \Aw \ge  \frac{8}{9}\frac{L_{\rm SN,pk}\rin^{\aph+2\alpha}}{\eB \beta_0^2 c^3}r^{-\aph-2\alpha}.
 \eqe
 {  This should be considered as a strict lower limit, since it is derived for a SN optical light curve peaking at $\rin$.}
\item The emission from  secondary electrons (at all energies) will be negligible, unless $\Qepp/Q_{\rm e}\gtrsim 1$.  A ratio equal to unity corresponds to
\eqb
\label{eq:cond3}
\Aw = 24^s\frac{4\pi \mpr \beta_0 \kep \rin^{-\alpha}}{10 \spp}r^{1+\alpha},
\eqe
where we used \eqn{ratio}. In the following, we consider $0.1 \le \Qepp/Q_{\rm e}\le 10$ as a range of values with possible interest for the secondary radiative signatures.
\end{enumerate}
The parameter space $\Aw-r$ calculated for $\alpha=0$ and $\aph=0$ is presented in \fig{space1}. The blue colored region denotes the parameter regime where {  a collisionless shock may be formed, thus allowing particle acceleration}, and at the same time  $U_{\rm B}\ge U_{\rm ph}$. The red colored region shows the parameter space where  $1 \le \Qepp/Q_{\rm e}\le 100$. For a given shock velocity at the breakout radius ($\rin$), the value of the respective $\Aw$ can be read from the blue dashed line. Each horizontal line starting from a point $(\Aw,\rin)$ and extending to $\rd$ (orange line) denotes the evolutionary path of the system that we will consider. Interestingly, the two different evolutionary paths shown in \fig{space1} have initially the same ratio of secondary to primary injection rates. As long as the shock breakout occurs in the CSM with a wind-like density profile, the ratio $\Qepp/Q_{\rm e}$ at $r=r_{\rm in}$ is independent of $\Aw$ and $\rin$
(see also \eqn{ratio-wind-2}).
\begin{figure}
 \centering
 \includegraphics[width=0.5\textwidth]{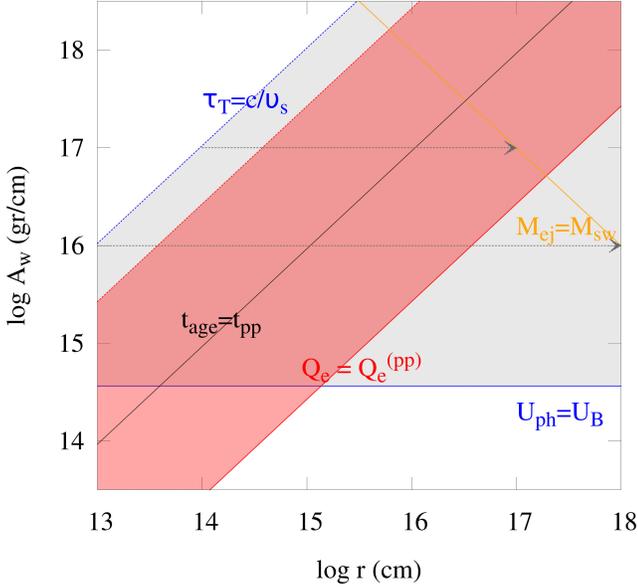}
 \caption{The $\Aw-r$ parameter space for a wind CSM density profile.
 The constraints \eqn{cond1} and \eqn{cond2} are shown with blue solid and dashed lines, respectively.
 The condition \eqn{cond3} is plotted with a red solid line. A ratio $\Qepp/Q_{\rm e}$ equal to {  1 and 100}
 corresponds to the red {  solid and dashed} lines, respectively. The deceleration
 radius for different $\Aw$ (see \eqn{rdec-general}) is overplotted with an orange line. The black solid line corresponds to $t_{\rm age}=t_{\rm pp}$.
 Each arrow (grey dashed line) denotes an ``evolutionary path'' of the system under consideration.  Other parameters used: $L_{\rm sn,pk}=10^{41}$~erg s$^{-1}$, $M_{\rm ej}={  10}M_\odot$, $\kep=10^{-3}$, $\eB=0.01$, $\vo={  0.03c}$, $\alpha=0$, $\aph=0$, and $s=2$.}
 \label{fig:space1}
\end{figure}
The free model parameters and their values for the default case, that represents a SN shock propagating with a constant velocity in a wind-like CSM, are summarized in Table~\ref{tab0}.

\section{A semi-analytical framework for the particle evolution}
\label{sec:evolution}
In this section, we detail our semi-analytical formalism to solve for the evolution of the electron and proton spectrum \citep[see also][]{finkedermer12}. This will be later applied to the early phases of SNe that expand in a CSM with non-uniform density profile. We stress, however, that the formalism we present here can be applied to a generic phase of supernova evolution and to an arbitrary power-law profile of the external medium.

We will be solving the evolutionary equations for protons and electrons (both primary and secondary) in the presence of losses and of a given injection term, which is provided by the particle escape downstream of the shock \citep[e.g.][]{mastichiadis96,kirkrieger98}. The general equation to be solved is  
\eqb
\label{eq:dist}
\frac{\partial \tilde{N}_{\rm i}}{\partial t}+\frac{\partial}{\partial \gamma}
\left(\dot{\gamma}\tilde{N}_{\rm i}\right)=\tilde{Q}_{\rm i}(\gamma, t)
\eqe
where $i=e,p$, $\dot{\gamma}\equiv {\rm d}\gamma/{\rm d}t$ is the energy loss term and $\tilde{N}_{\rm i}(\gamma,t)$ is the total number of particles in the shell at a given time $t$  with Lorentz factors between $\gamma$ and $\gamma+{\rm d}\gamma$. In the absence of the injection term on the right hand side and/or of a particle sink term on the left hand side, \eqn{dist} simply describes the conservation of the particle number. Spatial diffusion and advection are not considered here, since the shell is assumed to be homogeneous and all the material to be contained in the shell. Diffusion in momentum space due to moving Alfv{\'e}n waves turns to be negligible and is, therefore, neglected.
% {  [Lorenzo] however, diffusion in momentum space due to moving alfven waves is to be checked. \ls{no, negligible}} 

It is more convenient to solve the equation above with respect to the shock radius $r$ instead of time $t$. Since ${\rm d}r/{\rm d}t=\vs(r)$, where $\vs$ is the shock velocity, the particle kinetic equation can be recast in the form
\eqb
\frac{\partial N_{\rm i}}{\partial r}+\frac{\partial}{\partial \gamma}\left(\gamma^\prime \, N_{\rm i}\right)=
Q_{\rm i}(\gamma, r),
\label{eq:kinetic-general}
\eqe
where $\gamma^\prime\equiv {\rm d}\gamma/{\rm d}r$, $Q_{\rm i}(\gamma, r)\equiv {\tilde{Q}_{\rm i}(\gamma, t)}/{\vs(r)}$ and $N_{\rm i}\equiv N_{\rm i}(\gamma,r)$. 
% \ls{[The following paragraph can also go into a (long) footnote, but I want to say that Sturner got it wrong.]}
In the case where the particle-containing volume  undergoes adiabatic expansion, it is the total number of particles $N_{\rm i}$ that should enter the equation above instead of  the volume-averaged number density $n_{\rm i}\equiv {\rm d}^2N/{\rm d}V{\rm d}\gamma$.
In fact, the volume-averaged number density $n_{\rm i}$ would not be conserved due to the volume expansion of the shell. The kinetic equation written in terms of differential particle densities would then read (to be contrasted with the commonly-used eq.~(9) in \cite{sturner97})
\eqb
\frac{\partial n_{\rm i}}{\partial r}+\frac{\partial}{\partial \gamma}
\left(\gamma^\prime n_{\rm i}\right)+3\frac{n_{\rm i}}{r}=\frac{{Q}_{\rm i}(\gamma,r)}{V}
\eqe
where the third term on the left hand side accounts for the expansion of the shell volume $V$. 
% $V=4\pi r^2 h$. We remind that $h$ is the shell thickness and is a fraction of the shock radius $r$. 
The following sections describing the semi-analytical formalism can be skipped from readers that are interested only in the astrophysical implications of our results (see \sect{radio}). The method described below is also detailed in e.g. \citet{stawarz08} and \citet{petromast09}, albeit in a different context.
\subsection{The proton distribution}
\label{sec:evol-proton}
% \subsection{The cooling terms for protons}
The cooling term for relativistic protons due to adiabatic expansion is given by
\eqb
\label{eq:adcool}
{\gamma^\prime}_{\rm ad}=-\frac{\gamma}{3}\, \frac{{\rm d} \ln V}{{\rm d}r}=-\frac{\gamma}{r},
\eqe
where the volume of the shocked shell is $V\approx 4\pi r^2 h$ and $h$ is the width of shell, which is assumed to be a fraction of the shock radius.  Inelastic \pp collisions can be approximated as catastrophic energy losses, since the proton in one collision loses a significant fraction ($\kappa_{\rm pp}$) of its energy \citep[e.g.][]{sturner97}. In this approximation, the \pp collisions are treated as an escape term in the kinetic equation of protons  (\eqn{kinetic-general}), which is of the form $-N_{\rm p}/t_{\rm pp}$ \citep[e.g.][]{schlickeiser02}. In this case, \pp  losses do not affect the evolution of the high-energy cutoff of the distribution \citep[e.g.][]{mastkirk95}, but just the number of available protons. The energy loss terms due to photohadronic interactions are less important than the aforementioned cooling processes and can be safely neglected. For sufficiently high density in the shocked shell, trans-relativistic protons may also cool due to Coulomb collisions with background thermal 
electrons. However, at these low energies 
the exact shape of the proton distribution is not crucial, since their non-thermal 
emission is irrelevant to radio observations and the 
cross section for \pp collisions decreases significantly.
The evolution of the proton distribution function will follow
\eqb
\label{eq:pdist}
\frac{\partial N_{\rm p}}{\partial r}+\frac{\partial}{\partial \gamma}\left({\gamma^\prime}_{\rm ad} N_{\rm p}\right)+
\frac{N_{\rm p}}{\vs(r) t_{\rm pp}}=Q_{\rm p}(\gamma,r)
\eqe
where $t_{\rm pp}$ is defined in \eqn{tpp}. In the case that the energy dependence of the source term can be described by a power law, i.e. $Q_{\rm p}\propto \gamma^{-s}$, the proton distribution is also a power law with the same exponent. The evolution of the proton Lorentz factor, which is governed by \eqn{adcool},  is written as 
\eqb\label{eq:peq1}
\gamma_0=\gamma\left(\frac{r}{r_0}\right).
\eqe
In the absence of a source term, the solution to \eqn{pdist} is
\eqb
\label{eq:phomo}
N_{\rm p}(\gamma, r)=N_{\rm p}(\gamma_0, r_0)\frac{\gamma_0}{\gamma}\funp(r,r_0),
\eqe
where $\gamma_0 > \gamma$ is  particle Lorentz factor at an arbitrary radius $r_0<r$ and
\eqb
\label{eq:fp}
\funp(r,r_0)=\exp\left[-\frac{A_{\rm pp}}{\dpp}\left(\frac{\rin}{r_0}\right)^{\dpp} \left(1-\left(\frac{r_0}{r}\right)^{\dpp}\right)\right].
\eqe
In the above we introduced  $\dpp=w-1+\alpha$ and 
\eqb
A_{\rm pp}=\frac{\rin}{t_{\rm pp} \vo} \simeq 10 (w-1)\,\beta_{0,-1}^{-2},
\label{eq:Kpp}
\eqe
where we used \eqn{rho}. The catastrophic \pp losses reduce the number of protons in the source through the exponential term, while it does not affect the radial evolution of a single proton energy. Roughly speaking, the \pp loss term will dominate the adiabatic loss term in \eqn{pdist}, if
$r/\rin \lesssim A_{\rm pp}^{1/\dpp} \simeq [10(w-1)]^{1/\dpp} \beta_{0,-1}^{-2/\dpp}$. For the default case, where $\dpp=1$, we find that the adiabatic losses will be controlling the proton evolution for $r\gtrsim 10\rin$. 

For a generic source term of protons, $\Qp(\gamma, r)$ the proton distribution as a function of shock radius is given by
\eqb
\label{eq:ptot}
N_{\rm p}(\gamma,r)=\int_{r_{\min}}^r  \Qp(\gamma_0,r_0)\, \frac{\gamma_0}{\gamma}\funp(r,r_0) \,{\rm d} r_0,
\eqe
where $\gamma_0$ in the integral is considered a function of $r_0,r$ and $\gamma$ (see \eqn{peq1}).  The lower integration limit, $r_{\min}$, depends on the injection radius as well as on the cooling history of particles (for details, see \app{app3}).
% The solution for the case of an instantaneous injection event, which gives insight to the solutions obtained for  general injection profiles, is presented in \app{app2}. 
\subsection{The electron distribution}
\label{sec:evol-electron}
Besides cooling due to adiabatic expansion (see \eqn{adcool}), relativistic electrons lose energy due to synchrotron radiation and IC scattering. Using eqs.~(\ref{eq:B}) and (\ref{eq:vshock}) the synchrotron cooling term is written as
\eqb
\label{eq:syncool}
\gamma^\prime_{\rm syn}=
\frac{1}{\vs}\left(\frac{{\rm d}\gamma}{{\rm d}t}\right)_{\rm syn}=-K_{\rm syn}\frac{\gamma^2}{\rin} \left(\frac{\rin}{r}\right)^{2\ab+\alpha}
\eqe
where $K_{\rm syn}=\sth B_0^2 \rin/(6\pi \mel c \vo)$ is a dimensionless constant and $q_{\rm syn}=2\ab+\alpha$. 
For the calculation of the IC cooling term, we assume that the dominant photon field that is present in the shocked shell is the SN optical radiation.
The IC scatterings  between electrons and photons with characteristic energy $\eph=1$~eV take place in the Thomson regime,
as long as $\gamma \lesssim \gamma_{\rm KN} \equiv \mel c^2 /\eph \sim 5\times 10^5$. For electrons with $\gamma < \gamma_{\rm KN}$ the IC cooling term reads, 
in complete analogy with \eqn{syncool}, as
\eqb
\label{eq:iccool}
\gamma^\prime_{\rm ic}=
\frac{1}{\vs}\left(\frac{{\rm d}\gamma}{{\rm d}t}\right)_{\rm ic}=-K_{\rm ic} 
\frac{\gamma^2}{\rin}\left(\frac{\rin}{r}\right)^{q_{\rm ic}}
\eqe
where $K_{\rm ic}=4 \sth U_0 \rin /(3 \mel c \vo)$ and $q_{\rm ic}=2+\aph+\alpha$. For $\gamma \gg \gamma_{\rm KN}$ the cooling rate is reduced due to the Klein-Nishina suppression of the cross section \citep[e.g.][]{blumenthalgould70}. Although approximate expressions of the cooling rate that interpolate between the Thomson and Klein-Nishina regime can be found \citep[e.g.][]{moderski05}, 
these do not allow for analytical solution of the electron kinetic equation. However, if the IC cooling is not important compared to other loss processes, then the complications of the Klein-Nishina effects can be ignored. Indeed, for dense CSM, where the emission signatures of secondary electrons will be important, synchrotron cooling is expected to dominate over IC cooling (see also \fig{space1}).   
% \ls{[The question is whether the electrons that are emitting now in the radio could have come from electrons that initially cooled in the Klein-Nishina regime. in principle, we can experiment with it: take the full evolution with included KN, and compare with the case in which you neglect KN corrections. does it make any difference for the purpose of this paper?]} 
Thus, we may confidently use  \eqn{iccool}, as long  as the electrons cool predominantly due to synchrotron radiation (see condition \eqn{cond2} and \fig{space1}). 

The evolution of the electron Lorentz factor $\gamma$ in radius, under the effect of adiabatic, synchrotron and IC losses is described by
% \footnote{The implications of electron cooling in a decaying magnetic field on the prompt GRB spectra have recently presented in \cite{uhm14}.} 
\eqb
{\gamma^\prime}=-\frac{\gamma}{r}-K_{\rm syn} \frac{\gamma^2}{\rin}\left(\frac{\rin}{r}\right)^{\qsyn}-K_{\rm ic} \frac{\gamma^2}{\rin}\left(\frac{\rin}{r}\right)^{\qic}
\eqe
which is conveniently rewritten as 
\eqb
\frac{{\rm d}[(\gamma r)^{-1}]}{{\rm d}r}=\frac{K_{\rm syn}}{\rin^2}\left(\frac{\rin}{r}\right)^{1+\qsyn}+\frac{K_{\rm ic}}{\rin^2} \left(\frac{\rin}{r}\right)^{1+\qic}
\eqe
whose general solution, for $\qsyn, \qic \ne 0$, is 
\eqb
\label{eq:geq}
\gamma=\gamma_0\left(\frac{r_0}{r}\right)\frac{1}{1+\gamma_0\, (r_0/\rin)\, \fune(r,r_0)}.
\eqe
The synchrotron and IC losses are incorporated in the function 
\eqb
\label{eq:funce}
\fune(r,r_0)=\frac{K_{\rm syn}}{\qsyn}\left(\frac{\rin}{r_0}\right)^{\qsyn}\left[1-\left(\frac{r_0}{r}\right)^{\qsyn}\right]+\frac{K_{\rm ic}}{\qic}\left(\frac{\rin}{r_0}\right)^{\qic}\left[1-\left(\frac{r_0}{r}\right)^{\qic}\right].
\eqe
Alternatively, $\gamma_0$ can be explicitly written as a function of $\gamma$, $r$ and $r_0$ as
\eqb\label{eq:geq1}
\gamma_0=\gamma\left(\frac{r}{r_0}\right)\frac{1}{1-\gamma\, (r/\rin)\, \fune(r,r_0)}
\eqe
In the absence of a continuous source of particles and/or a sink term of particles, \eqn{kinetic-general}
for the evolution of the electron spectrum {\sl is a conservation equation} for the number of electrons.
It follows that the electron spectrum $N_{\rm e}(\gamma,r)$ at radius $r$ can be simply related to the 
electron spectrum at an arbitrary radius $r_0$, namely with  $N_{\rm e}(\gamma_0,r_0)$, via
\eqb
\label{eq:ehomo}
N_{\rm e}(\gamma,r)=N_{\rm e}(\gamma_0,r_0)\left|\frac{\partial \gamma_0}{\partial \gamma}\right|
\eqe
where $|\partial \gamma_0/\partial \gamma|$ can be computed from \eqn{geq1}
\eqb
\left|\frac{\partial \gamma_0}{\partial \gamma}\right|= \left(\frac{r_0}{r}\right)\frac{\gamma_0^2}{\gamma^2}
\eqe
In the presence of a source term of particles, which can be described by any generic function $\Qe(\gamma, r)$, the electron distribution
at each radius is given by
\eqb\label{eq:etot}
N_{\rm e}(\gamma,r)=\int_{r_{\rm min}}^r  \Qe(\gamma_0,r_0)\, \left|\frac{\partial \gamma_0}{\partial \gamma}\right| \,{\rm d} r_0,
\eqe
 where $\gamma_0 \equiv\gamma_0(r_0;\gamma,r)$ is given by \eqn{geq1}. All the information regarding the cooling
of electrons is carried by the term $\left|\frac{\partial \gamma_0}{\partial \gamma}\right|$, which when
convolved with the specific source term of particles, provides us with  a self-consistent
description of the electron evolution with radius and energy. 

\subsubsection{Secondary electrons}
The evolution of the secondary electron distribution, $\Nepp(\gamma,r)$, is calculated using \eqn{etot} for a source term  that depends on the proton distribution (see e.g. \eqn{qpion-inst}) and  for $\gamma_0$ defined by \eqn{geq1}. 
In \sect{model} an approximate expression for $\Qepp$ has been used, since it was sufficient for a rough estimate. In what follows, a more accurate expression for the production rate of secondary electrons is adopted \citep[][henceforth, KAB06]{kelner06}.  Using the KAB06 formalism and our notation, the injection rate of secondaries can be written
as 
\eqb
\label{eq:qe-pp-KA}
\Qepp(\gamma,r) =\frac{c n}{\vs M}\int_0^1 \frac{{\rm d}x}{x}\spp\left(y\right)
F_{\rm e}\left(x,y \right)N_{\rm p}\left(y, r\right),
\eqe
where $M=\mpr/\mel$, $y=E_{\rm p}/\mpr c^2$, $x=E_{\rm e}/E_{\rm p}=\gamma / M y$, $E_{\rm e}$, $E_{\rm p}$ are the electron and proton energies, and $F_{\rm e}(x,E_{\rm p})$ is defined by eqs.~(62)-(65) in KAB06. We note
that $n$ and $\vs$ have, in principle, a radial dependence. The expression above is valid for $E_{\rm p}\ge 0.1$~TeV and 
$x>10^{-3}$ (or, $\gamma \gtrsim 200$). An accurate continuation of the calculations to proton energies close to the threshold energy for \pp collisions has been presented by \cite{dermer86}. For the purposes of our study, it is sufficient to adopt the $\delta$-function approximation for the pion production rate for lower proton energies (see discussion in KAB06). In this case, the secondary electron injection rate can be calculated by  
\eqb
\label{eq:qe-pp-delta}
\Qepp(\gamma,r) =2\frac{c n}{\vs M}\frac{\tilde{n}}{\tilde{\kappa}_{\rm pp}}\int_{E_{\min}}^{\infty} \frac{{\rm d}E_\pi}{E_\pi}\spp\left(y\right)
\tilde{f}_{\rm e}\left(x\right)N_p\left(y, r\right),
\eqe
where  eqs.~(36)-(39) and (77) of KAB06 were used. In the above equation, $E_{\min}= E_{\rm e}+ m_\pi c^2 / 4 E_{\rm e}$, $x=E_{\rm e}/E_{\pi}$, $E_{\pi}$ is the charged pion energy, $y={  1+ E_\pi/(\mpr c^2}\tilde{\kappa}_{\rm pp})$, $\tilde{\kappa}_{\rm pp}=0.17$ is the proton inelasticity used in KAB06, $\tilde{f}_{\rm e}$ is a function defined by eqs.~(36)-(39) in KAB06, and $\tilde{n}$ is the pion production multiplicity. This depends on the power-law index of the proton distribution as well as on the species of secondary particles. KAB06 provide the values for $s=2,2.5$ and $3$. These are respectively $\tilde{n}=0.77, 0.62$ and 0.67. Since protons suffer only from adiabatic losses, the power-law index of the distribution will be the same as at injection.
Thus, by choosing  $s$ to be 2, 2.5 or 3 at injection, we can adopt the values for $\tilde{n}$ for the calculation of the secondary injection rate.
% Despite of being useful for the calculating the injection of secondaries at low energies,
As the $\delta$-function approximation is not accurate for high proton energies, namely $E_{\rm p}\gg 0.1$~TeV, in our calculations we combine both approaches. For $E_{\rm p} \lesssim 100$~TeV,  we use the rate given by \eqn{qe-pp-delta} while the rate of \eqn{qe-pp-KA} is used, otherwise.

In brief, we have presented a semi-analytical model for calculating the time evolution of primary and secondary particle distributions. The adopted formalism
\begin{itemize}
\item  takes into account the cooling history of all particles that have been injected into the emission region until a given radius/time. 
\item  provides the shape of the spectrum at all energies. More precisely, even if at radius $r$ particles are injected with $\gamma\geq \gmin(r)$, the  computed spectrum extends below $\gmin(r)$,  as a result of the cooling of particles injected at earlier times. This is particularly important for 
secondary electrons, since the minimum injected momentum in that case is ultra-relativistic, namely $\gamma_{\rm e, min}^{(\rm pp)} \simeq 76$. Furthermore,   the cooling break in the particle distributions is a natural outcome of the self-consistent treatment of particle evolution; there is no need to introduce the breaks by hand.
\item allows us to calculate self-consistently the evolution of the low-and high-energy cutoffs of the particle distributions. These can be used, for example, 
to calculate the time evolution  of the minimum/maximum characteristic synchrotron frequencies,  as well as the synchrotron self-absorption frequency (see \sect{radio}). 
\item can be equally employed to source terms with a different energy dependence than the usual power-law  $\propto \gamma^{-s}$; this lies upon the
fact that the calculation of $N_{\rm i}(\gamma,r)$ is simplified into a one-dimensional integral (see eqs.~(\ref{eq:pdist}) and (\ref{eq:etot})). This last point justifies our choice of the term {\sl semi-analytical model}.
\end{itemize}

\section{The evolution of particle distributions}
\label{sec:distributions}
It has been already noted by \cite{murase14} that a combination of $\dot{M}_{\rm w}$ and $\vw$ leading to high values of the mass load parameter $\Aw$ is  necessary for the secondary synchrotron emission to  be observable at high  ($>100$~GHz) radio frequencies.  Since the magnetic field strength scales as $n^{1/2}$, electron synchrotron cooling\footnote{Synchrotron cooling does not affect the proton spectrum, even for high values of $\Aw$, where the \pp losses become dominant.} becomes important for high values of $\Aw$, unless $\epsilon_B\ll 10^{-4}$. Using the formalism presented in \sect{model} and \sect{evolution} we can derive the evolution of primary and secondary electron distributions for $s=2$.

In particular, if the particles radiating at radio frequencies belong to the cooled part of the distribution, i.e. $\gamma \gg \gcool$, where the cooling break $\gcool$ is given by
\eqb
\label{eq:gbr-0}
\gcool \approx \frac{\qsyn}{K_{\rm syn}}\left(\frac{r}{\rin} \right)^{\qsyn-1} \propto r^{w-\alpha-1},
\eqe
it can be shown that (for the derivation, see \app{app3}) 
\eqb
\label{eq:sol-primary-cool}
N_{\rm e, >\gcool}(\gamma,r) & \approx & \frac{Q_{\rm 0e}\rin}{K_{\rm syn}\gamma^3}\left(\frac{r}{\rin}\right)^{2+\alpha} \\ \nonumber \\
N_{\rm e, >\gcool}^{(\rm pp)}(\gamma,r) & \approx & \frac{Q_{\rm 0e}^{(\rm pp)}\rin}{K_{\rm syn}\gamma^3}\left(\frac{r}{\rin}\right)^{3-w},
\label{eq:sol-secondary-cool}
\eqe
where the subscript $\gtrless \gcool$ should be interpreted as $\gamma \gtrless \gcool$.
It is noteworthy that, for a constant shock velocity, the number of accelerated electrons that have undergone synchrotron cooling increases as $r^2$, whereas the number of secondary electrons increases at a slower rate, i.e. $\Nepp\propto r$. However, if the shock encountered a thick CSM shell with a flatter\footnote{The expressions above have been derived for $\qsyn=w-\alpha \neq 0$ and cannot be directly applied to the case of a uniform medium, unless $\alpha\neq0$. This requires modification of \eqn{funce} and a similar analysis as the one presented in \app{app3}.} density profile, e.g. $w=1/2$, then a faster increase of the number of secondary electrons is expected. 
For completeness, the results for the uncooled ($\gamma \ll \gcool$) primary and secondary electron distributions are listed below:
\eqb
\label{eq:sol-primary-uncool}
N_{\rm e, <\gcool}(\gamma,r)& \approx & \frac{Q_{\rm 0e} \rin}{\gamma^2 (4-w+2\alpha)}\left(\frac{r}{\rin}\right)^{3-w+2\alpha} \\ \nonumber  \\
N_{\rm e, <\gcool}^{(\rm pp)}(\gamma,r) & \approx &  \frac{Q_{\rm 0e}^{(\rm pp)}\rin}{\gamma^2 (5-2w+\alpha)}\left(\frac{r}{\rin} \right)^{4-2w+\alpha}.
\label{eq:sol-secondary-uncool}
\eqe
As synchrotron cooling is not important in this regime, we find, as expected, no explicit dependence of the electron number on the magnetic field strength. In fact, the uncooled electron part of the distribution traces directly the injection rate.
For the default case, we find $N_{\rm e}\propto r$ and $\Nepp\propto$ const, which should be respectively compared to eqs.~(\ref{eq:sol-primary-cool}) and (\ref{eq:sol-secondary-cool}).  
\begin{figure*}
  \includegraphics[width=0.48\textwidth]{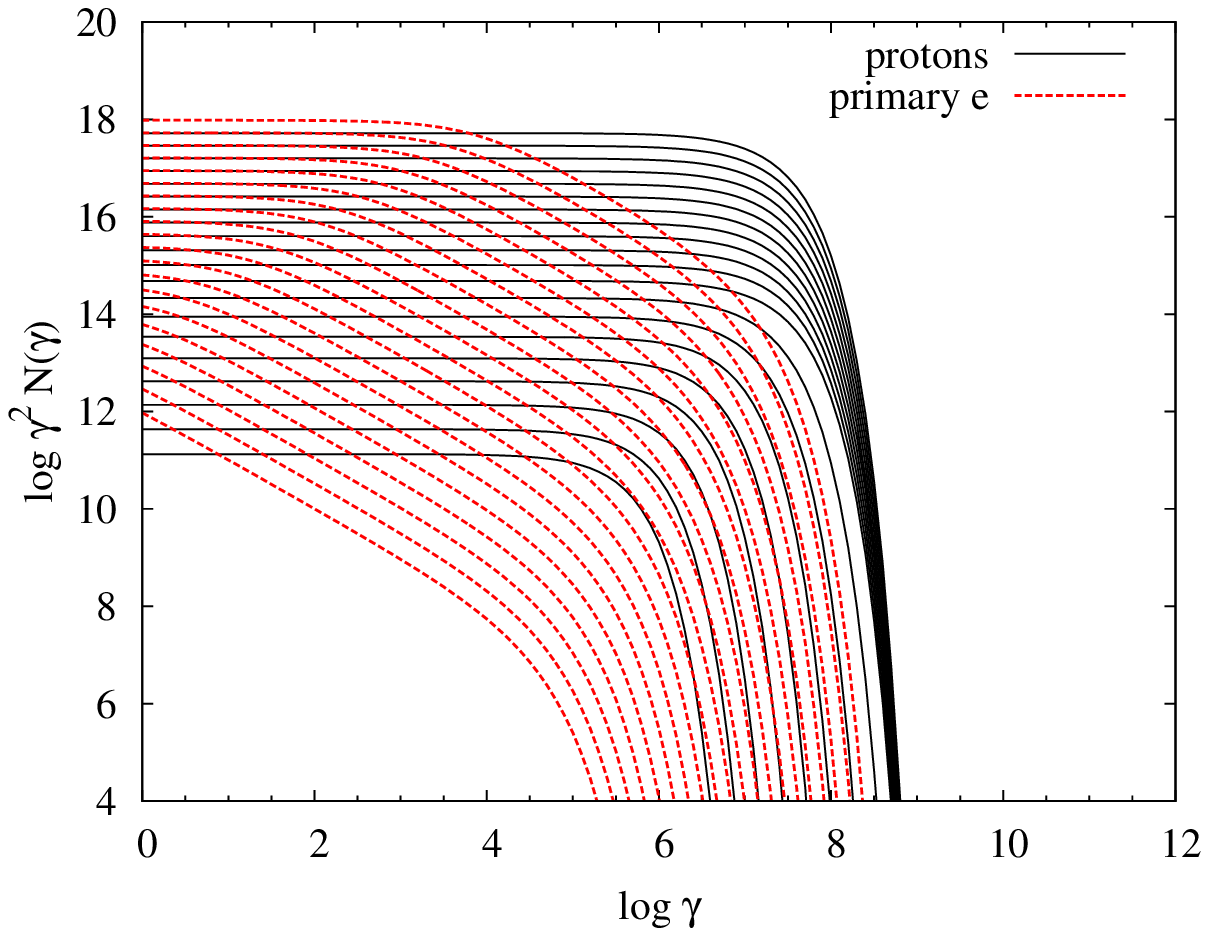}
  \hspace{0.1in}
 \includegraphics[width=0.48\textwidth]{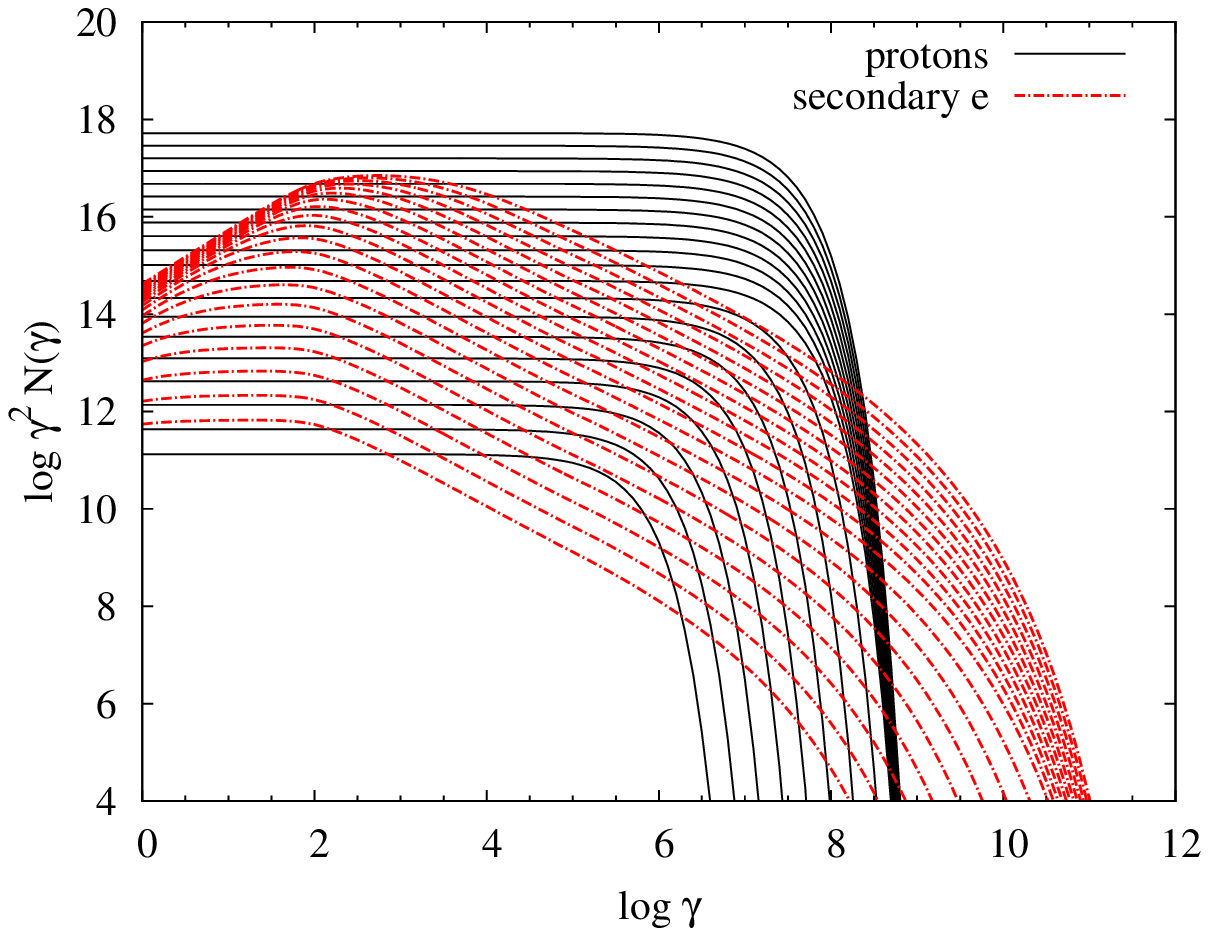}
 \caption{Snapshots of the particle distributions starting from $r=10^{{  0.1}}\rin$ and increasing
 by {  0.25} in logarithmic units over its previous value up to a radius $r=10^{  5}\rin$. {  Left and right panels show the primary and secondary electron distributions (red) compared to the proton distribution (black).}
 The accelerated proton and electron distributions at injection are modelled as  $N_{\rm i}(\gamma) \propto \gamma^{-p} \exp(-\gamma/\gmax)$. The adopted parameters correspond to the default case described in text:
 $w=2$, $\alpha=0$, $\aph=0$, $s=2$, $\vw=0.01c$, $\vo={  0.03}c$, $\rin=10^{13}$~cm \, (i.e., $\Aw=10^{  16}$~gr cm$^{-1}$), $L_{\rm SN}=10^{41}$~erg s$^{-1}$, $M_{\rm ej}={  10}M_\odot$,  $\eB=0.01$, $\epsilon_{\rm p}=0.1$, $K_{\rm ep}=10^{-3}$, and $Q_{\rm 0p}=1$. All results  can be linearly scaled for a realistic value of $Q_{\rm 0p}$.}
 \label{fig:example}
\end{figure*}
In the general case of $s\ne2$, the radial dependence of the total number of cooled and uncooled electrons can also be derived using basic scaling arguments, as we now detail. The number of primary electrons not affected by synchrotron cooling is 
$N_{\rm e} \propto \epsilon_p \kep \vs^2 r^{3-w}$, where the factor $r^{3-w}$ is related to the total accreted mass. As $\vs \propto r^\alpha$ it follows that, quite generally, the number of uncooled primary electrons will be
\eqb
N_{\rm e,<\gcool} \propto  r^{3-w+2\alpha} \gamma^{-s}
\eqe
The number of primary electrons with $\gamma>\gcool$ can then be  simply written as  
\eqb
N_{\rm e,>\gcool}=N_{\rm e,<\gcool}\left(\gcool \right) \left(\frac{\gamma}{\gcool}\right)^{-s-1}.
\eqe
Since $\gcool \propto r^{w-\alpha-1}$  (see \eqn{gbr-0}), we derive the scaling relation 
\eqb
N_{\rm e,>\gcool}\propto r^{3-w+2\alpha} r^{w-\alpha-1} \gamma^{-s-1} \propto r^{2+\alpha}\gamma^{-s-1}.
\eqe
Similar considerations apply to the case of secondary electrons.  Their number is given by $N_{\rm p} t_{\rm dyn}/t_{\rm pp}$, or equivalently
\eqb
N_{\rm e,<\gcool}^{(\rm pp)} \propto r^{3-w+2\alpha} \gamma^{-s} r^{1-\alpha-w},
\eqe
which results in the differential electron distribution 
\eqb
N_{\rm e,<\gcool}^{(\rm pp)}\propto r^{4+\alpha-2 w}\gamma^{-s}.
\eqe
The distribution above the cooling break is then given by
\eqb
N_{\rm e,>\gcool}^{(\rm pp)}\propto r^{3-w}\gamma^{-s-1}.
\eqe

\subsection{An indicative example of particle evolution}
An indicative example for the evolution of the particle distributions as obtained for the default case (see Table~\ref{tab0}) is presented in \fig{example}.
{  For clarity reasons, the primary and secondary electron distributions are displayed in the left and right panels with red dashed lines, respectively.}
For the adopted parameters, synchrotron radiation is the dominant cooling process for electrons (primary and secondary) for a wide range of shock radii, i.e. $r\approx 10^{14}-10^{17}$~cm, which corresponds to  $t\approx {  1.3-1300}$~days for $\vo={  9\times10^3}$~km s$^{-1}$. Here, the deceleration radius is $\rd\simeq 1$~pc. 

At early times, the distribution of electrons is cooled due to synchrotron losses down to $\gamma\sim \gemin$ with the cooling break (see \eqn{gbr-0}) progressively moving to larger values as the magnetic field decreases; the cooling break is identified by the location where the  slope in electron spectrum steepens from $-2$ to $-3$. On the contrary, the proton distribution at all radii has the same shape as at injection, since it is affected only by adiabatic losses. The proton maximum Lorentz factor at injection increases at early times according to \eqn{gmax4} (see the first three black curves from the bottom) and saturates at the value defined by \eqn{gmax1}. The spacing between the curves shown in \fig{example} reflects the radial (or, equivalently time) evolution of the different particle species. In particular, we find that $N_{\rm e}$ at a given Lorentz factor increases faster than $\Nepp$, since the latter depends also on the CSM density profile that decreases with radius. In addition, the radial 
dependence of $N_{\rm i}$ as obtained numerically agrees quantitatively  with the analytical predictions  (see eqs.~(\ref{eq:sol-primary-cool})-(\ref{eq:sol-secondary-uncool})).

Figure \ref{fig:densities} shows with solid lines the evolution of the proton (black lines), primary electron (blue lines), secondary electron (red lines),
magnetic (orange lines), and post-shock thermal ({  cyan} line) energy densities, for the same parameters as in \fig{example}\footnote{{  A comparison  against the energy densities of the thermal radiation fields can be found in \app{app4}.}}. The ratios of the particle (magnetic) to thermal energy densities are overplotted with dashed lines. The ratio $U_{\rm B}/U_{\rm th}$, where $U_{\rm B}=B^2/8\pi$, is constant and equal to $\eB$, as expected. Similarly, we find that $U_{\rm p}/U_{\rm th} \rightarrow 0.1\epsilon_{\rm p,-1}$, where $U_{\rm p} = (\mpr c^2/V) \int {\rm d} \gamma \, \gamma\, N_{\rm p}(\gamma)$ and $V\approx \pi r^3$ is the volume of the shocked shell. The early increase of the ratio $U_{\rm p}/U_{\rm th}$ is the result of the increasing proton maximum energy at small radii. 

% \ls{it may be worth reminding here that our definition of $\epsilon_{\rm p}$ assumes $\gamma_{\rm M,p}\gg1$ for all values of $s>2$}
The ratio $U_{\rm e}/U_{\rm th}$ --- where $U_{\rm e}$ is the primary electron energy density,  defined in a similar way as $U_{\rm p}$ --- increases, albeit with a slow rate. For all the considered radii, we find $U_{\rm e}/U_{\rm th} \ll 1$. This ratio is similar to the parameter $\epsilon_{\rm e}$ that is commonly used in the literature. However, it differs in essence, since $U_{\rm e}$ in our approach takes into account the cooling history of all 
electrons that have been injected at smaller radii. In the absence of cooling, $U_{\rm e}/U_{\rm p} \approx 10^{-3} K_{\rm ep,-3}I_{\rm e}/I_{\rm p}$, where 
$I_{\rm j} = \int_{\gamma_{\rm j, m}}^{\gamma_{\rm j, M}} {\rm d}\gamma \gamma^{-s+1}$. At large radii,  where synchrotron cooling becomes less important, \fig{densities} shows indeed that $U_{\rm e}/U_{\rm p}\approx 10^{-3}$ (see solid black and blue lines). The increase of $U_{\rm e}/U_{\rm th}$  (blue dashed line) is another sign of less efficient electron cooling.
Finally, the energy density of secondary electrons decreases faster compared to all other energy densities and its ratio to $U_{\rm th}$ is the lowest one at large radii.
\begin{figure}
  \includegraphics[width=0.49\textwidth]{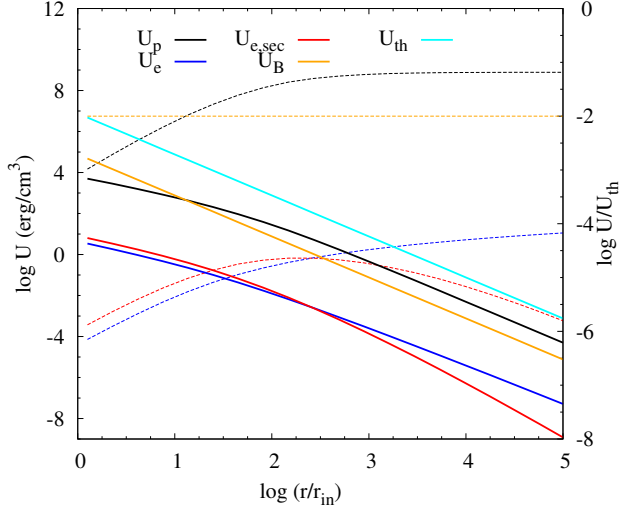}
  \vspace{-0.2in}
 \caption{Evolution of the proton (black lines), primary electron (blue lines), secondary electron (red lines),
magnetic (orange lines), and post-shock thermal ({  cyan} line) energy densities (solid lines), for the same parameters as in \fig{example}. The ratios of the particle (magnetic) to the thermal energy densities are overplotted with dashed lines of the corresponding color. All parameters are the same as in \fig{example} except for $Q_{\rm 0p}$. Here, the exact value as obtained from \eqn{Qop} was used.}
 \label{fig:densities}
\end{figure}
\section{Synchrotron emission}
\label{sec:synchrotron}
Having derived the evolution of the particle distributions as the forward shock propagates in the CSM, the synchrotron emission from both primary and secondary electrons, as well as from relativistic protons can be derived. 
Synchrotron emission at radio frequencies may be suppressed because of absorption processes, such as synchrotron self-absorption and free-free absorption. Moreover, in the presence of a background plasma, the synchrotron emission of a relativistic particle with Lorentz factor $\gamma$ will be suppressed at low enough frequencies, since the beaming of the radiation is not  as strong at frequencies $\ll \gamma \omega_{\rm p}$ , where $\omega_{\rm p}$ is the plasma frequency; this is known as the Razin effect \citep[][]{razin60}.

The intensity of synchrotron radiation at a given radius is given by
\eqb
\label{eq:Issa}
I_\nu = S_\nu(1-{\rm e}^{-\tau_\nu^{\rm ssa}})\rm{e}^{-\tau_\nu^{\rm ff}},
\eqe
where $I_\nu$ is in units of erg cm$^{-2}$ s$^{-1}$ sr$^{-1}$ Hz$^{-1}$.  In the above expression, $S_\nu=j_\nu/\alpha_\nu^{\rm ssa}$, $j_\nu$ is the synchrotron emissivity,  $\alpha_\nu^{\rm ssa}$ is the synchrotron self-absorption coefficient (see e.g. eq.~(6.50) in \citealt{Rybicki79}), $\tau_\nu^{\rm ssa} = \alpha_{\nu}^{\rm ssa} h$ is the optical depth for synchrotron self-absorption, and  $\tau_\nu^{\rm ff} \approx \alpha_{\nu}^{\rm ff} r$ is the optical depth for free-free absorption in the progenitor wind. Assuming that the ionized wind is composed of protons, $T_{\rm e}$  and $n_{\rm csm}$ are the temperature and number density of the unshocked wind, and $hv \ll kT_{\rm e}$,  the free-free absorption coefficient can be written as \citep{rybickilightman86}
\eqb
\alpha_\nu^{\rm ff}(r) = 0.018 T_{\rm e}^{-3/2}n_{\rm csm}^2(r)\bar{g}_{\rm ff} \nu^{-2},
\eqe
where $\bar{g}_{\rm ff}\sim1$ is the velocity averaged Gaunt factor for a Maxwellian distribution and $T_{\rm e} = 10^{5}$ K is a typical value \citep[see e.g.][]{Fransson1996,Bjornsson2014}. The luminosity per unit frequency is then given by 
\eqb
\frac{L_\nu}{4\pi}= 4 \pi r^2 I_\nu ,
\eqe
for $\nu \ge \nu_{\rm Rz}$, where $\nu_{\rm Rz}=2ecn_{\rm csm}(r)/B(r) \propto r^{-w/2-\alpha}$ is the Razin frequency  \citep[e.g.][]{rybickilightman86}. At even lower frequencies the synchrotron emission essentially shuts down.
% This is given by \ls{we can erase this equation, lets refer to rybicki. the equation that follows can also be neglected.}
% \eqb
% \alpha_\nu^{\rm ssa}(r)=-\frac{c^2}{8 \pi \nu^2 m_{\rm i} c^2}\int d\gamma P(\nu;\gamma) \gamma^2 \frac{\partial}{\partial \gamma} \left[\frac{n_{\rm i}(\gamma,r)}{\gamma^2}\right],
% \eqe
% where $P(\nu;\gamma)$ is the single particle synchrotron emissivity (see e.g. eq.~(6.18) in \cite{Rybicki79})\footnote{We note that for the proton synchrotron radiation
% the single particle emissivity as well as the characteristic synchrotron frequency  should be multiplied by $\mel/\mpr$.} and $n_{\rm i}$ is the  volume-averaged number density at a given radius, namely
% \eqb
% n_{\rm i}(\gamma,r)=\frac{N_{\rm i}(\gamma,r)}{4 \pi r^2 h(r)}.
% \eqe

%  The respective optical depth
% is  $\tau_\nu^{\rm ff} = \alpha_{\nu}^{\rm ff} h$, and the synchrotron intensity given by \eqn{Issa} should be
% multiplied by the factor $\exp(-\tau_\nu^{\rm ff})$. 
\subsection{Characteristic frequencies}
\label{sec:frequencies}
The typical synchrotron frequency of electrons with $\gamma=\gcool$, which is defined as $\vcool = 3eB \gcool^2/4\pi \mel c$, reduces to
\eqb
\vcool \approx {  0.03}\, \frac{(w-\alpha)^2}{(w-1)^{3/2}}\beta_{0, {  -1.5}}^{1/2}\epsilon_{\rm B, -2}^{3/2} r_{\rm in, 13}^{-1/2} \left(\frac{r}{\rin}\right)^{3w/2-2-\alpha}\unit{GHz} ,
\eqe
for parameters relevant to this study.  {  The Razin effect becomes important at frequencies below}
\eqb
\nu_{\rm Rz} \simeq 96\, \epsilon_{\rm B, -2}^{-1/2} \beta_{\rm 0, -1.5}^{-2} A_{\rm w, 16}^{-1/2}\frac{\rin}{r}\unit{GHz},
\label{eq:razin}
\eqe
{  where we used the default parameters listed in Table~\ref{tab0}. For most parameters, we find that the Razin effect is not relevant for the radio emission, since $\nu_{\rm Rz}$ lies below the other characteristic absorption frequencies, which we calculate below.}
A characteristic free-free absorption frequency can also be defined by $\tau_{\rm ff}=1$. This is given by 
\eqb
% \nu_{\rm ff} = 0.067 r^{1/2} T_{\rm e}^{-3/4}n_{\rm csm}(r),
\nu_{\rm ff} = 0.134\, r^{1/2} T_{\rm e}^{-3/4}n_{\rm csm}(r),
\label{eq:vff}
\eqe
which scales as $r^{-w+1/2}$. For $w=2$, the free-free absorption frequency reduces to
\eqb
% \nu_{\rm ff} \simeq 2\times10^4 \, A_{\rm w, 15} r_{\rm in, 13}^{-3/2} T_{\rm e, 5}^{-3/4} \left(\frac{r}{\rin}\right)^{-3/2} \unit{GHz}.
\nu_{\rm ff} \simeq {  4\times 10^5} \, A_{\rm w, {  16}} r_{\rm in, 13}^{-3/2} T_{\rm e, 5}^{-3/4} \left(\frac{r}{\rin}\right)^{-3/2} \unit{GHz}.
\label{eq:vff-wind}
\eqe
Typically, the synchrotron emission at $1-50$~GHz, where most radio observing facilities currently operate (e.g. JVLA, WSRT, GMRT etc.) 
will be free-free absorbed at early times, unless $\Aw\sim10^{12}-10^{13}$~g cm$^{-1}$ \citep[for a detailed analysis, see][]{murase14}. 

Similarly, the synchrotron self-absorption frequency can be calculated through  $\alpha_\nu^{\rm ssa} h =1$. In particular, an explicit solution of the synchrotron self-absorption frequency $\nu_{\rm ssa}$ can be derived using the $\delta$-function approximation for the single particle synchrotron emissivity and the expressions of $N_{\rm e}$ and $N_{\rm e}^{(\rm pp)}$ derived for $s=2$ and $\gamma \gtrless \gcool$ (see \sect{distributions}).
For typical parameter values, the electron distribution is expected to be cooled down to Lorentz factors close to the minimum one. For the sake of simplicity we, therefore, present the expression for $\nu_{\rm ssa}$ as obtained using eqs.~(\ref{eq:sol-primary-cool}) and (\ref{eq:sol-secondary-cool}), i.e. for the cooled electron distributions:
\eqb
\label{eq:vssa}
\nu_{\rm ssa}(r) = \nu_0 X(r) \kep^{2/7} \ep^{2/7} \eB^{1/14}\beta_0^{9/14}\rin^{-5/14}
\eqe
where $X(r)$ is a function defined as 
\eqb
\label{eq:X}
X(r)=\left(\frac{r}{\rin}\right)^{-5w/14+\alpha}\left[1+\frac{3\times10^{-3}(w-1)}{(4-w+2\alpha)\kep \beta_0^2}\left( \frac{r}{\rin}\right)^{-w+1-\alpha}\right]^{2/7}
\eqe
and $\nu_0$ is a numerical constant given by
\eqb
\nu_0^{7/2}=\frac{45\mpr}{2^{10}\mel\pi}\frac{\left(\pi \mu \mpr c^4\right)^{1/4}}{\ln\left(\frac{\gpmax}{\gpmin}\right)}\left(\frac{4(w-1)c^2}{\sth}\right)^{5/4}\left(\frac{3\mel c^2}{2h \Bcr}\right)^{1/2}
\eqe 
with $\Bcr=4.4\times10^{13}$~G being the critical magnetic field strength. For the sake of simplicity, $\gpmax$ is treated as a constant with a typical value $10^7$. The second term in the parenthesis in \eqn{X} corresponds to the secondary electrons, and is the dominant one at small radii, where the contribution of secondary synchrotron emission is expected to be more significant. 
For the default case (see Table~\ref{tab0}), \eqn{vssa} results in  
\eqb 
\nu_{\rm ssa}(r) \simeq & {  2.5}\times10^3\, \epsilon_{\rm B,-2}^{1/14}\epsilon_{\rm p,-1}^{2/7}K_{\rm ep,-3}^{2/7}\beta_{0,{  -1.5}}^{-9/14}r_{\rm in, 13}^{-5/14} \times\\ \nonumber
&\left(\frac{r}{\rin}\right)^{-5/7}\!
\left[1+\frac{{  1.5\times 10^3}}{K_{\rm ep, -3} \beta_{0,{  -1.5}}^2}\left( \frac{r}{\rin}\right)^{-1}\right]^{2/7} \unit{GHz},
\eqe
which is at least one order of magnitude lower than the free-free absorption frequency given by \eqn{vff-wind}. The peak synchrotron luminosity 
is thus expected at $\nu_{\rm ff}$, unless the density of the CSM is much lower than what is considered here; in this case, however, secondary electrons will not play a major role in the radio emission. 
% \ls{I understand this is an estimate that the referee asked, but it looks a bit out of place here. maybe we can put it at the beginning, when we define the magnetic field, saying that for observations at 100 days the field is .... Maria, you choose, not a big deal anyway.}

\subsection{Synchrotron light curves}
\label{sec:lc}
In this section we present analytical expressions for the power-law decay index of the optically thin synchrotron flux, using the $\delta$-function approximation for the single particle synchrotron emissivity. Starting from relations that apply to the cooled part of the electron distribution ($\gamma > \gcool$) with $s=2$, we find the optically thin synchrotron luminosity at $\nu > \vcool$ to be
\eqb
\label{eq:Lv-primary-cool}
L_{\nu > \vcool}=\frac{1}{2}Q_{\rm 0e}\mel c^2 \vo  \left(\frac{r}{\rin} \right)^{2-w+3\alpha}\nu^{-1}
\eqe
for shock- accelerated primary electrons, and 
\eqb
\label{eq:Lv-secondary-cool}
L_{\nu>\vcool}^{({\rm pp})}=\frac{1}{2} Q_{\rm 0e}^{(\rm pp)} \mel c^2 \vo  \left(\frac{r}{\rin} \right)^{3-2w+2\alpha}\nu^{-1}
\eqe
for secondary electrons. At frequencies below the cooling break frequency ($\nu < \nu_{\rm c}$), the above expressions become
\eqb
\label{eq:Lv-primary-uncool}
L_{\nu < \vcool}=\frac{Q_{\rm 0e} \rin \sth c B_0^{3/2}}{12\pi \left(4-w+2\alpha\right)}\left(\frac{h \Bcr}{\mel c^2}\right)^{1/2}\left(\frac{r}{\rin}\right)^{3-7w/4+7\alpha/2}\nu^{-1/2}
\eqe
for shock-accelerated primary electrons, and 
\eqb
\label{eq:Lv-secondary-uncool}
L_{\nu<\vcool}^{({\rm pp})}=\frac{Q_{\rm 0e}^{(\rm pp)}\rin \sth c  B_0^{3/2}}{12\pi \left(5-2 w+\alpha\right)}\left(\frac{h \Bcr}{\mel c^2}\right)^{1/2} \left(\frac{r}{\rin} \right)^{4-11w/4+5\alpha/2}\nu^{-1/2}
\eqe
for secondary electrons. For the derivation of the above, we made use of eqs.~(\ref{eq:sol-primary-uncool}) and (\ref{eq:sol-secondary-uncool}).
Moreover, \eqn{Lv-secondary-cool} and \eqn{Lv-secondary-uncool} are derived for the case where the evolution of the proton distribution is dictated by adiabatic losses, which holds for most shock radii (see discussion after \eqn{Kpp}). 

The  power-law index of the optically thin synchrotron light curves    depends on the temporal evolution of the shock velocity and on the CSM density profile. Interestingly, for the default choice of $w=2$ and $\alpha=0$, the synchrotron flux due to accelerated electrons is expected to be constant  or slightly decaying as $r^{-1/2}$ depending on whether $\nu > \vcool$ or $\nu < \vcool$, respectively. If the  secondary synchrotron radiation dominates in the radio frequencies, the light curve will decay faster due to the decreasing injection rate. We also note that the radial dependences presented in eqs.~(\ref{eq:Lv-primary-cool})-(\ref{eq:Lv-secondary-uncool}) apply also to cases where the accelerated distributions have $s> 2$. 

A simple relation between the power-law decay exponents of the primary ($t^{-\chi_1}$) and secondary ($t^{-\chi_2}$) synchrotron light curves can be derived
\eqb
\chi_2 & = & \left\{ \begin{array}{cc}
                       2\chi_1 +\frac{1+4\alpha}{1-\alpha}, &  \nu > \vcool \\ \nonumber \\
                      \frac{11}{7}\chi_1+\frac{\frac{5}{7}+3\alpha}{1-\alpha},& \nu < \vcool 
                     \end{array}
\right.
\eqe
where we made use of eqs.~(\ref{eq:Lv-primary-cool}) - (\ref{eq:Lv-secondary-uncool}) and the relation $t \propto r^{1-\alpha}$.
% \begin{subnumcases}{\chi_2=}
%                             2\chi_1 +\frac{1+4\alpha}{1-\alpha}, \, \nu > \vcool \label{eq:lc-relation-cool}  \\ \nonumber \\
%                             \frac{11}{7}\chi_1+\frac{\frac{5}{7}+3\alpha}{1-\alpha},\, \nu < \vcool.  \label{eq:lc-relation-uncool}
% \end{subnumcases}
The transition from primary-dominated to secondary-dominated synchrotron emission at a given frequency band will be associated with a change
in the decay slope of the radio light curve. Interestingly, the change $\Delta \chi \equiv \chi_2 - \chi_1$ is independent of the electron cooling regime, namely
\eqb
\Delta \chi = \frac{w-1+\alpha}{1-\alpha}.
\label{eq:Deltachi}
\eqe
Thus, the predicted break in the light curve for a wind-type CSM and a constant shock velocity is  $\Delta \chi=1$. This break should be observable as long as the synchrotron luminosities of primary and secondary electrons are comparable at the transition time. As we show next (see e.g. \fig{lc}), this condition is satisfied at high frequencies ($>100$~GHz) and early times.

\subsection{Synchrotron peak luminosity}
\label{sec:peak}
We present next the expressions for the peak synchrotron luminosity, whenever $\nu_{\rm ff}$ is the peak frequency. Although these are not applicable in all cases, e.g. if $\nu_{\rm ssa} > \nu_{\rm ff}$, they can be used as a quick reference whenever the  mass loading parameter that is inferred from the observations is $\gtrsim 10^{14}$~gr cm$^{-1}$.

The respective peak luminosities can be obtained by inserting $\nu_{\rm ff}$ in the expressions of \sect{lc}. These are given by
\eqb
\label{eq:Lpk-primary-cool}
L_{\rm pk, \nu_{\rm ff} > \vcool} \simeq {  1.8\times 10^{22}}\frac{K_{\rm ep,-3}\epsilon_{\rm p,-1}\beta_{0,{  -1.5}}^{3}T_{\rm e,5}^{3/4} r_{\rm in,13}^{3/2}}{\ln\left(\gamma_{\rm p,M,8}\right)}\left(\frac{r}{\rin}\right)^{3/2+3\alpha}\!\!\frac{\unit{erg}}{\unit {Hz\,s}}
\eqe
and
\eqb
\label{eq:Lpk-secondary-cool}
L_{\rm pk, \nu_{\rm ff}>\vcool}^{(\rm pp)} \simeq \frac{6 \times 10^{  25}(w-1)}{4-w+2\alpha} \frac{\epsilon_{\rm p,-1}\beta_{0,{  -1.5}}T_{\rm e,5}^{3/4} r_{\rm in,13}^{3/2}}{\ln\left(\gamma_{\rm p,M,8}\right)}\left(\frac{r}{\rin}\right)^{5/2-w+2\alpha}\!\!\!\!\!\frac{\unit{erg}}{\unit {Hz\,s}}.
\eqe
If $\nu_{\rm ff} \le \vcool$ similar expressions for the peak luminosity can be derived. These read 
\eqb
\label{eq:Lpk-primary-uncool}
L_{\rm pk, \nu_{\rm ff}<\vcool} &\simeq & {  8.2\times10^{24}}\frac{K_{\rm ep,-3}\epsilon_{\rm p,-1} \epsilon_{\rm B, -2}^{3/4}\beta_{0, {  -1.5}}^{9/4} r_{\rm in, 13}^{3/2}T_{\rm e,5}^{3/8}}{\ln\left(\gamma_{\rm p,M,8}\right)} \times \nonumber \\ \nonumber \\
            & & \frac{(w-1)^{5/4}}{4-w+2\alpha} \left(\frac{r}{\rin}\right)^{11/4-5w/4+7\alpha/2} \frac{\unit{erg}}{\unit {Hz\,s}} 
\eqe
and 
\eqb
L_{\rm pk, \nu_{\rm ff}<\vcool}^{({\rm pp})} & \simeq & {  2.5}\times 10^{28} \frac{\epsilon_{\rm p,-1}\epsilon_{\rm B,-2}^{3/4}T_{\rm e,5}^{3/8}r_{\rm in, 13}^{3/2}\beta_{0,{  -1.5}}^{1/4}}{\ln\left(\gamma_{\rm p,M,8}\right)}\times \nonumber \\ \nonumber \\
                       & & \frac{(w-1)^{9/4}}{5-2w+\alpha}\left( \frac{r}{\rin}\right)^{15/4-9w/4+5\alpha/2} \frac{\unit{erg}}{\unit {Hz\,s}}.
\label{eq:Lpk-secondary-uncool}
\eqe
A few things worth commenting follow:
\begin{itemize}
 \item the radial dependence of the peak luminosity is different if the synchrotron emission is dominated by primary or secondary electrons. In particular, for the case of a CSM wind density profile and a constant shock velocity, we find $L_{\rm pk, \nu_{\rm ff}>\vcool}\propto r^{3/2}$ $\left(L_{\rm pk, \nu_{\rm ff}>\vcool}^{(\rm pp)}\propto r^{1/2}\right)$ and $L_{\rm pk, \nu_{\rm ff}<\vcool} \propto r^{1/4}$ $\left(L_{\rm pk, \nu_{\rm ff}<\vcool}^{(\rm pp)}\propto r^{-3/4}\right)$ for primary (secondary) electrons.
 \item at early times (small radii) observations of the peak luminosity may generally probe the cooled secondary electron distribution (see \eqn{Lpk-secondary-cool}), while at larger radii the peak luminosity tracks the uncooled primary electron distribution (see \eqn{Lpk-primary-uncool}). 
 \item for the parameter values of interest in this study, the electron distributions are expected to be cooled due to synchrotron losses, at least at early times. Thus, a characteristic time $t_{\rm pk,eq}$ can be defined by $L_{\rm pk, \nu_{\rm ff}>\vcool}=L_{\rm pk,\nu_{\rm ff}>\vcool}^{(\rm pp)}$. This is written as 
 \eqb
 \label{eq:teq-cool}
 \frac{t_{\rm pk, eq}}{t_{\rm bo}} \simeq \frac{1}{1-\alpha} \left(\frac{Q_{\rm 0e}^{(\rm pp)}}{Q_{\rm 0e}} \right)^{\frac{1-\alpha}{w-1+\alpha}}\!\! =  \left[\frac{{  3.2\times10^3}(w-1)}{4-w+2\alpha}K_{\rm ep, -3}^{-1} \beta_{0,{  -1.5}}^{-2}\right]^{\frac{1-\alpha}{w-1+\alpha}},
 \eqe
 where eqs.~(\ref{eq:Lpk-primary-cool}) and (\ref{eq:Lpk-secondary-cool}) were used. In the above, $t_{\rm bo}\approx \rin/\vo$ is the time of the shock breakout as long as this happens in the CSM wind. Interestingly, $t_{\rm pk, eq}$ depends only on two parameters, namely $\kep$ and $\vo$. As the shock velocity at breakout can be inferred from observations \citep[e.g.][]{Waxman2007,katzbudnik10}, 
the transition time  from secondary to primary synchrotron emission, {  at $\nu_{\rm pk}$},  depends solely on {  one microphysical parameter related to the acceleration process, namely $\kep$.} For $w=2$ and $\alpha=0$, one finds
 \eqb
 \label{eq:teq-wind-cool}
 \frac{t_{\rm pk, eq}}{t_{\rm bo}} \simeq {  1.6\times10^3}\, K_{\rm ep, -3}^{-1} \beta_{0,{  -1.5}}^{-2}.
 \eqe
\end{itemize}
The peak synchrotron luminosity is therefore expected to be dominated by the radiation of secondary electrons for
\eqb
% \label{eq:t_sec1}
% t & \lesssim &  {  190} \, K_{\rm ep,-3}^{-1} r_{\rm in, 13}/\beta^{  3}_{0,{  -1.5}}\unit{d}, \, {\rm or} \\
t\le t_{\rm pk, eq} \simeq {  190} \, K_{\rm ep,-3}^{-1} A_{\rm w, 16}/\beta^{  2}_{0,{  -1.5}}\unit{d}.
\label{eq:t_sec2}
\eqe
If the peak frequency lies below the cooling break frequency, i.e. $\nu_{\rm ff} < \vcool$, then this characteristic time is obtained from eqs.~(\ref{eq:Lpk-primary-uncool}) and (\ref{eq:Lpk-secondary-uncool}), and  has the same dependence on the shock velocity and $\kep$ (but a different coefficient). 

At the transition time the peak frequency is expected at 
\eqb
\nu_{\rm pk, eq} \simeq {  5.6} \, T_{\rm e, 5}^{-3/4} A_{\rm w, {  16}}^{-1/2}K_{\rm ep,-3}^{3/2}\beta_{0, {  -1.5}}^{3/2}\unit{GHz}
\label{eq:vpk-wind-cool}
\eqe
where we assumed that $\nu_{\rm ff}$ is the peak synchrotron frequency and made use of \eqn{vff} and \eqn{teq-wind-cool}. {  The analytical estimates
for $t_{\rm pk, eq}$ and $\nu_{\rm pk}$ are in agreement with the numerical results presented in the next section (see e.g. Fig.~\ref{fig:spectra}). }
\begin{figure}
 \centering 
\includegraphics[width=0.48\textwidth]{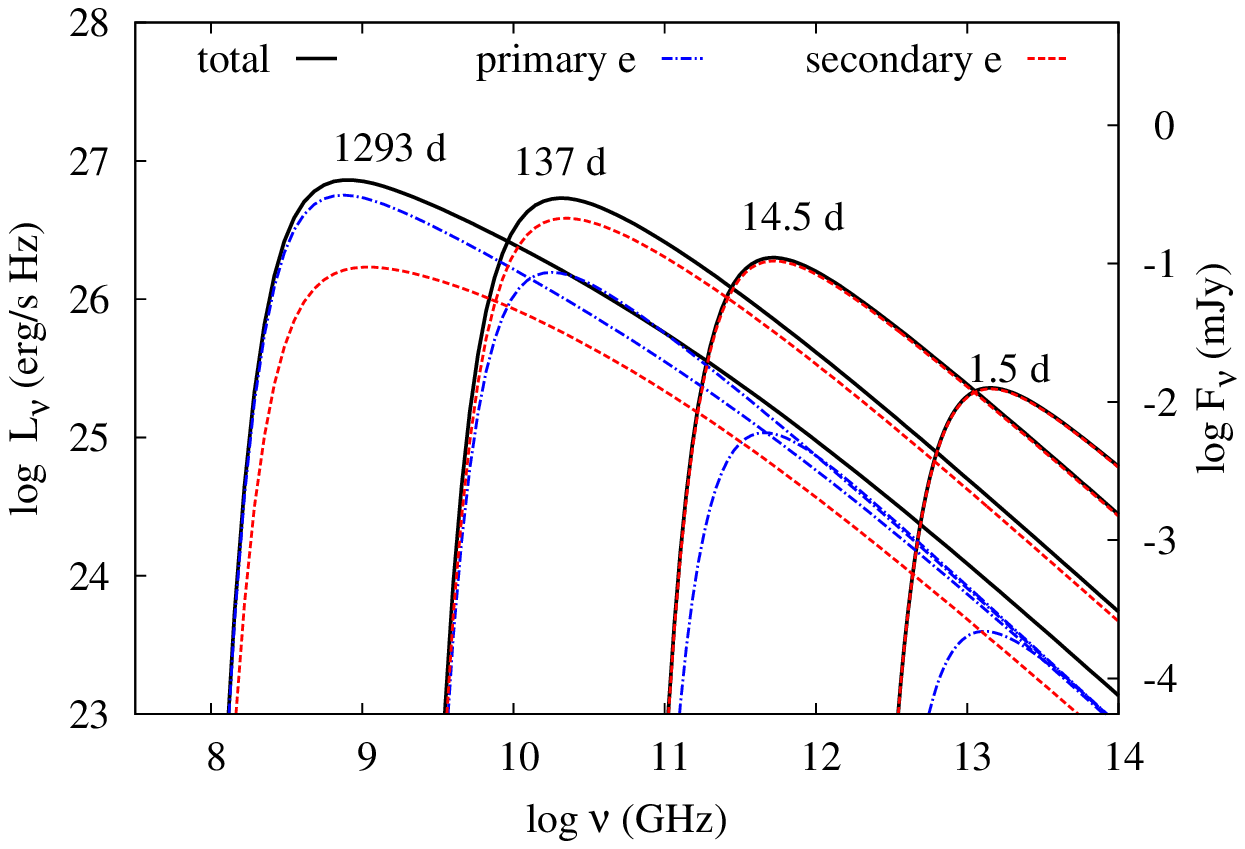} 
\hspace{0.2in}
\includegraphics[width=0.48\textwidth]{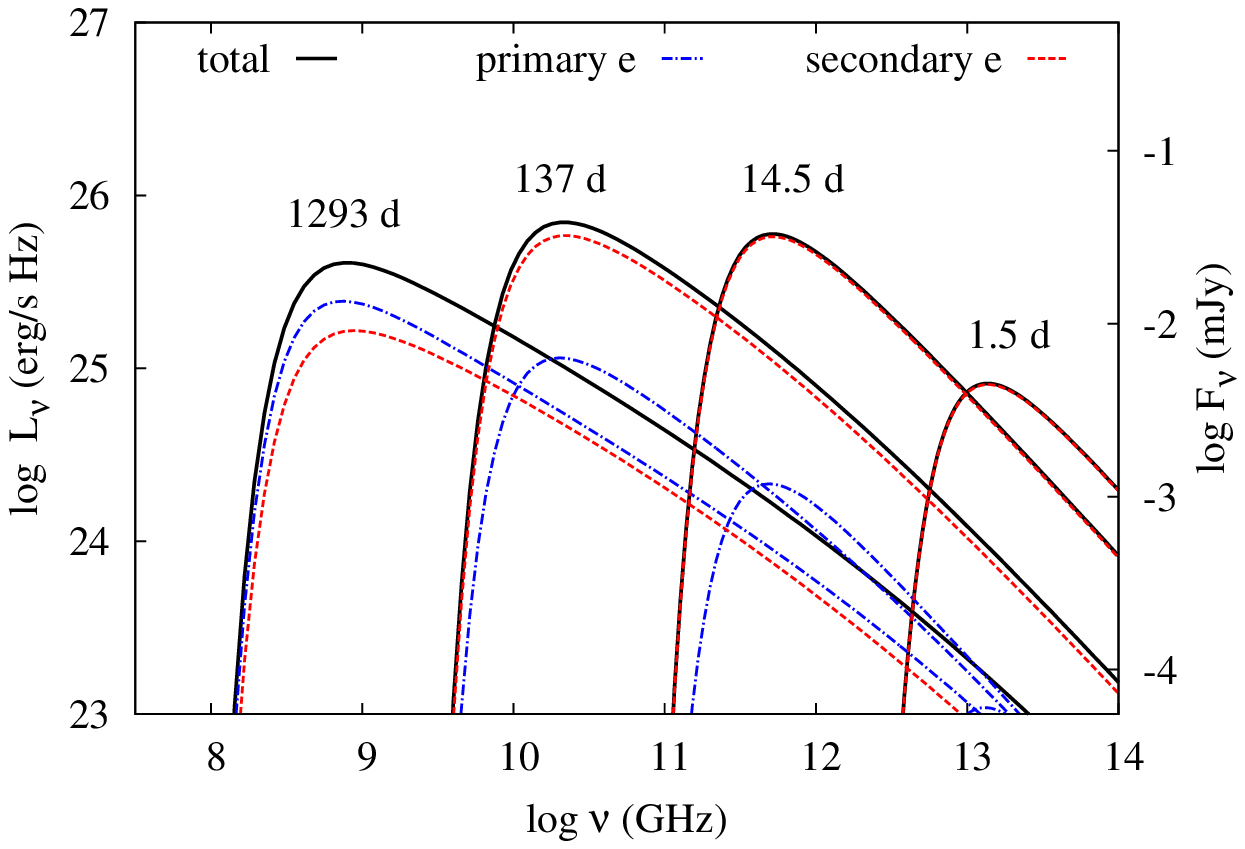} 
\caption{Snapshots of the synchrotron spectra (thick black lines) calculated for a wind-like CSM density profile ($w=2$) with mass loading parameter $\Aw=10^{  16}$~g cm$^{-1}$ and a constant shock velocity $\vo={  0.03}c$, after taking into account the Razin suppression, synchrotron self-absorption, and free-free absorption due to an electron plasma with temperature $T_{\rm e}=10^5$~K. {  Top} and {  bottom} panels are obtained for $\eB=10^{-2}$ and $10^{-4}$, respectively. The contribution of the primary (blue dashed-dotted lines) and secondary (red dashed lines) electrons to the total synchrotron emission is also shown. The proton synchrotron contribution is negligible and for that reason not shown. For the flux conversion, a luminosity distance $d_{\rm L}=${  40}~Mpc was assumed. For the other parameters, see Table~\ref{tab0}.}
\label{fig:spectra}
\end{figure}

\section{The effects of model parameters}
\label{sec:numerical}
In the following paragraphs we investigate the effect of the most important model parameters ($\eB, \alpha, \vo$, and $w$) on the radio spectra and light curves. Each case will be compared against the default one with parameters summarized in Table~\ref{tab0}.  The results that follow are numerically calculated based on the formalism presented in \sect{evolution} and should be compared to the analytical expressions derived in the previous sections.
\subsection{The role of $\eB$}
Figure~\ref{fig:spectra} shows snapshots of the synchrotron spectra including the Razin suppression, synchrotron self-absorption, and free-free absorption, which is the dominant absorption process for the considered radii. {  Top} and {  bottom} panels are obtained for $\eB=10^{-2}$ and $10^{-4}$, respectively.  
% A few remarks on \fig{spectra} follow.
% \begin{itemize}
In both cases we find that at early times the synchrotron emission from the SN shock is dominated by the emission of secondaries electrons, whereas the opposite situation is realized at late times. For $\eB=10^{-2}$ ({  top} panel), the spectral index of the optically thin synchrotron spectrum is $\beta=s/2=1$  ($L_{\nu}\propto \nu^{-\beta}$) at all radii, regardless whether the emission is dominated by primary or secondary electrons. The {  bottom} panel of \fig{spectra} illustrates the effect of a lower value of $\eB$ on the synchrotron spectra. At early times, the magnetic field is still strong enough for the electrons to cool, thus leading to a spectral index $\beta=s/2=1$. At later times though, synchrotron cooling becomes less efficient and the synchrotron spectral index becomes $\beta=(s-1)/2=1/2$. The transition between the two cooling regimes is evident in the spectra of both primary and secondary electrons. In both cases, however, the spectral slope alone cannot be used to differentiate between 
primary and 
secondary synchrotron radiation, unless the shock-accelerated proton and electron distributions have different power-law slopes at injection. We note also that the peak luminosity in the right panel remains constant during the transition  from secondary-dominated to primary-dominated synchrotron emission.
\begin{figure}
 \centering 
\includegraphics[width=0.45\textwidth]{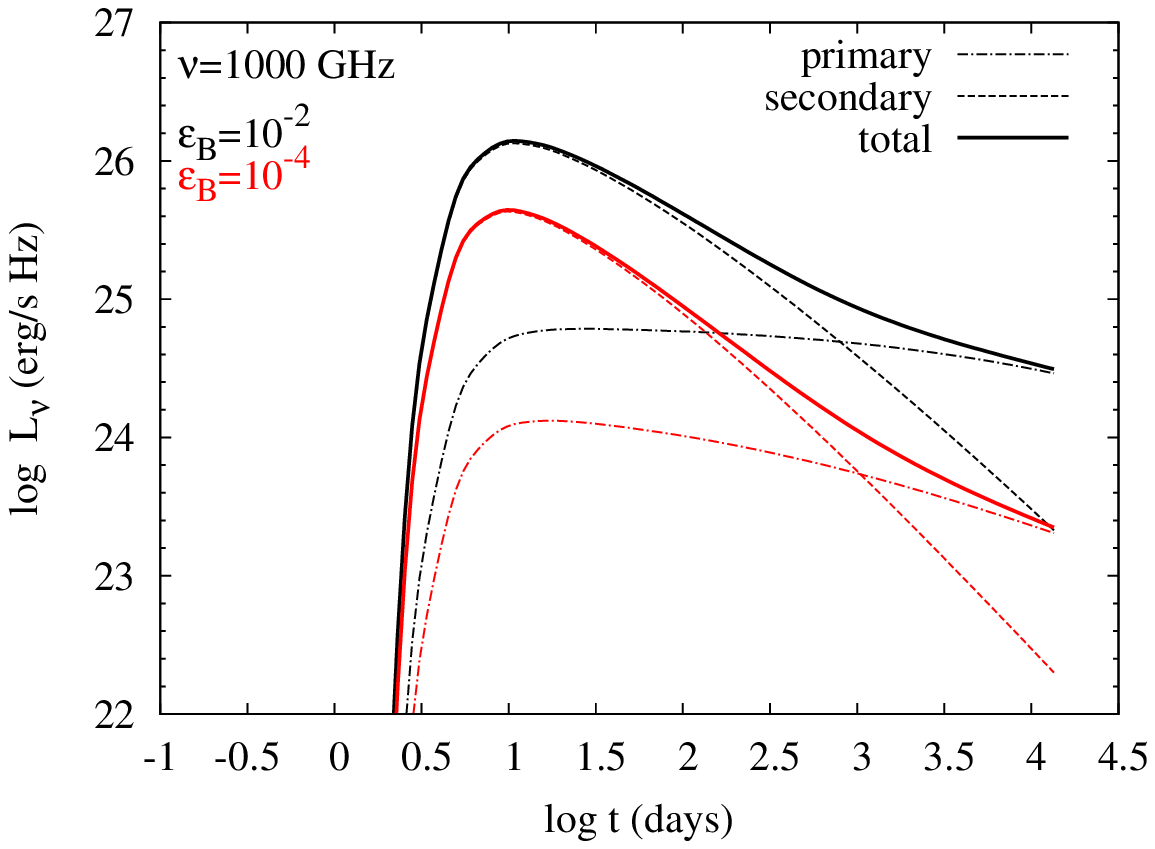} 
\includegraphics[width=0.45\textwidth]{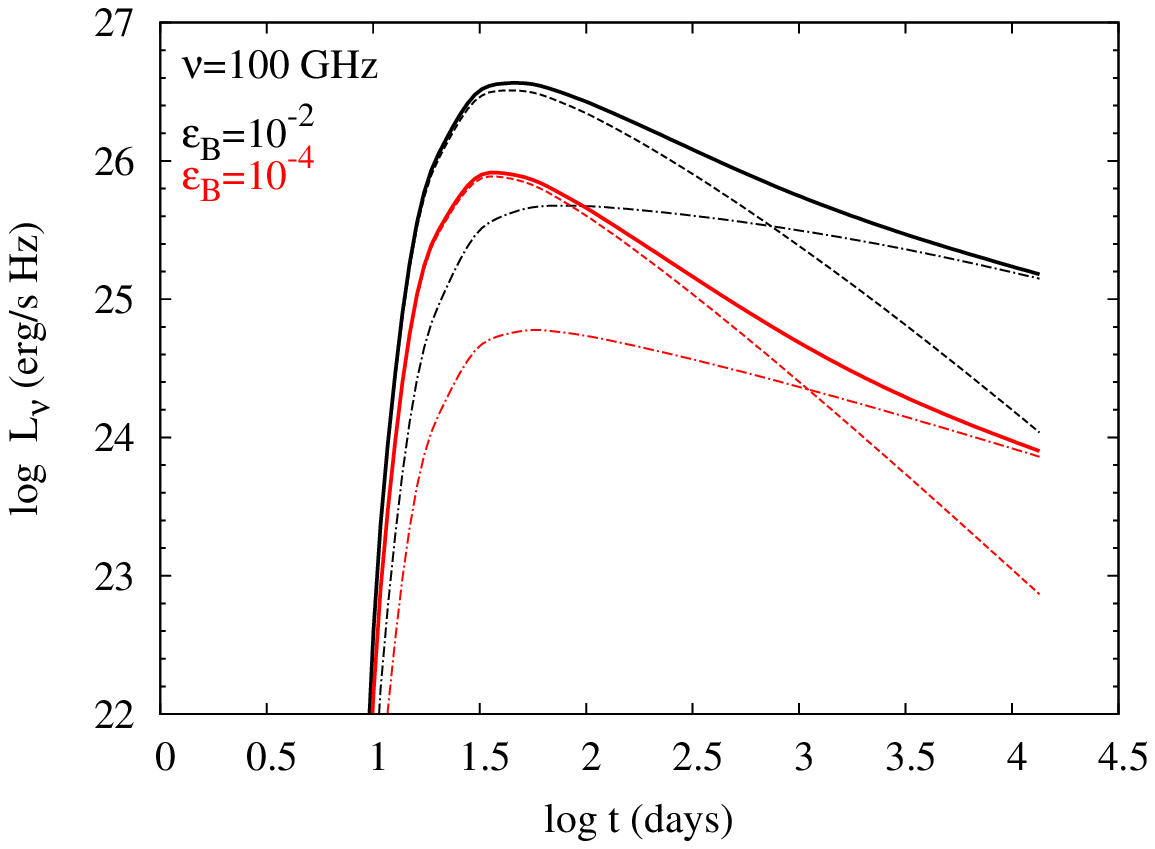} 
\includegraphics[width=0.45\textwidth]{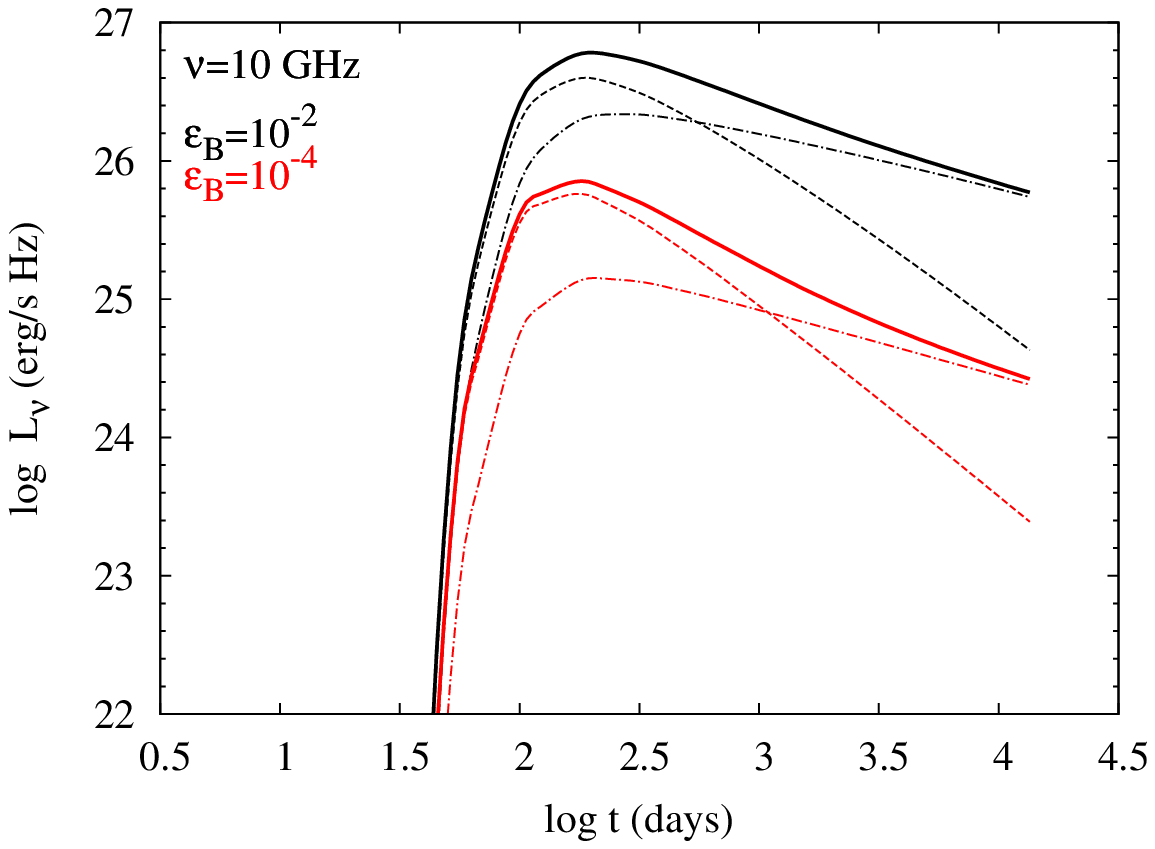} 
\caption{Radio light curves (thick solid lines) calculated at 1000~GHz, 100~GHz, and 10~GHz (from {  top to bottom}) for the cases presented in \fig{spectra}. Black and red colored lines correspond to $\eB=10^{-2}$ and $\eB=10^{-4}$, respectively. Different types of lines show the contribution of primary and secondary electrons to the total synchrotron emission, as indicated in the legend. All other parameters used are listed in Table~\ref{tab0}.}
\label{fig:lc}
\end{figure}

\begin{table}
\centering
\caption{Power-law decay index of the primary and secondary synchrotron light curves as determined numerically for $\alpha=0$ and two values of $\eB$. All other parameters are the same as in \fig{spectra} and \fig{lc}. For reference, the analytically derived values for $\nu > \vcool$ ($\nu<\vcool$) are $\chi_1=0$ and $\chi_2=1$ ($\chi_1=0.5$, $\chi_2=1.5$).}
% \begin{threeparttable}
\begin{tabular}{c|cc|cc}\hline
\diaghead{\theadfont ColumnmnHead}%
{$\nu$\,(GHz)}{$\eB$}& \multicolumn{2}{c}{\thead{$10^{-2}$}}&\multicolumn{2}{c}{\thead{$10^{-4}$}} \\   \hline
     & $\chi_1$ & $\chi_2$ & $\chi_1$ & $\chi_2$ \\ \hline
1000 &  {  0.1}        &  {  1.0}   & {  0.3}    &      {  1.1}      \\   
100  &  {  0.2}        &  {  0.9}  & 0.4   &      {  1.3}     \\    
10   &  {  0.3}        &  {  1.0}   & {  0.4}    &      {  1.3}     \\    
\hline
\end{tabular}
%   \tnote{a} For the 2009 data, a flatter proton distribution with $s_{\rm p}=0.6$ was assumed.
%  \end{threeparttable}
\label{tab1}
\end{table} 
The radio light curves for the cases exemplified in \fig{spectra} have been calculated at three characteristic frequencies (1000, 100 and 10~GHz) and presented in \fig{lc} (from {  top to bottom}). The transition from secondary-dominated to primary-dominated synchrotron emission is demonstrated by a change
in the decay slope of the light curve at 1000~GHz and 100~GHz. Moreover, the time of the transition does not depend on $\eB$. All these features are in agreement with our analytical predictions in \sect{lc} and \sect{peak}. 

The power-law decay indices, whenever these can be defined, of the primary and secondary synchrotron fluxes are listed in Table~\ref{tab1}. These should be compared to the analytical predictions for $\alpha=0$, namely $\chi_1=0$ and $\chi_2=1$ ($\chi_1=0.5$ and $\chi_2=1.5$, respectively) for $\nu > \vcool$ ($\nu < \vcool$, respectively). Indeed, for $\eB=10^{-2}$ we find that the numerically derived indices are closer to the values given by eqs.~(\ref{eq:Lv-primary-cool}) and (\ref{eq:Lv-secondary-cool}), while $\chi_{1,2}$ approach the asymptotic values 0.5 and 1.5, respectively, for $\eB=10^{-4}$. 
Small deviations from the analytical predictions are expected, since these are valid asymptotically, i.e. the distribution of electrons radiating at a given radio frequency should be described as $\gamma^{-s}$ or $\gamma^{-s-1}$.

\subsection{The role of $\Aw$}
To exemplify the role of the mass loading parameter we calculated the 10~GHz radio light curves for the default case 
of a wind-type CSM ($w=2$) and a constant shock velocity ($\alpha=0$). The results for $\Aw$ ranging between $10^{15}$~gr cm$^{-1}$ and $10^{17}$~gr cm$^{-1}$
are presented in \fig{lc_Aw}. The light curves of the primary (dashed-dotted lines) and secondary (dashed lines) synchrotron emission at the chosen frequency are also shown.  The calculations have been performed up to deceleration radius, which increases for lower values of $\Aw$ (see \eqn{rdec-general}).  Both the peak synchrotron luminosity and peak time increase for higher $\Aw$ values. The latter is the result of a higher free-free absorption frequency due to the denser CSM. This also leads to a higher magnetic field strength and a higher particle injection rate, which explains the increase of the peak luminosity. As $\Aw$ increases, the contribution of secondary electrons to the observed emission becomes larger, since the dependence of the secondary injection rate $\Qepp$ on the CSM density is stronger (see e.g. eqs.~(\ref{eq:Qop}), (\ref{eq:Qoe}) and (\ref{eq:ratio})). 
\begin{figure}
\centering 
\includegraphics[width=0.48\textwidth]{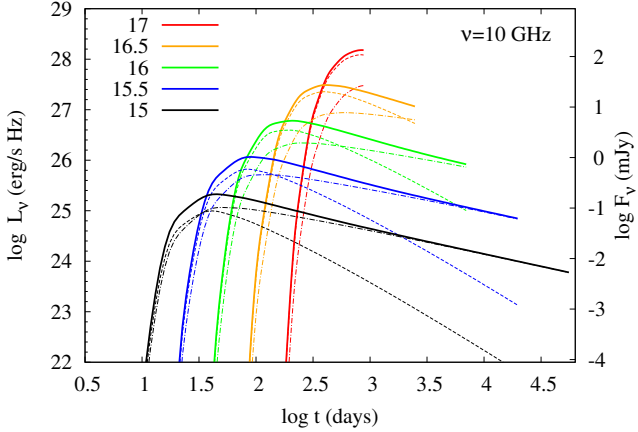}
\caption{Radio light curves at {  10~GHz} for different values of the mass loading parameter $\Aw$. The (logarithmic) values (in g cm$^{-1}$ units) are listed in the inset legend. The contribution of primary and secondary synchrotron emission to the total one (thick solid lines) is shown as dashed-dotted and dashed lines, respectively.  The results are plotted up to the deceleration time, which explains the abrupt interruption of the light curves. For the calculation of the radio flux, a source at a fiducial distance of 10~Mpc was adopted. All other parameters used are listed in Table~\ref{tab0}. Here, the ejecta mass was assumed to be $5 M_{\odot}$.}
\label{fig:lc_Aw}
\end{figure}
\subsection{The role of $\vo$}
The shock-breakout velocity $\vo$ may be indirectly inferred from the optical light curves (rise time and peak luminosity) of interaction-powered SNe \citep[e.g.][]{ofek14}. As typical values lie in the range $5\times 10^3-10^4$~km s$^{-1}$ \citep{ofek14b}, we have so far presented results for $\vo=0.03c$ ($9\times 10^3$~km s$^{-1}$). Here, we demonstrate the effects of the shock velocity at breakout on the radio light curves by adopting $\vo=0.1c$ ($3\times 10^4$~km s$^{-1}$)\footnote{Such fast shocks are often inferred for SNe associated with GRBs  \citep[e.g.][]{Soderberg2010, margutti13}.}. As illustrated in Fig.~\ref{fig:lc_vo}, higher shock velocities result in more luminous radio emission (see also eqs.~(\ref{eq:Lpk-primary-cool})-(\ref{eq:Lpk-secondary-uncool})). This can be understood as an increase of the magnetic field strength and of the post-shock thermal energy density (see e.g. eq.~(\ref{eq:Qop})). Slower shocks, on the other hand, favour the production of secondary electrons, in agreement 
with the analytical predictions (\sect{secondary}). The transition time between secondary-dominated to primary-dominated synchrotron emission is strongly dependent on the shock velocity; for $\vo=0.1c$ the transition occurs at $\sim 10$~d, whereas secondaries dominate the 100~GHz radio emission until $10^3$~d for $\vo=0.03c$. 
\begin{figure}
\centering 
\includegraphics[width=0.48\textwidth]{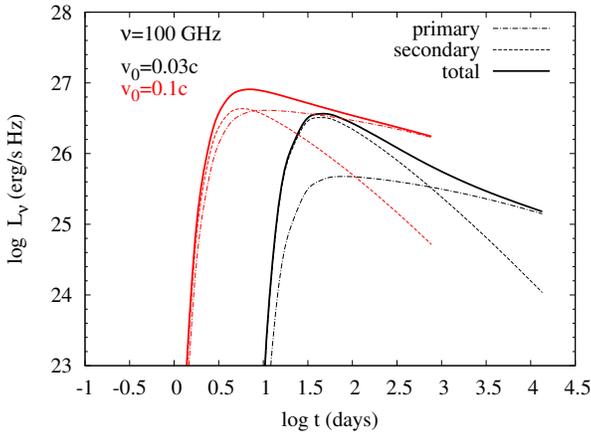}
\caption{Radio light curves at 100~GHz for $\vo=9\times10^3$~km s$^{-1}$ (black lines) and $\vo=3\times10^4$~km s$^{-1}$. All other parameters are listed in Table~\ref{tab0}.}
\label{fig:lc_vo}
\end{figure}
\subsection{The role of $T_{\rm e}$}
The free-free absorption frequency is sensitive to the temperature of the upstream ionized CSM ($\nu_{\rm ff} \propto T^{-3/4}_{\rm e}$).  This is exemplified in Fig.~\ref{fig:lc_Te}, where the 100~GHz and 10~GHz light curves are plotted for $T_{\rm e}=10^5$ (black, blue  lines) and $10^4$~K (red, orange lines). A lower temperature shifts the  peak time to later times, thus decreasing the time interval where the contribution of secondaries to the observed emission is significant. In particular, at 10~GHz the radio emission is expected to be dominated by the synchrotron emission of primary electrons at all times, unless $T_{\rm e} \gtrsim 10^4$~K (see blue and orange lines). 
\begin{figure}
\centering 
\includegraphics[width=0.48\textwidth]{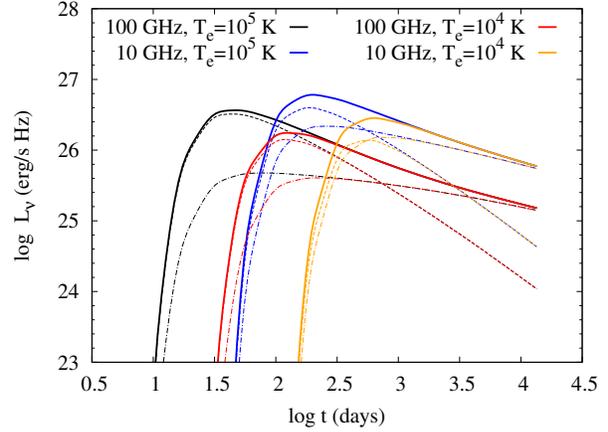}
\caption{Radio light curves at 100~GHz and 10 GHz for two temperature values of the ionized unshocked CSM as indicated on the plot. All other parameters are listed in Table~\ref{tab0}.}
\label{fig:lc_Te}
\end{figure}
\begin{figure}
 \centering 
\includegraphics[width=0.48\textwidth]{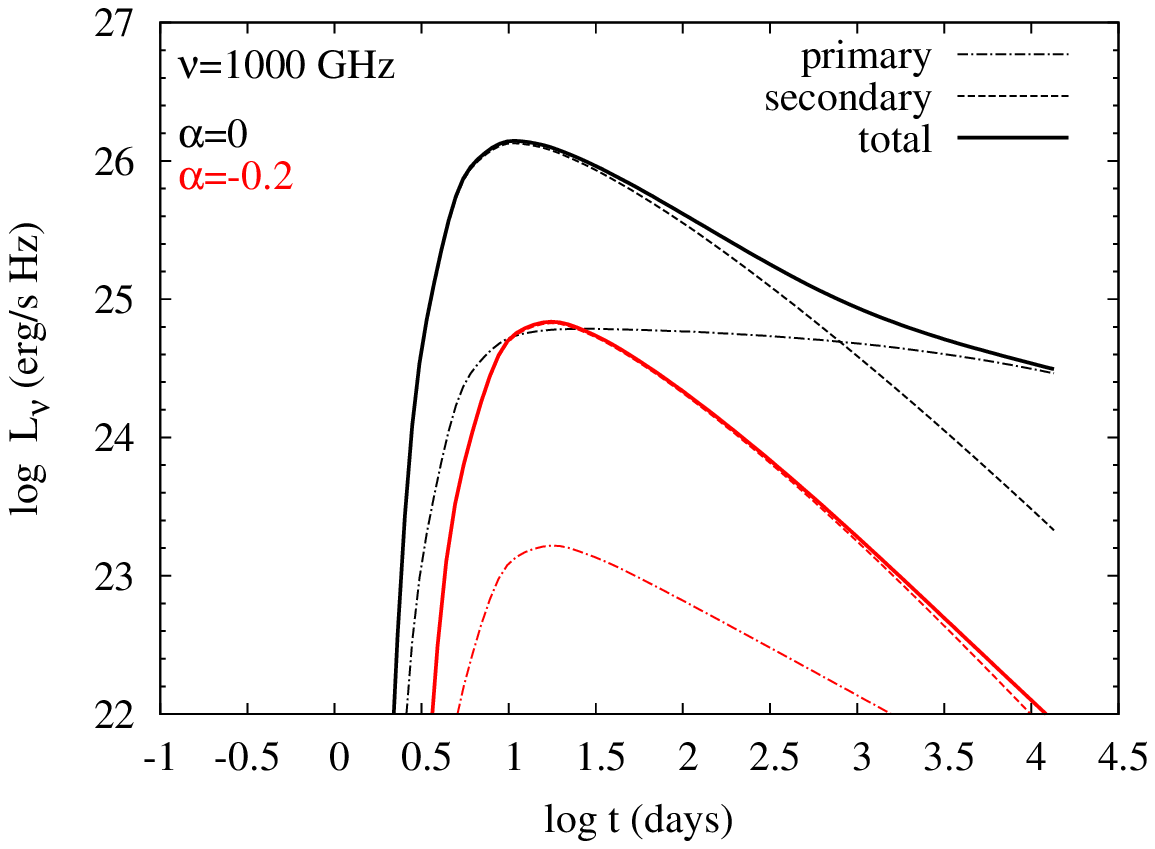} 
\includegraphics[width=0.48\textwidth]{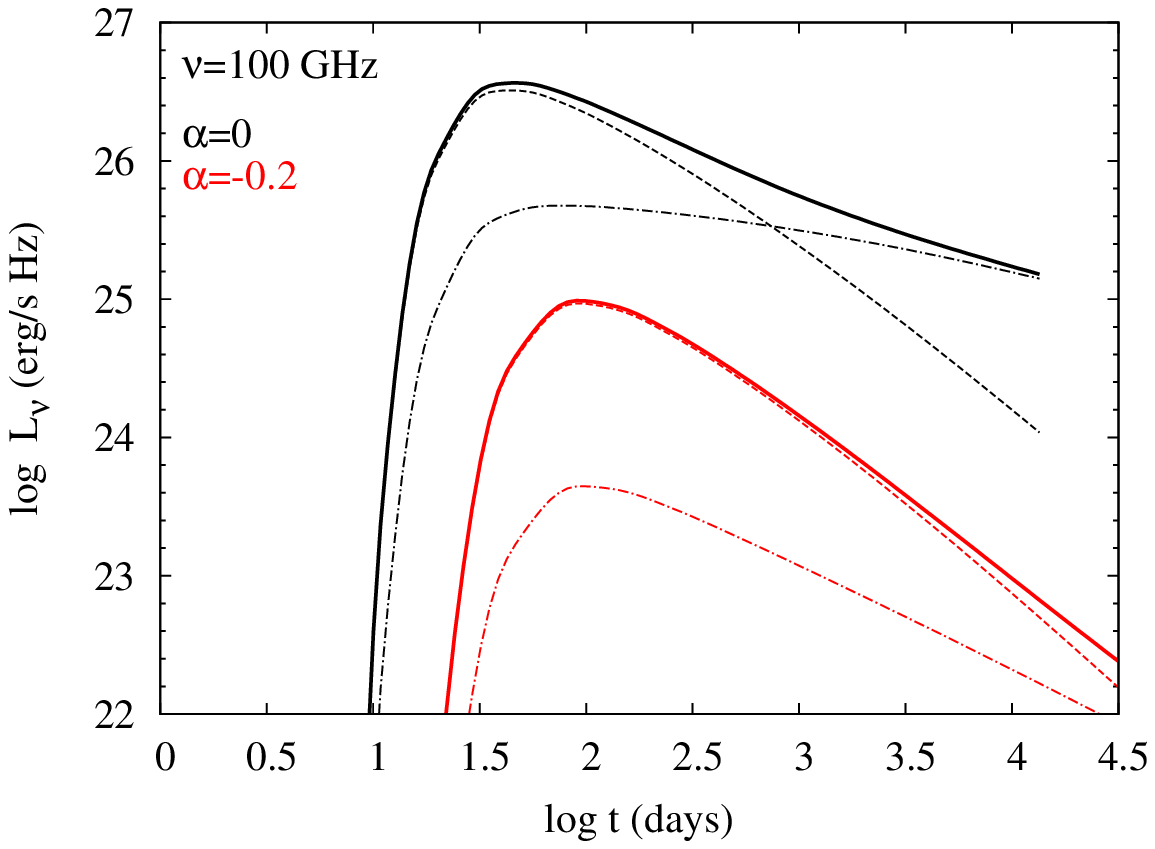} 
\includegraphics[width=0.48\textwidth]{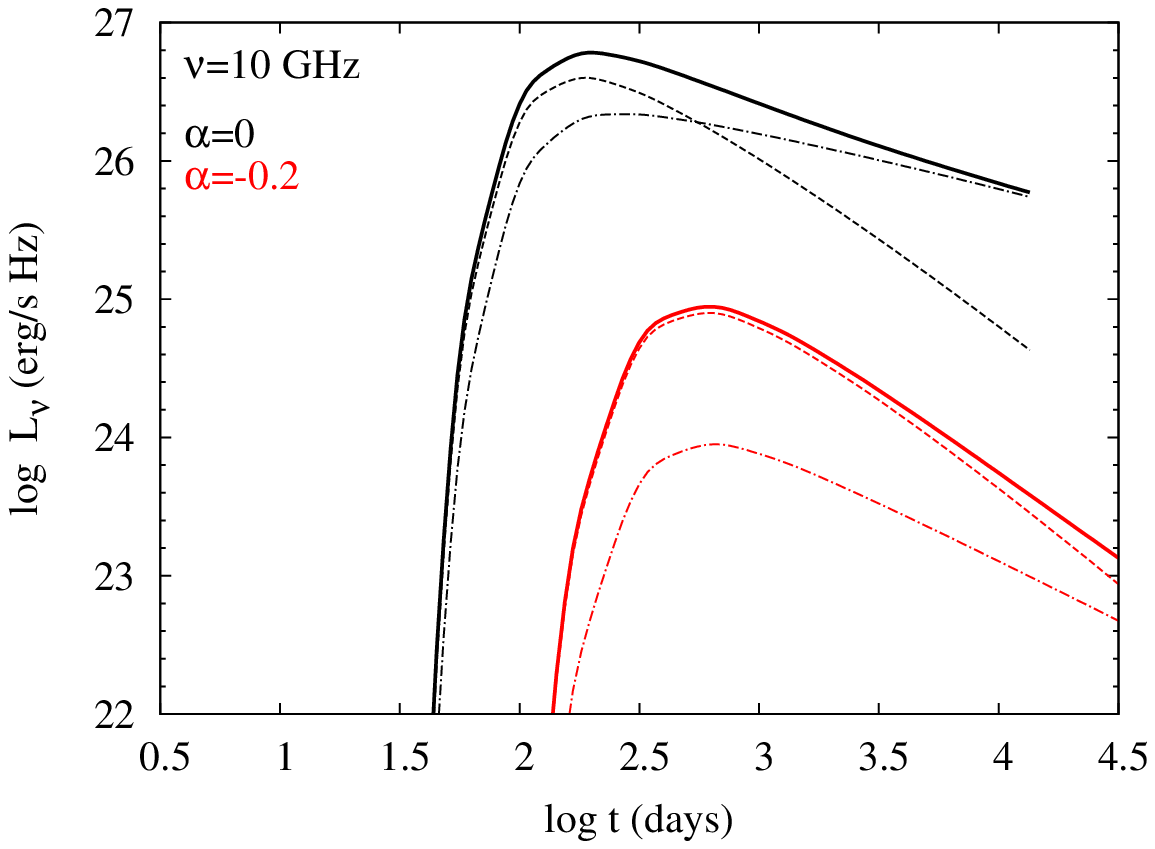} 
\caption{Radio light curves calculated at 1000~GHz, 100~GHz, and 10~GHz (from {  top to bottom}) for a weakly decelerating SN shock with $\alpha=-0.2$ (red lines).  In all panels, the results for the case of a constant shock velocity ($\alpha=0$) are plotted for comparison (black lines). Different types of lines show the contribution of primary and secondary electrons to the total synchrotron emission, as indicated in the legend. All other parameters used are listed in Table~\ref{tab0}. 
}
\label{fig:lc2}
\end{figure}
\subsection{The role of $\alpha$}
For specific combinations of the ejecta and CSM density profiles, a weakly decelerating SN shock ($\alpha \lesssim 0$) is a viable outcome \citep{Chevalier1982,Berger2002,Soderberg2008,Soderberg2010,Chakraborti2015,Kamble2014a,Kamble2015,Fransson1996}. 
%{  [Atish]: here you could add some examples, like the 2013Sn, that point to this direction}. 
The radio light curves at 10~GHz, 100~GHz and 1000~GHz for $\alpha=-0.2$ are presented in \fig{lc2}. In each panel, the respective light curves calculated for $\alpha=0$ are overplotted for a direct comparison. In both cases, the radio light curves of secondary synchrotron emission decay faster compared to those of primary synchrotron radiation. The total synchrotron luminosity is lower in the case of a weakly decelerating shock, since the injection rate of protons and, in turn, electrons decreases. It is also notable that in the case of $\alpha<0$, the secondary electrons contribute almost 100\% to the total emission until late times, {  for observing frequencies as low as 10~GHz}. At a given time, the weakly decelerating shock will be located at a smaller distance than that traveled by a shock moving with constant velocity, thus probing a dense CSM even at late times.

\section{Relevance to SN radio observations}
\label{sec:radio}
Several dozens of SNe within the local Universe (luminosity distance $d_{\rm L} < 200 \unit{Mpc}$)
have been detected in radio frequencies (see \citealt{Chevalier2006} and Fig. 6 in \citealt{Kamble2015}). All of these successful detections involve core-collapse SNe of all types with no detection of type Ia SN so far (see e.g. \citet{Chomiuk2015} and references therein). 

The SN radio luminosity has a wide distribution arising primarily from the dispersion in progenitor mass-loss rates and 
SN shock velocities. The measured mass-loss rates of type Ibc SNe, which are related to compact massive Wolf-Rayet stars, are low 
($\dot{M}_{\rm w} \sim 10^{-6} - 10^{-5}\, {\rm M_{\odot}/yr}$) with typical wind velocities $\vw = 1000\, \unit{km \,s}^{-1}$. 
With such low mass-loss rates these SN shocks cannot be  powered by interaction with previously ejected matter from the progenitor. Thus,  their wide range  in luminosity could be attributed to the wide dispersion of their shock velocities. Indeed, the shocks in SNe Ibc are among the fastest ranging 
from $\beta \simeq 0.1 - 0.5$ \citep{Kamble2014b}.

Core collapse SNe due to their massive supergiant progenitors, such  as type II SNe, are the best candidates for interaction-powered SNe. Progenitor mass-loss rates in SNe IIn can be as high as $\dot{M}_{\rm w} \sim 10^{-3} \, \unit{M_{\odot}/yr}$, while their stellar winds are typically slow with $\vw = 10$~km s$^{-1}$). Radio observations of type IIn SNe show evidence of free-free absorption due to the optically thick ionized wind in the progenitor environment.
As a result, the radio emission from these SNe rises slowly, with a typical rise time of $\sim 10^3$ days at $\sim$GHz frequencies.

A growing number of optically very luminous SNe with $L_{\rm bol} \geq 10^{43} \unit{erg s^{-1}}$, which may be also powered by the CSM-shock interaction,  is being detected by optical surveys \citep{Smith2007,Gal-Yam2009,Quimby2011}.  Such an interaction scenario requires unprecedentedly high mass-loss rates for the SN progenitors, approaching $\dot{M}_{\rm w} \sim 1\, {\rm M_{\odot}/yr}$. Only a few attempts to observe the radio emission from SLSNe have been made without any detection so far \citep{Chomiuk2011,Chomiuk2012}. Currently, it is not clear if there is a physical reason behind the absence of a radio signal or if it is an observational effect due to the large distances of SLSNe. 

\subsection{Synchrotron peak luminosity vs. peak time}
\label{sec:Lpk_vpk}
It is instructive to view our results on the peak synchrotron luminosities of primaries and secondaries in the context of radio SNe observations. The peak luminosity versus peak time plot offers the most informative way to project the radio SNe \citep{Chevalier1998, Chevalier2006, kamble15}.
The peak time $t_{\rm pk}$ corresponds to the time when the peak synchrotron frequency, namely $\nu_{\rm pk}\equiv \max(\nu_{\rm ff}, \nu_{\rm ssa})$, crosses a fixed observing frequency {  ($\nu$)}.
For the parameter values we are interested in, it is safe to assume that $\nu_{\rm pk}=\nu_{\rm ff}$. For the case of a wind-type CSM medium and a constant shock velocity, this happens at a shock radius 
\eqb
r_{\rm pk} \propto \Aw^{2/3} \nu^{-2/3}T_{\rm e}^{-1/2},
\label{eq:rpk}
\eqe
while the respective peak time is expressed as
\eqb
t_{\rm pk} \propto  \Aw^{{  2/3}} \beta_0^{-1} \nu^{-2/3} T_{\rm e}^{-1/2}.
\label{eq:tpk}
\eqe 
We remark that both the peak time and radius depend on $\Aw$: a denser CSM shifts $t_{\rm pk}$ to later times, having important consequences for the relative importance of secondary versus primary synchrotron emission. Moreover, the peak time is inversely proportional to the shock velocity, for a given mass loading parameter.

In the following, we assume that the electrons radiating at the peak synchrotron frequency belong to the cooled part of the distribution.
The primary and secondary  synchrotron luminosities at $\nu=\nu_{\rm pk}$ can be obtained by substitution of eqs.~(\ref{eq:rpk})-(\ref{eq:tpk}) into eqs.~(\ref{eq:Lpk-primary-cool}) and (\ref{eq:Lpk-secondary-cool}). Our results are summarized in Fig.~\ref{fig:Lpk} where the maximum of 
$L_{\rm pk, \nu_{\rm ff}>\vcool}$ (colored circles) and $L_{\rm pk,  \nu_{\rm ff}>\vcool}^{(\rm pp)}$ (colored triangles) is plotted as a function of $t_{\rm pk}$. Measurements of radio SNe of various types (for details, see figure caption) are overplotted with filled and open black symbols.  As most observations are performed at the $5$~GHz frequency band, the results shown in Fig.~\ref{fig:Lpk} are obtained for $\nu=5$~GHz.  The two colored curves are obtained from the semi-analytical model for $\vo=0.1c$ and $0.03c$, as marked on the plot.  The color coding corresponds to the value of the mass loading parameter $\Aw$ (see \eqn{rho-wind}), as indicated in the color bar at the top. The black solid line is the locus of points with $L_{\rm pk, \nu_{\rm ff}>\vcool}=L_{\rm pk,\nu_{\rm ff}>\vcool}^{(\rm pp)}$. It divides the plot in two regions where the peak luminosity  at $\nu_{\rm pk}=5$~GHz and $t_{\rm pk}$ is expected to be dominated by secondary (right to the line) or primary (left to the line) electrons.  
It is intriguing that the model-derived curve for $\beta_0=0.03c$ and $A_{\rm w} \gtrsim 10^{16}$ gr cm$^{-1}$ passes close to most of the type IIn observations (filled diamonds), suggesting that the peak synchrotron luminosity is dominated
by the radiation of secondary electrons. Type IIn SNe could be therefore serve as candidate sources for the detection of hadronic acceleration in SN shocks.

Higher values of the mass loading parameter push the peak synchrotron luminosity (independently from the nature of radiating electrons) to higher values. As noted previously, the passage of $\nu_{\rm pk}$ across the 5~GHz frequency band happens at later times as $\Aw$ increases. 
Our results suggest that it is possible to probe the emission from secondaries with observations at 5~GHz if $t_{\rm pk} \simeq 300-1000$~days for shock velocities $\vo = 0.03c-0.1c$ and dense CSM with $\Aw \gtrsim 10^{16}-10^{17}$ gr cm$^{-1}$.

The steep slope of the black solid line in Fig.~\ref{fig:Lpk} (that indicates the locus where $L_{\rm pk, \nu_{\rm ff}>\vcool}=L_{\rm pk,\nu_{\rm ff}>\vcool}^{(\rm pp)}$) implies a very strong dependence of the peak luminosity at the transition time $t_{\rm pk, eq}$ on $t_{\rm pk}$. Assuming that $\epsilon_{\rm p}$ and $T_{\rm e}$ are known parameters and using eqs.~(\ref{eq:Aw-wind}), (\ref{eq:Lpk-primary-cool}), (\ref{eq:vff-wind}) and (\ref{eq:teq-wind-cool}), it can be shown that 
\begin{equation}
L_{\rm pk, eq} \propto t_{\rm pk,eq}^6 \nu^9 K_{\rm ep}^{-8}~~.
\end{equation}
Interestingly, $L_{\rm pk, eq}$ depends only on a single parameter related to the acceleration process, that is $K_{\rm ep}$. The  determination of the latter could be, therefore, possible if $L_{\rm pk, eq}$, $t_{\rm pk, eq}$ and $\nu$ were independently inferred from the observations. For example, a {  flattening} of the light curve at $\nu=\nu_{\rm pk}$ by $\Delta \chi=1$ would denote $t_{\rm pk, eq}$, as illustrated in Figs.~\ref{fig:lc} and \ref{fig:lc2}.

% \ls{comments: 1) in this section, are we assuming that the peak is controlled by cooled electrons. lets state that clearly. it may be interesting to study also the case of uncooled electrons. 2) current observing frequency is 5 GHz, but future we will have higher frequencies. if we rescale the time and frequency as $t_{pk}\nu^{3/2}$, the theoretical plot will be the same, I believe. of course, the data points will only be applicable to the case of 5 GHz, but it is ok. 3) it may be worth adding my argument for the quasi-vertical line that separates primaries and secondaries. again, I have derived it only for the case of cooled electrons, but it will be straightforward to get it for uncooled electrons.}
\begin{figure}
 \centering
  \includegraphics[width=0.49\textwidth]{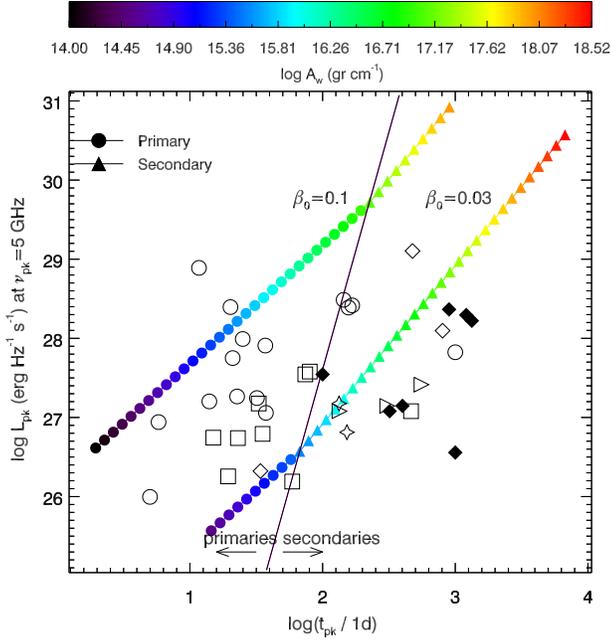}
 \caption{Peak spectral radio luminosity versus the peak time {  at a frequency} $\nu=\nu_{\rm ff}=5$~GHz and two values of the shock velocity $\vo=0.1c$  and $0.03c$. The colored symbols correspond to the analytical predictions for cooled primary electrons (circles) and secondary electrons (triangles) given by eqs.~(\ref{eq:Lpk-primary-cool}) and (\ref{eq:Lpk-secondary-cool}), respectively. Only the maximum of $L_{\rm pk,  \nu_{\rm ff}>\vcool}$  and $L_{\rm pk,  \nu_{\rm ff}>\vcool}^{(\rm pp)}$ is shown.  The black solid line is the locus of points with $L_{\rm pk,  \nu_{\rm ff}>\vcool}=L_{\rm pk,  \nu_{\rm ff}>\vcool}^{(\rm pp)}$.  For the analytical calculations,  we considered a wind-like CSM and a constant shock velocity. Color coding is used for the mass loading parameter $\Aw$, as indicated in the color bar at the top. Filled and open black symbols denote measurements from radio SNe of various types (data are from \citealt{kamble15}): Ib/c, open circles; IIb, open squares; IIn, filled diamonds; IIL, 
open triangles; and II (SN 1978K, 1981K \& 1982aa), open diamonds. Data for SN 1993J and 2013df are plotted as 
stars.}
 \label{fig:Lpk}
\end{figure}

\section{Discussion}
\label{sec:discussion}
We have presented a semi-analytical model for calculating the temporal evolution of primary and secondary particle distributions in the post-shock region of a SN forward shock propagating in a dense CSM. With the adopted formalism we were able to track the cooling history of all particles, which have been injected into the emission region up to a given radius, and to calculate the respective non-thermal radio emission. We have focused on the early phases of the SN evolution (i.e., before the Sedov phase) by presenting radio spectra and light curves for times prior to the deceleration time. The semi-analytical formalism can be, however, easily extended to the Sedov phase by adopting the adequate power-law index for the radial velocity profile. The light curves are expected to be steeper than those presented here, while at these later times, observations will probe the primary electron distribution.

So far, we have not discussed the origin of the magnetic field in the post-shock region of the SN shock. As the magnetic field of the unshocked progenitor wind is expected to be weak\footnote{Assuming magnetic flux-freezing, the magnetic field strength at a radius $r$, is estimated to be $B=10^{-3}\, B_{\star,1} \left(r_{\star,11}/r_{13}\right)^2$~G, where $B_\star$ is the strength on the stellar surface \citep[e.g.][]{barvainis87} and $r_\star$ is the stellar radius.}, magnetic field amplification is required for the particle acceleration. Non-resonant two-stream instabilities driven by cosmic-ray protons propagating ahead of the shock have been proposed as the amplification process \citep[e.g.][]{bell_lucek01, bell04, bell05, caprioli_spitkovsky14, cardillo2015}. We caution, however, that if the large-scale magnetic field of the unshocked CSM is preferentially toroidal, the resulting shock will be quasi-perpendicular (i.e., with the magnetic field perpendicular to the shock normal), and particle 
acceleration will be inefficient. Thus, the relevant assumption is that the unshocked CSM field is weak or radial \citep[e.g.][]{sironi_spitkovsky13, caprioli_spitkovsky14b}, which may be, however, questionable.

% The results presented in Sections \ref{sec:distributions}-\ref{sec:numerical} have been obtained under the  assumption of immediate cosmic-ray acceleration after the shock breakout in the CSM. This translates into an injection radius for relativistic particles (see e.g. eqs.~(\ref{eq:ptot}) and (\ref{eq:etot})) that is equal to $\rin$. This is evidently a simplifying assumption, as the formation of a collisionless shock is not expected to take place just after the shock breakout \citep[see][]{chevalier_fransson08}. 

We have mostly focused on the case of a SN shock expanding in a smooth progenitor wind. However, our calculations are applicable to a generic CSM density profile. The environment of the progenitor star does not always have a smooth density profile. Density enhancements in the CSM may occur due to various reasons, such as variable stellar winds and interactions with the companion star, in case of a binary system. Interaction of the shock wave with such density enhancements would result in radio light curves exhibiting sudden and abrupt enhancements in (or dimming of) their brightness. Indeed, several radio SNe have been observed to show such features as early as a few weeks to as late as several months. Some examples of extreme variability include SN\,1996cr \citep{Bauer2008,Meunier2013}, 2001em \citep{Schinzel2009}, 2003gk, 2004cc, 2004dk, 2004gq \citep{Wellons2012}, 2007bg \citep{Salas2013}, PTF11qcj \citep{Corsi2014} and SN\,2014C \citep{Kamble2014c}. Depending on the width and the mass of the intervening 
shell, the interaction of the shock wave with the shell  could complicate the dynamics of the interaction \citep{chevalier89,dwarka05,dwarka07,Chugai2006,pan13}.  Provided that the shock velocity can still be approximated as a power-law in radius  and that the emission from the reverse shock is negligible, our formalism may still be employed to assess 
the radio synchrotron emission resulting from the interaction of the SN shock with a dense shell of uniform density.

Currently, the modelling of radio SNe involves the emission from primary electrons that are accelerated in the vicinity of the SN shock. We explicitly showed that secondary electrons may contribute significantly to the observed radio synchrotron emission, with their contribution becoming more important at higher radio frequencies (i.e. $\gtrsim 10-100$~GHz), for denser CSM (e.g. $A_{\rm w}\gtrsim 10^{16}$~gr cm$^{-1}$) and lower shock velocities (e.g. $\vo \lesssim 10^4$~km s$^{-1}$). As the radio brightness of the SN shock depends mainly on its radius and the number of radiating electrons, the inclusion of secondary electrons in the emission process could result in different estimates of the physical parameters, such as the mass-loss rate and/or the shock radius. The latter could be measured independently with high angular resolution imaging (e.g., Very Large Baseline Interferometry -- VLBI) of nearby bright young SNe; this has been successfully carried out for about a dozen SNe and GRBs \citep{
Bietenholz2014,
Pihlstrom2007,Taylor2004}. Since the derived physical parameters from the  synchrotron emission model depend on various microphysical parameters (e.g., $\kep$ and $\ep$), independent radius measurements with VLBI observations could potentially constrain the synchrotron model \citep{deWitt2016}. Considering that the secondary electrons would contribute significantly at early times, a nearby bright and young SN will be a good candidate for direct assessment of this effect.

% 
% If the contribution of secondary electrons was to be included in the fitting of radio SNe observations, the inferred mass-loss rate required for explaining the observed luminosity would be less. It might be therefore possible to explain the SN radio emission using either primary electrons or both primary and secondary electrons, but with different estimated model parameters; this will be the subject of a future publication.

A robust prediction of our model is the {  flattening} of the radio light curve by $\Delta \chi = (w-1+\alpha)/(1-\alpha)$ at high radio frequencies (i.e. $\gtrsim 10-100$~GHz) during the transition from secondary-dominated to primary-dominated synchrotron emission. Interestingly, $\Delta \chi$ is the same for all frequencies above or below the cooling break frequency.  The change in the decay slope of the light curve is expected to be smooth (see e.g. left panels in Figs.~\ref{fig:lc} and \ref{fig:lc2}). The transition time {  at the peak frequency} should lie in the range $\sim 6-60$~d {  ($\sim$ 60$-$600~d, respectively)} for $\vo=3\times 10^4$~km s$^{-1}$ {  ($\vo=9\times10^3$ km s$^{-1}$, respectively)} and $A_{\rm w} \sim 3\times 10^{15}-3\times 10^{16}$ gr cm$^{-1}$ (see also Fig.~\ref{fig:lc_Aw}), while the peak frequency is, respectively, expected at $\sim 18-56$~GHz {  (3-10~GHz)}. However, it is not trivial to search for such breaks in existing radio data, {  especially if the shock is 
fast}
. The reason being that most of the radio observations are typically performed at 5-10 GHz, while some of the brightest SNe have been observed at higher frequencies, e.g. 40~GHz.
% Observations at  $\sim$ 300 GHz are occasionally performed only for nearby bright SNe at very late times, where an extended follow up is not possible due to the lack of good sensitivity. 
The Atacama Large Millimeter Array (ALMA), with significant improvement in sensitivity in the millimeter waveband, would be an ideal telescope for the early detection of the SNe and the precise estimate of the secondary electron contribution to the SN brightness. ALMA would also be able to search for the predicted light curve {  flattening} due to the rapid decline of the secondary electrons, thus providing a direct test of the model.

% {  [All]} role of the reverse shock
 
\section{Summary}
\label{sec:summary}
% Several type IIn SNe have been observed in the radio band, as the SN shock wave sweeps up the dense wind of the progenitor star. The SN shock wave, which is assumed to accelerate the radio-emitting electrons, can in principle accelerate high-energy protons as well.  
We have investigated the role of inelastic \pp collisions as the SN shock sweeps through the dense CSM,  focusing on the radio signatures from secondary electrons that are produced in the decay chain of charged pions. We have presented a semi-analytical one-zone model for calculating the temporal evolution of primary and secondary particle distributions in the post-shock region of the SN forward shock. With the adopted formalism, we were able to track the cooling history of all particles that have been injected into the emission region up to a given time, and calculated the radio spectra and light curves. These, upon comparison to the observations, can be used to constrain the acceleration efficiency of protons and electrons in fast supernova shocks.

We showed that, for a given SN shock expanding in a wind-like medium, secondary electrons control the early radio signatures, but their contribution decays faster than that of primary electrons, which dominate at later times. We derived the decay slope of radio light curves in the case of secondary-dominated ($t^{-\chi_2}$) and primary-dominated ($t^{-\chi_1}$) synchrotron emission and showed that, {  at a given frequency}, a break of $\Delta \chi=\chi_2-\chi_1= (w-1+\alpha)/(1-\alpha)$ is expected during the transition between the two regimes. This simplifies to $\Delta \chi=1$ for the case of wind-type CSM and a constant shock velocity. The transition time {  at the peak frequency, in particular,} was found to depend only on the mass-loading parameter for a wind-like CSM, the shock velocity $\vo$ and the ratio  $K_{\rm ep}$ as $t \sim {  190} \, K_{\rm ep,-3}^{-1} A_{\rm w, {  16}}/\beta_{\rm 0,{  -1.5}}^2$~d; the peak frequency at that time is expected at  $\nu \sim {  5.6} \, \unit{GHz} \, T_{
\rm e, 5}^{-3/4} A_{\rm w, {  16}}^{-1/2}K_{\rm ep,-3} ^{3/2}\beta_{0, {  -1.5}}^{3/2}$. Thus, radio observations performed at high frequencies  $\sim {  3-60}$~GHz and at times  ${  \sim 6-600}$~d may be used to probe the presence of secondary electrons for 
shocks with ${  \vo=9\times10^3-3\times 10^4}$~km s$^{-1}$ and dense CSM ($A_{\rm w} \sim 3\times 10^{15}-3\times 10^{16}$ gr cm$^{-1}$).  Besides the transition time, the peak time of the light curve at a given observing frequency was also shown to be a strong predictor of the relative role of secondary versus primary electrons. We showed that 
early peak times imply a low-density CSM and a primary-dominated synchrotron emission. Late peak times  (i.e., $t_{\rm pk}\gtrsim250-1000$~d) at 5~GHz, on the contrary,  suggest a fast ($\vo = 9\times 10^3 -3\times 10^4$~km s$^{-1}$) shock wave propagating in a dense medium ($\Aw \gtrsim 10^{16}-10^{17}$ gr cm$^{-1}$), where secondary electrons are likely to power the peak flux.

\section*{Acknowledgments}
{  We thank the referee for useful comments that helped to improve the manuscript.}
M.P. acknowledges support for this work by NASA through Einstein Postdoctoral 
Fellowship grant number PF3~140113 awarded by the Chandra X-ray 
Center, which is operated by the Smithsonian Astrophysical Observatory
for NASA under contract NAS8-03060. 
%A.K. .. and L.S. ...

\appendix
% \section[]{Model parameters}
% \label{sec:app0}

\section[]{Minimum momentum of accelerated particles}
\label{sec:app1}
For typical velocities of 
early supernovae, $\vs \sim 10^4 \unit{km\,s}^{-1}$, the thermal momentum of post-shock 
ions will be $p_{\rm th,p}\sim \mpr \vs$, where factors of order unity are dropped. Thus, the bulk of the shocked ions
will be non-relativistic. Recent hybrid and fully-kinetic simulations \citep[e.g.][]{park2015} suggest that the power law of accelerated particles starts at $p_{\rm p, m}\sim 3 p_{\rm th,p}$, namely the minimum momentum of the power-law distribution still falls in the non-relativistic regime. For a power-law distribution in momentum with power-law index  $2<s<3$ extending from $p_{\rm p,m}$ up to  ultra-relativistic energies most of the kinetic energy is carried by particles 
having $\gamma\sim 2$. On the contrary, for $s>3$, most of the energy is contributed by particles at $p_{\rm p, m}$. 
In fact, if $x=\gamma-1$, the particle distribution is $dN/d x\propto x^{-(s+1)/2}$ for $x\ll1$ and the usual $dN/dx\propto x^{-s}$ for $x\gg1$. Indeed, for $2<s<3$ the peak of $x^2 dN/dx$, i.e., where most of the energy resides,  occurs at $x\sim 1$, or equivalently $\gamma\sim 2$. Throughout the present study we assume that $2<s<3$.  For a constant (or mildly decelerating) shock velocity, as expected in the free expansion phase, the injection momentum $p_{\rm p,m}$ does not change over time (or slightly decreases), which implies that the fraction of  energy in non-relativistic -- yet, non-thermal -- protons, relative to the bulk of the particles, does not change (significantly) over time. Therefore, the assumption that a constant fraction of shock energy is channeled into non-thermal protons (either non-relativistic or relativistic) implies that a constant fraction of shock energy ($\ep$, in our formalism) is channeled into ultra-relativistic protons. 
\section[]{Bethe-Heitler energy loss timescale}
\label{sec:BHloss}
 Protons with Lorentz factors $\gamma_{\rm p, BH} \gtrsim  \mel  c^2/\eph = 5\times 10^5/(\eph/1\,{\rm eV})$ may lose energy through Bethe-Heitler pair production. Here, we derive the respective loss timescale, which may be recast in a similar form as $t_{\rm pp}$. 
 The inverse of the Bethe-Heitler energy loss timescale is given by \citep{blumenthal70}
% For our default case of a wind-like CSM, we find $\gpmax^{(1)}\simeq 8\times10^6 \beta_{0,-1}^2 A_{\rm w,13}^{1/2} \epsilon_{\rm B,-2}^{1/2}$. 
\eqb
t^{-1}_{\rm BH}(\gamma,r) = \frac{3}{8 \pi \gamma}\sth c\af \frac{\mel}{\mpr} \int_{2}^{\infty} \!\! d\kappa \ n_{\rm ph}\left(\frac{\kappa}{2\gamma}\right)\frac{\phi(\kappa)}{\kappa^2},
\label{eq:tpe-general}
\eqe
where $\af$ is the fine structure constant, $n_{\rm ph}$ is the differential photon number density in the shocked shell, $\kappa = 2\gamma \epsilon/\mel c^2$, $\epsilon$ is the photon energy and  $\phi(\kappa)$  is a function defined by a double integral (see eq.~(3.12) in \cite{chodorowksi92}). By approximating the photon field as $n_{\rm ph}(\epsilon) = (n_{\rm ph}/\eph)\epsilon \delta(\epsilon-\eph)$, where $n_{\rm ph}=U_{\rm ph}/\eph$, and by substitution in \eqn{tpe-general} we find
\eqb
t_{\rm BH}(\gamma,r) = \frac{4\pi \mpr}{3 \mel \af}\frac{f^{-1}(x)}{c\sth n_{\rm ph}(r)},
\label{eq:tpe-0}
\eqe
where $x\equiv2\gamma\eph$ and $f(x)\equiv \phi(x)/x^2$ that has its maximum $f_{\max} \sim 1$ at $x \simeq 47$ (see Fig.~2 in \cite{chodorowksi92}). Setting $f(x)\sim 1$, an effective cross section $\kappa_{\rm BH}\sigma_{\rm BH} \simeq 6\times 10^{-31}$~cm$^2$ can be defined and the timescale can be written as 
\eqb
t_{\rm BH}= \tau_{\rm BH}\left(\frac{r}{\rin}\right)^{\aph+2},
\eqe
	where $\tau^{-1}_{\rm BH}= \kappa_{\rm BH} \sigma_{\rm BH}  c U_0/\eph$ and \eqn{uph1} was also used. By requiring $t_{\rm acc,p}=t_{\rm BH}$ and for $\gamma>\gamma_{\rm p, BH}$, a limiting Lorentz factor of the protons can be derived
\eqb
\label{eq:gmax5}
\gpmax^{(3)} = \frac{eB_0\beta_0^2 \eph}{6 \kappa_{\rm BH} \sigma_{\rm BH} U_0 \mpr c^2}\left(\frac{r}{\rin}\right)^{2+\aph-\ab+2\alpha},
\eqe
which is similar to \eqn{gmax4}.

\section{Asymptotic expressions for the primary and secondary electron distributions}
\label{sec:app3}
Let us consider the following injection rate for primary electrons, with power-law energy slope $s=2$
\eqb
\label{eq:qo-primary}
\Qe(\gamma,r)=\frac{Q_{\rm 0e}}{\gamma^2}\left( \frac{\rin}{r}\right)^{w-2-2\alpha} H[\gamma-\gemin]H[\gemax(r)-\gamma]H[r-\rin],
\eqe
where $\gemax(r)$ is obtained from $t_{\rm acc,e}=t_{\rm syn}$ (see \eqn{gmax1}) and the factor $r^{-w+2+2\alpha}$ accounts for
a general CSM density profile and a radially dependent shock velocity (see eqs.~(\ref{eq:n}) and (\ref{eq:vshock}), respectively). 
The evolution of an electron's Lorentz factor is given by \eqn{geq}, or equivalently 
\eqb
\label{eq:g0}
\gamma_0 =\gamma\frac{r}{r_0}\left[1-\frac{K_{\rm syn}\gamma}{\qsyn}\left(\frac{\rin}{r}\right)^{\qsyn-1}\left(\left(\frac{r}{r_0}\right)^{\qsyn} -1 \right) \right]^{-1}.
\eqe
where $\qsyn=2\ab+\alpha=w-\alpha$. As we are interested in the synchrotron cooling dominated regime, the IC cooling term has dropped in the equation above. Using \eqn{etot} we find that the electron distribution is given by
\eqb
\label{eq:sol}
N_{\rm e}(\gamma,r) = \frac{Q_{\rm 0e}}{\gamma^2}\frac{\rin^{w-2-2\alpha}}{(4-w+2\alpha)r} \left(r^{4-w+2\alpha}-r_{\min}^{4-w+2\alpha} \right),
\eqe
where $r_{\min}=\max(\rin, r_{\rm cr})$ and $r_{\rm cr}$ is defined by the condition $\gemax(r_0)-\gamma_0 >0$. Using \eqn{g0} this condition leads to 
\eqb
\label{eq:general-rcr}
1> \frac{\gamma}{\gemax(r_0)}\frac{r}{r_0} + 
\frac{K_{\rm syn}\gamma}{\qsyn}\left(\frac{\rin}{r} \right)^{\qsyn-1}\left(\left(\frac{r}{r_0}\right)^{\qsyn}-1\right),
\eqe
which should be solved with respect to $r_0$.  The first term in the r.h.s. of the above relation contains information for the adiabatic cooling of electrons, whereas the second term is related to the synchrotron cooling. Since we are interested in deriving the evolution of the electron distribution when the dominant cooling process is synchrotron radiation,  we can solve \eqn{general-rcr} for $r_0$ after dropping the first term in the r.h.s. We find
% \eqb
% \label{eq:rcr}
% r_0 > r_{\rm cr} \equiv r \left(1+\frac{\qsyn}{K_{\rm syn}\gamma}\left(\frac{r}{\rin} \right)^{\qsyn-1} \right)^{-\frac{1}{\qsyn}}.
% \eqe
\eqb
\label{eq:rcr}
r_0 > r_{\rm cr} \equiv r \left(1+\frac{\qsyn}{K_{\rm syn}\gamma}\left(\frac{r}{\rin} \right)^{\qsyn-1} \right)^{-\frac{1}{\qsyn}},
\eqe
which can also be expressed as 
\eqb
\label{eq:rcr-1}
\frac{r_{\rm cr}}{r} \approx \left  \{ \begin{array}{ll}
                           1-\frac{1}{K_{\rm syn}\gamma} \left(\frac{r}{\rin} \right)^{\qsyn-1}, & \gamma \gg \gamma_{\rm br}\\ \\
                           \left[\frac{K_{\rm syn}\gamma}{\qsyn} \left(\frac{r}{\rin} \right)^{\qsyn-1}\right]^{1/\qsyn}, &  \gamma \ll \gamma_{\rm br}
                            \end{array}
\right.
\label{eq:rcr-2}
\eqe
where 
\eqb
\label{eq:gbr}
\gamma_{\rm c} \equiv  \frac{\qsyn}{K_{\rm syn}}\left(\frac{r}{\rin} \right)^{\qsyn-1} \propto r^{w-\alpha-1}.
\eqe
% 
% For $\gamma \gg \qsyn / K_{\rm syn} \left(r/\rin \right)^{\qsyn-1}$, the above expression simplifies to $r_{\rm cr} \approx r \left(1-\frac{1}{K_{\rm syn}\left( \right)^{\qsyn-1}$
We turn now to the lower integration limit $r_{\min}$, which is given by $r_{\min}\equiv\max[r_{\rm cr}, \rin]$.  For $\gamma \gg \gcool$ it can be shown that $r_{\min} = r_{\rm cr}$, whereas $r_{\min}=\rin$ for $\gamma \ll \gcool$.
% If $\rin < r_{\rm cr}$, $\gamma > \gamma_{\rm c}$, otherwise $\gamma\le \gamma_{\rm c}$, where $\gamma_{\rm c}$ is written as
% \eqb
% \gamma_{\rm c} = \frac{\qsyn}{K_{\rm syn}}\left(\frac{r}{\rin}\right)^{\qsyn-1}\left[ \left(\frac{r}{\rin} \right)^{\qsyn}-1\right]^{-1}.
% \eqe
For most parameter values and at the early phases of the SN shock evolution, the radio emitting electrons belong to the cooled part of the spectrum ($\gamma > \gcool$). For high values of $\Aw$, in particular, the electron distribution may cool due to synchrotron losses down to  $\gamma \sim\gemin$. 
Substitution of the high-$\gamma$ branch in \eqn{rcr-1} to \eqn{sol} leads to
% \eqb
% \label{eq:sol1}
% N_{\rm e}(\gamma,r)=\frac{Q_{\rm 0e}}{\gamma^2} \frac{r^{3-w+2\alpha}\rin^{w-2-2\alpha}}{(4-w+2\alpha)} 
% \left[1-\left(1+\frac{\qsyn}{K_{\rm syn}\gamma}\left(\frac{r}{\rin}\right)^{\qsyn-1}\right)^{\frac{-4+w-2\alpha}{\qsyn}} \right]
% \eqe
% In the limit where 
% \eqb
% \label{eq:small}
% \gamma \gg  \frac{\qsyn}{K_{\rm syn}}\left(\frac{r}{\rin}\right)^{\qsyn-1},
% \eqe
% the expression \eqn{sol1} simplifies to
\eqb
\label{eq:sol1}
N_{\rm e}(\gamma,r)\approx \frac{Q_{\rm 0e}\rin}{K_{\rm syn}\gamma^3}\left(\frac{r}{\rin}\right)^{2+\alpha}.
\eqe
Interestingly, the number of accelerated electrons with a given Lorentz factor does not dependent on the CSM density profile, e.g. $N_{\rm e} \propto \gamma^{-3} r^2$, for a constant shock velocity ($\alpha=0$). For $\gamma \ll \gcool$, the electron distribution is simply given by \eqn{sol} with $r_{\min}=\rin$. We remark that these results are valid as long as $\gcool \gtrsim \gemin$.

The above analysis can be directly applied to the secondary electrons, using the appropriate expression for the injection rate, which depends linearly on the distribution of relativistic protons (see e.g. eqs.~(\ref{eq:qe_pp}) and (\ref{eq:qpion})). The evolution of the proton distribution can be explicitly calculated for the generic injection profile of \eqn{Qop}, if it is dictated by the adiabatic losses (i.e., the catastrophic loss term in \eqn{pdist} can be neglected). This is a good assumption, for all radii $r\gtrsim 10^{w+\alpha-1}\beta_{0,-1}^{-2(w+\alpha-1)}$ (see discussion after \eqn{Kpp}). In this case, the proton distribution is given by
\eqb
N_{\rm p}(\gamma,r) = \frac{Q_{\rm 0p}\rin\gamma^{-s}}{s-w+2\alpha+2}\left( \frac{r}{\rin}\right)^{-w+3+2\alpha}\left[1- \left(\frac{\rin}{r}\right)^{s-w+2+\alpha}\right],
\eqe
where the term in the parenthesis is $\simeq 1$ for $r>\rin$ and $s-w+2+\alpha>0$. Substitution of the above expression in \eqn{qpion-inst} results in 
\eqb
\label{eq:qo-secondary}
\Qepp(\gamma,r)=\frac{Q_{\rm 0e}^{(\rm pp)}}{\gamma^{s}}\left(\frac{r}{\rin}\right)^{3-2w+\alpha}\!\!\!\!H[\gamma-\gemin]H[\gemax-\gamma]H[r-\rin],
\eqe
where the minimum and maximum Lorentz factors are now given by \eqn{minmax}. 
% We note that the above
% expression is a generalization of \eqn{qe-inst}, which was obtained for the default case $\beta=2$. In this case,
The electron distribution can be then explicitly derived for $s=2$. This is written as 
\eqb
\label{eq:sol-secondary}
N_{\rm e}^{(\rm pp)}(\gamma,r) = \frac{Q_{\rm 0e}^{(\rm pp)}}{\gamma^2}\frac{\rin^{2w-3-\alpha}}{(5-2w+\alpha)r} \left(r^{5-2w+\alpha}-r_{\min}^{5-2w+\alpha} \right),
\eqe
where $r_{\min}=\max(r, r_{\rm cr})$ and  $r_{\rm cr}$ is same as in \eqn{rcr}.  
% as we are interested in the part of the distribution where synchrotron cooling dominates
% over adiabatic cooling. 
Making the the same approximations as before, the asymptotic expression for the secondary electron distribution in the synchrotron cooling dominated regime is 
%This allows us to drop the first term in the r.h.s. of \eqn{general-rcr} containing % $\gemax$, which is different for primaries and secondaries.
% In the asymptotic regime defined by \eqn{small}, we thus find
\eqb
\label{eq:sol1-secondary}
N_{\rm e}^{(\rm pp)}(\gamma,r)\approx \frac{Q_{\rm 0e}^{(\rm pp)}\rin}{K_{\rm syn}\gamma^3}\left(\frac{r}{\rin}\right)^{3-w}.
\eqe
The synchrotron luminosity $L_{\nu}$ in the optically thin regime  scales as $N_{\rm e}(r) B(r)^{(s^\prime+1)/2}$ \citep[see e.g.][]{Rybicki79}, where 
$s^\prime=s+1$, if the emitting electrons belong to the cooled part of the distribution ($\gamma \gg \gcool$) and $s^\prime=s$, otherwise.
Using \eqn{B} and \eqn{sol1} for $s=2$ we find 
\eqb
 L_\nu(r) \propto  r^{2-w+3\alpha}, \, {\rm or} 
 \eqe
 \eqb
 L_\nu(t)  \propto  t^{(2-w+3\alpha)/(1-\alpha)}.
\eqe
If the luminosity at the particular frequency is dominated by
secondary synchrotron emission, we find using \eqn{B} and \eqn{sol1-secondary} that
\eqb
L^{(\rm pp)}_\nu(r) \propto r^{3-2w+2\alpha}, \, {\rm or}
\eqe
\eqb
 L^{(\rm pp)}_\nu(t) \propto t^{(3-2w +2\alpha)/(1-\alpha)}.
\eqe
\section{Comparison of thermal and non-thermal energy densities}
\label{sec:app4}
For certain SNe types that are characterized by a dense CSM,  the shock breakout is expected to take place within the progenitor's wind. After the shock breakout a collisionless shock may be formed, thus allowing the acceleration of particles \citep{katz11, murase11}. The X-ray photons produced at the shock soon after the shock breakout will be Compton down-scattered and thermalized, since $\tau_{\rm T} \sim c/\vs = 10-30$ at shock breakout \citep[e.g.][]{chevalier_irwin11, svirski12}. 
As long as $\tau_{\rm T} \gtrsim 1$, the energy density of thermal radiation may be written as \citep[see e.g.][]{murase14}:
\eqb
U_{\rm r, th} \simeq \tau_{\rm T} \frac{\xi_{\rm r}L_{\rm k}}{4\pi c r^2} \simeq \frac{c}{\vo}\frac{\xi_{\rm r}L_{\rm k}}{4\pi c \rin^2}\left(\frac{\rin}{r}\right)^{w+1},
\eqe
where $\xi_{\rm r}$ is the fraction of the SN kinetic luminosity $L_{\rm k}$ that emerges as thermal radiation and $\tau_{\rm T} = (c/\vo)\left(\rin/r\right)^{w-1}$. Typical values of the thermal electron temperature in the shocked CSM are $T^{\rm (d)}_{\rm e}\sim 10^8 -10^9$~K, although the exact values depend on whether thermal equilibrium between ions and electrons was achieved,  or not \citep{Fransson1996}. If $T_{\rm e}^{\rm (d)} \gtrsim 2 \times 10^7$~K, then thermal electrons cool via thermal bremsstrahlung, producing X-ray radiation. Hard ($\gtrsim 1-2$~keV) X-rays can escape without severe attenuation, if $\tau_{\rm T}\lesssim 1$. For a wind-like density profile ($w=2$),  the energy density of X-ray photons is given by
\eqb
U_{\rm X, ff} = \frac{L_{\rm ff}}{4\pi c \rin^2}\left(\frac{\rin}{r}\right)^2,
\eqe
where 
\eqb
L_{\rm ff} \simeq 1.5 \times 10^{40}\,\frac{A_{\rm w, 16} \left(T^{\rm (d)}_{\rm e, 8}\right)^{1/2}}{\beta_{0, -1.5}}\frac{\rin}{r} \unit{erg \, s}^{-1},
\eqe
where we used eq.~(5.15b) by \citet{Rybicki79}, assuming an ionized CSM composed of pure hydrogen. It is interesting to compare $U_{\rm r, th}$ and $U_{\rm X, ff}$ with the non-thermal particle and magnetic energy densities in the downstream region of the SN forward shock. This is exemplified in Fig.~\ref{fig:fig_app} for $L_{\rm k}=10^{42}$~erg s$^{-1}$, $\xi_{\rm r}=0.1$ and $T_{\rm e}^{(d)}=10^8$~K. All other parameters are the same as in Fig.~\ref{fig:densities}. At least for this indicative example, it is the magnetic energy density that dominates over $U_{\rm r, th}$ and $U_{\rm X, ff}$ at almost all radii. Thus, synchrotron cooling is expected to be the main energy loss process for electrons. 
\begin{figure}
 \centering
 \includegraphics[width=0.5\textwidth]{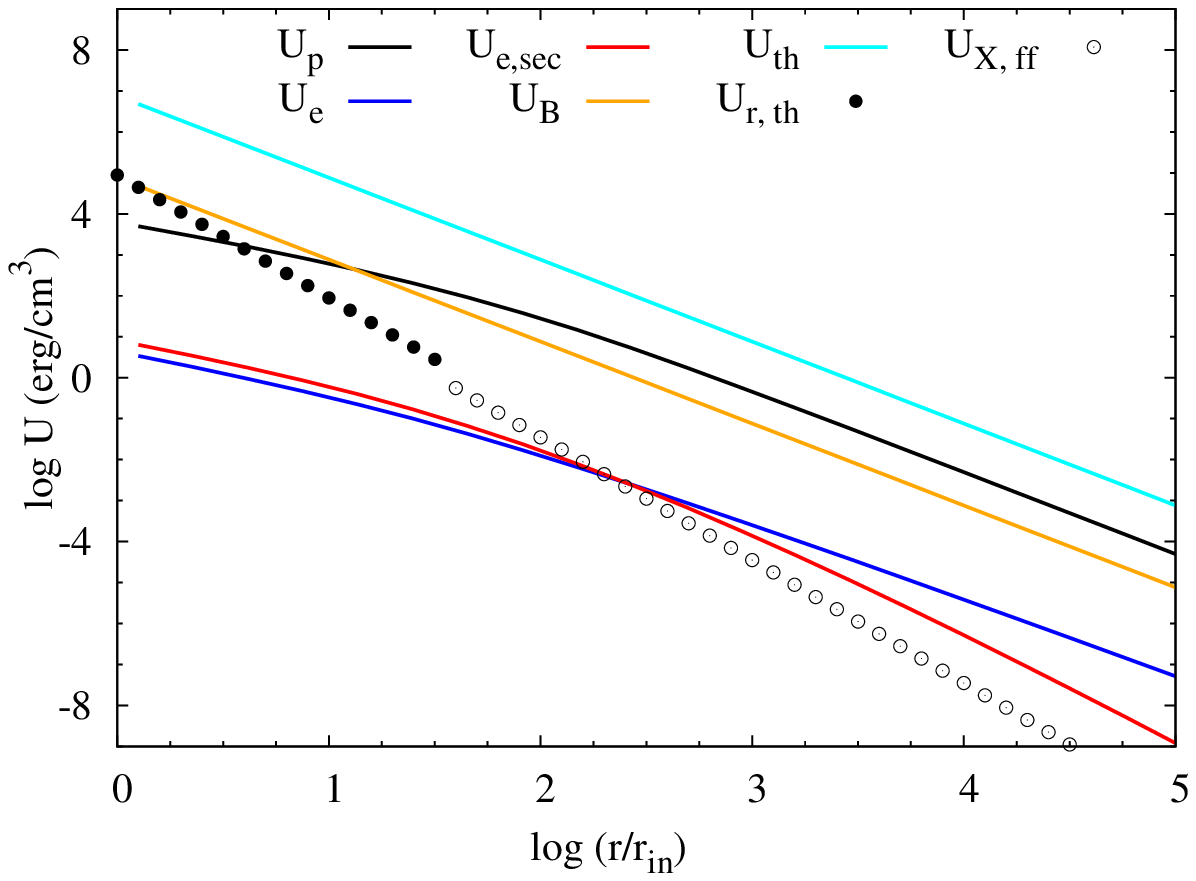}
 \caption{Same as in Fig.~\ref{fig:densities}. In addition, $U_{\rm r, th}$ (filled symbols) and $U_{\rm X,ff}$ (open symbols) are shown for radii where
 $\tau_{\rm T} \ge 1$ and $< 1$, respectively. Other parameter used are:   $L_{\rm k}=10^{42}$~erg s$^{-1}$, $\xi_{\rm r}=0.1$ and $T_{\rm e}^{(d)}=10^8$~K. }
 \label{fig:fig_app}
\end{figure}
Yet, the aforementioned radiation fields can, in principle, contribute to electron cooling via inverse Compton scattering. 
$U_{\rm r, th}$ turns out to be a more important source for electron cooling, since most of the inverse Compton scatterings between the electrons and X-ray bremsstrahlung photons take place in the Klein-Nishina regime. We note that these additional cooling channels for electrons were not taken into account in our calculations; we would have to rely on many assumptions regarding the thermal processes, which were not the focus of this study.  

% \section[]{}
% \label{sec:app2}
\bibliographystyle{mn2e} % style mn2e.bst
\bibliography{pairs}

\end{document}